\documentclass{aa}

\usepackage{graphicx}
\usepackage{epsfig}
\usepackage{natbib}
\usepackage{pifont}

\usepackage[varg]{txfonts}
\usepackage[font=footnotesize,labelfont={bf,sc,small},figurename=Fig.]{caption}
\usepackage[colorlinks,allcolors=blue]{hyperref}
\usepackage{subfig}
\usepackage{rotate}
\usepackage{appendix}

\usepackage{varioref}

\newcommand\rs[1]{_\mathrm{#1}}

\newcommand{\casa}{Cas~A}

\begin{document} 

\title{The fully developed remnant of a neutrino-driven supernova:}
\subtitle{Evolution of ejecta structure and asymmetries in SNR Cassiopeia A}

\author{S.\ Orlando\inst{1}
   \and A.\ Wongwathanarat\inst{2}
   \and H.-T.\ Janka\inst{2}
   \and M.\ Miceli\inst{3,1}
   \and M.\ Ono\inst{4,5}
   \and S.\ Nagataki\inst{4,5} \and \\
   F.\ Bocchino\inst{1}
   \and G.\ Peres\inst{3,1}
}

\offprints{S. Orlando}

\institute{INAF -- Osservatorio Astronomico di Palermo, Piazza del Parlamento 1, I-90134 Palermo, Italy\\ 
\email{salvatore.orlando@inaf.it}
\and Max-Planck-Institut f\"ur Astrophysik, Karl-Schwarzschild-Str. 1, D-85748 Garching, Germany
\and Dip. di Fisica e Chimica, Universit\`a degli Studi di Palermo, Piazza del Parlamento 1, 90134 Palermo, Italy
\and Astrophysical Big Bang Laboratory, RIKEN Cluster for Pioneering Research, 2-1 Hirosawa, Wako, Saitama 351-0198, Japan
\and RIKEN Interdisciplinary Theoretical \& Mathematical Science Program (iTHEMS), 2-1 Hirosawa, Wako, Saitama 351-0198, Japan
  }

\date{Received date / Accepted date}

\abstract
{The remnants of core-collapse supernovae (SNe) are probes of the
physical processes associated with their parent SNe.}
{Here we aim at exploring to which extent the remnant keeps memory
of the asymmetries that develop stochastically in the neutrino-heating
layer due to hydrodynamic instabilities (e.g., convective overturn
and the standing accretion shock instability; SASI) during the
first second after core bounce.}
{We coupled a 3D hydrodynamic model of a neutrino-driven SN explosion
which has the potential to reproduce the observed morphology of the
Cassiopeia A (\casa) remnant, with 3D (magneto)-hydrodynamic
simulations of the remnant formation.  The simulations cover $\approx
2000$~years of expansion and include all physical processes relevant
to describe the complexities in the SN evolution and the subsequent
interaction of the stellar debris with the wind of the progenitor
star.}
{The interaction of large-scale asymmetries left from the earliest
phases of the explosion with the reverse shock produces, at the age
of $\approx 350$~years, an ejecta structure and a remnant morphology
which are remarkably similar to those observed in \casa. Small-scale
structures in the large-scale Fe-rich plumes created during the
initial stages of the SN, combined with hydrodynamic instabilities
that develop after the passage of the reverse shock, naturally
produce a pattern of ring- and crown-like structures of shocked
ejecta. The consequence is a spatial inversion of the ejecta layers
with Si-rich ejecta being physically interior to Fe-rich ejecta.
The full-fledged remnant shows voids and cavities in the innermost
unshocked ejecta (in most cases physically connected with ring-like
features of shocked ejecta in the main shell) resulting from the
expansion of Fe-rich plumes and their inflation due to the decay
of radioactive species. The asymmetric distributions of $^{44}$Ti
and $^{56}$Fe (mostly concentrated in the northern hemisphere,
pointing opposite to the kick velocity of the neutron star) and
their abundance ratio are both compatible with those inferred from
high-energy observations of Chandra and NuSTAR. Finally, the simulations
show that the fingerprints of the SN can still be evident $\approx
2000$~years after the explosion.}
{The main asymmetries and features observed in the ejecta
distribution of \casa\ can be explained by the interaction of the
reverse shock with the initial large-scale asymmetries that developed
from stochastic processes (convective overturn and SASI activity)
that originate during the first seconds of the SN blast.}


\titlerunning{The fully developed remnant of a neutrino-driven supernova}
\authorrunning{S. Orlando et~al.}

\maketitle

\section{Introduction}
\label{sec:intro}

Core-collapse supernovae (SNe), the final fate of massive stars,
play a major role in the dynamical and chemical evolution of galaxies
by driving, for instance, the chemical enrichment and the heating
of the diffuse interstellar gas. However, despite the central role
played by SNe in the galactic ecosystem, studying the physical
processes that govern SN engines is a rather difficult task due to
the rarity of these events in our galaxy (about one every $\approx
50$~years) and due to severe limitations in observations of
extragalactic SNe (because of their large distances from us). This
makes it extremely challenging to extract key information on the
explosion processes from observations in the immediate aftermath
of a SN, namely when the structure of the rapidly expanding stellar
debris keep full memory of the explosion mechanism.

Nonetheless, there has been a growing consensus in the literature
that the fingerprints of SN engines can still be found hundreds to
thousands of years after the explosion in the leftover of SNe, the
supernova remnants (SNRs). These appear as extended sources which
emit both thermal and non-thermal emission in different spectral
bands. Spatially resolved spectroscopy has allowed astronomers to
investigate in detail the structure of nearby SNRs and distribution
of chemical elements in their interior, revealing the complexity
of their morphology and fine-scale ejecta structures that are
impossible to observe in unresolved extragalactic sources
(e.g.~\citealt{2017hsn..book.2211M}).  In a few particular cases,
a three-dimensional (3D) reconstruction of the chemical distribution
and structure of stellar debris has been possible as, for instance,
in the SNRs Cassiopeia A (\casa; e.g.  \citealt{2010ApJ...725.2038D,
2013ApJ...772..134M, 2014Natur.506..339G, 2015Sci...347..526M,
2017ApJ...834...19G}), SN~1987A (e.g.  \citealt{2017ApJ...842L..24A,
2019ApJ...886...51C}) and, more recently, N132D (e.g.
\citealt{2020ApJ...894...73L}). These studies have revealed very
asymmetric distributions of the ejecta that might reflect pristine
structures and features of the parent SN explosions, possibly arising
from (magneto)-hydrodynamical (MHD/HD) instabilities developed at
the launching of the anisotropic blast wave (e.g.
\citealt{2006A&A...453..567M, 2011ApJ...732..114L, 2018SSRv..214...44L,
2020ApJ...889..144H}).  Particularly relevant are the distributions
of radioactive nuclei freshly synthesized during the SN, such as
$^{56}$Ni and $^{44}$Ti, and their decaying products, as $^{56}$Co
and $^{56}$Fe, which originate from the innermost regions of the
star where the explosion was launched.  Hundreds of years after the
SN, these nuclei might still encode the fingerprints of the physical
processes dominating the earliest phases of the SN in asymmetries
that occurred during the explosion, thus allowing us to probe the
physics of SN engines.

Information on the explosion dynamics and matter mixing that might
be extracted from observations of SNRs can be essential for
constraining sophisticated models that describe the complex phases
of SN evolution (e.g. \citealt{2013A&A...552A.126W, 2016ARNPS..66..341J,
2017hsn..book.1095J, 2018ApJ...865...81O, 2019MNRAS.485.3153B}).
SN models, however, describe the early phases from the core-collapse
up to a timescale of days only, although the age of young nearby
SNRs is typically of hundreds of years\footnote{A unique exception
is SN~1987A whose remnant was resolved about ten years after the
SN event thanks to observations in different wavelength bands, from
radio (e.g. with the {\it Australia Telescope Compact Array},
\citealt{2010ApJ...710.1515Z}), to  infrared (e.g. with {\it Spitzer},
\citealt{2010ApJ...722..425D}), to optical (e.g.  with {\it Hubble
Space Telescope}, \citealt{2000ApJ...537L.123L, 2011Natur.474..484L}),
to X-ray bands (e.g. with {\it Chandra} and {\it XMM-Newton},
\citealt{2006A&A...460..811H, 2013ApJ...764...11H, 2016ApJ...829...40F}).},
at which age the remnants have interacted with the
circumstellar/interstellar medium (CSM/ISM). This makes it
very difficult to disentangle the effects of the CSM/ISM interaction
on the observed remnant from those of the initial phases of the
explosion itself.

A strategy to link observed asymmetries and geometry of the SNR's
bulk ejecta with core-collapse SN simulations is to perform long-term
simulations that evolve core-collapse SNe to the age of SNRs (hundreds
or thousands of years), and compare the results of these simulations
with observations of SNRs. However, this is a rather challenging
task that requires a multi-scale, multi-physics, and multi-dimensional
approach to describe: the very different temporal and spatial scales
involved through the different phases of evolution, the structure
and chemical stratification of the progenitor star at collapse, the
explosive nucleosynthetic processes, the effects of post-explosion
anisotropies (inherently 3D), the interaction of the SNR with the
surrounding inhomogeneous (magnetized) medium, and the synthesis
of emission in different wavelength bands (necessary for comparison
of the model results with observations). The simulations have to
follow the entire life cycle of elements from the synthesis in the
progenitor star, through the reprocessing by nuclear reactions
during the SN, and the subsequent mixing of chemically inhomogeneous layers
of the ejecta with the CSM. An
additional difficulty stems from the need to disentangle the effects
of the SN explosion and of the structure of the progenitor star,
from those of the interaction of the blast with the inhomogeneous
CSM/ISM. First attempts of long-term 3D MHD/HD simulations confirmed
that the above approach is very effective in gaining a deep physical
insight of the phenomena that occurred during a SN (e.g.
\citealt{2015ApJ...810..168O, 2016ApJ...822...22O, 2017ApJ...842...13W,
2019A&A...622A..73O, 2019ApJ...877..136F, 2020ApJ...888..111O,
2020A&A...636A..22O, 2020ApJ...895...82V, 2020A&A...642A..67T,
2020arXiv200801763G}).

Here, we present the complete 3D evolution of a neutrino-driven SN
explosion from the core-collapse to the development of its full-fledged
remnant interacting with the CSM. To this end, we coupled an elaborate
3D hydrodynamic model of a neutrino-driven explosion
(\citealt{2017ApJ...842...13W}) with state-of-the-art 3D MHD/HD
simulations of the remnant formation (\citealt{2016ApJ...822...22O}).
Going beyond previous studies, the present one represents a significant
step forward that allows us to link, for the first time, modeling
attempts that have been carried out independently so far, because
they are either constrained to the early phase of the SN up to days
only (e.g. \citealt{2017ApJ...842...13W}), or starting the long-time
remnant evolution with artificial initial conditions (e.g.
\citealt{2016ApJ...822...22O}). This allowed us to describe the
development and evolution of remnant anisotropies self-consistently,
as a result of the neutrino-driven mechanism, and to identify ejecta
structures of the SNR that encode the imprint of large-scale
asymmetries left from the earliest moments of the explosion.

Our study focusses on a SN model that produces a remnant compatible
with the observed structure of \casa, one of the best studied young SNRs of
our galaxy (at a distance of $\approx 3.4$~kpc;
\citealt{2014ApJ...789....7L}). The 3D structure of \casa\ has been
characterized in excellent details by multi-wavelength observations
(e.g.  \citealt{2010ApJ...725.2038D, 2013ApJ...772..134M,
2015Sci...347..526M, 2014Natur.506..339G, 2017ApJ...834...19G}).
One of the outstanding characteristics of the \casa\ morphology is
its overall clumpiness and the presence of large-scale anisotropies
(most notably three Fe-rich regions, ``crowns'' and ring-like
structures, voids reaching even into the innermost unshocked ejecta,
and Si-rich ``sprays'' also referred to in the literature as
jet-like features). Since observations suggest that the remnant is
still expanding through the spherically symmetric wind of the
progenitor star (e.g.~\citealt{2014ApJ...789....7L}), the large-scale
anisotropies evident in the remnant are most likely due to processes
associated to the SN explosion. All the above
considerations make \casa\ an attractive laboratory for studying
the SN-SNR connection. Our aim is to investigate how \casa's final
morphology reflects the characteristics of the parent SN explosion
and, in particular, the asymmetries that developed by nonradial
hydrodynamic instabilities connected to the onset of the explosion.

The paper is organized as follows. In Sect.~\ref{sec:setup} we
describe the initial neutrino-driven SN model, the SNR model, and
the numerical setup; in Sect.~\ref{sec:results} we discuss the
results for the modeled structure of the ejecta in the full-fledged
remnant in comparison with observations of \casa; and in
Sect.~\ref{sec:concl} we draw our conclusions; in Appendices
\ref{app:vel_distr}$-$\ref{app:nh_em}, we discuss the effects of
radioactive decay and magnetic field on the evolution of the remnant
and on the spatial distribution of the ejecta.

\section{Problem description and numerical setup}
\label{sec:setup}

Our computational strategy is analogous to that described in
\cite{2020A&A...636A..22O} and consists in the coupling between
elaborate 3D HD models of the SN explosion and state-of-the-art 3D
MHD/HD simulations of the remnant formation. For the purpose of the
present paper, we considered a core-collapse SN simulation from the
blast-wave initiation by the neutrino-driven mechanism, computed
until about one day after the SN initiation. This model developed
an asymmetric morphological structure that is compatible with that
of \casa\ (\citealt{2017ApJ...842...13W}). This
simulation provided the initial conditions for our 3D SNR simulations
soon after the shock breakout (see Sect.~\ref{sec:sn_model}). Then
we followed the transition from the early SN phase to the emerging
SNR and the subsequent expansion of the remnant through the wind
of the stellar progenitor (see Sect.~\ref{sec:snr_model}) as in
\cite{2016ApJ...822...22O}.

We note that our simulations are not expected to reconstruct every
detail of the structure of \casa\ since the 3D SN model adopted
here was selected from a set of models of \cite{2015A&A...577A..48W}
and no fine-tuning was performed, neither on the progenitor star
nor on the explosion properties, to match the morphology and structure
of \casa\ (see discussion in \citealt{2017ApJ...842...13W}).
Nevertheless, our simulations allowed us to investigate if fundamental
chemical, physical, geometric properties of \casa\ can be explained
in terms of the processes associated to the asymmetric beginning
of a SN explosion and a sequence of subsequent hydrodynamic
instabilities that lead to fragmentation and mixing in the ejecta.
The simulations were extended till the age of 2000 years
to explore how and to which extent the remnant keeps memory of the
anisotropies that emerged from violent non-radial flows during the
early moments after the core-collapse.

\subsection{The initial neutrino-driven supernova model}
\label{sec:sn_model}

The initial conditions of our SNR simulations are provided by a SN
model presented and fully analyzed in \cite{2017ApJ...842...13W},
where it was denoted as W15-2-cw-IIb. Despite the fact that it was not fine-tuned
for a perfect match with \casa, its most notable characteristics
are the ability to produce, at about one day after the core-collapse,
spatial distributions of $^{44}$Ti and $^{56}$Ni that are compatible
with \casa. In particular the three pronounced Ni-rich fingers may
correspond to the extended shock-heated Fe-rich regions observed
in \casa. Moreover, the model predicts that most of the $^{44}$Ti
moves in the direction opposite to the kick velocity of the central
compact object\footnote{The evidence that species as Si, S, Ar, and Ca are
predominantly ejected opposite to the direction of neutron star
motion is common to many core-collapse SNRs, as inferred from the
analysis of a sample of remnants observed with {\em Chandra} and
{\em XMM-Newton} (e.g. \citealt{2017ApJ...844...84H,
2018ApJ...856...18K}).} (CCO).  These findings support the idea
that \casa\ is the remnant of a neutrino-driven SN explosion and
that its structure is the result of hydrodynamic instabilities which
have developed in the aftermath of the core-collapse. In the
following, we summarize the main features of this model (see
\citealt{2017ApJ...842...13W} for details).

Model W15-2-cw-IIb was developed on the basis of the results obtained
in a series of previous studies (\citealt{2010ApJ...725L.106W,
2013A&A...552A.126W, 2015A&A...577A..48W}), where SN models for
different progenitor stars and explosion energies were presented.
In particular, W15-2-cw-IIb derives from one of these previous
models: W15-2-cw (\citealt{2015A&A...577A..48W}). The latter considers
a $15\,M_{\odot}$ progenitor star (denoted as s15s7b2 in
\citealt{1995ApJS..101..181W} and W15 in \citealt{2015A&A...577A..48W})
which is characterized by a massive H envelope. The 3D
supernova simulation of model W15-2-cw was started at about 15
milliseconds after core bounce, and neutrino-energy deposition was
tuned to power an explosion with an energy of $1.5\times 10^{51}$~erg
$= 1.5$~bethe $= 1.5$~B (see \citealt{2013A&A...552A.126W,
2015A&A...577A..48W}). After the shock breakout at the stellar
surface, this energy is almost entirely the kinetic energy of the
ejecta, being the internal energy only a small percentage of the
total energy. The evolution was followed until shock breakout at
about 1 day after the core-collapse.

On the other hand, observations of light echoes showed that \casa\
is the remnant of a type IIb SN (\citealt{2008Sci...320.1195K,
2011ApJ...732....3R}) thus implying that its progenitor star has
shed almost all of its H envelope (see also
\citealt{1976ApJS...32..351K, 1978ApJ...219..931C}). In the light
of this, the original stellar model used for W15-2-cw was modified by
removing artificially most of its H envelope down to a rest
of $\approx 0.3\,M_{\odot}$ (the modified stellar model is termed
W15-IIb in \citealt{2017ApJ...842...13W}); in this way, the stellar
radius of the modified progenitor star reduces to $R_* = 1.5\times
10^{12}$~cm ($R_* = 21.4\,R_{\odot}$). Since the early phases in the
SN evolution are not affected by the structure of the H
envelope as long as the blast wave does not interact with it, model
W15-2-cw-IIb was calculated using as initial conditions the output
of model W15-2-cw at a post-bounce time of $t = 1431$~s, namely when
the SN shock has nearly reached $R_*$.

Model W15-2-cw is characterized by large-scale asymmetries in the
distribution of Fe-group elements, Si, and O. These
asymmetries were not imposed by hand but developed stochastically
mainly by convective overturn in the neutrino-heating layer (the
dominant hydrodynamic instability in this simulation during the
first second after core bounce; \citealt{2013A&A...552A.126W}). The
onset of convection was triggered at the beginning of the simulation
by random seed perturbations with an amplitude of 0.1\% (and
cell-to-cell variations) of the radial velocity. In model W15-2-cw,
the shock is strongly decelerated in the H envelope. This
produces a crossing of the density and pressure gradients in a dense
shell of ejecta that builds up between the forward shock and a
reverse shock that moves backward into the He layer and slows down
the swept-up ejecta. The crossing density and pressure gradients
trigger the growth of secondary Rayleigh-Taylor (RT) instability, so that 
initial, large-scale, metal-containing plumes and asymmetric
structures are massively affected by fragmentation into smaller
filaments and associated mixing with mantle and envelope material.
In contrast, in model W15-2-cw-IIb, thanks to the nearly complete
removal of the H envelope, the supernova shock moves outward
without strong deceleration in an extended H envelope of a stellar
progenitor. Therefore RT instability does not occur at the He/H
interface, and the large-scale asymmetries that emerge from the
early phases of the explosion, which contain high concentrations of
radioactive species, most notably $^{56}$Ni and $^{44}$Ti, evolve
without further deceleration and without fragmentation and mixing
with envelope matter.

The SN model accounts for the effects of gravity, both self-gravity
of the SN ejecta and the gravitational field of a central point
mass representing the neutron star that has formed after core bounce
at the center of the explosion. During the long-time simulation of
the supernova
explosion the neutron star was excluded from the computational grid
and replaced by a central point mass (our ``CCO'') and an inner
grid boundary with an outflow condition. The radius of this grid
boundary was successively increased for computational efficiency
of the time stepping in the course of the simulation (see
\citealt{2015A&A...577A..48W}). The fallback of material on the CCO
(the total mass that falls through the inner grid boundary and is
assumed to accrete onto the CCO during the evolution) was added to
the point mass. The model describes the stellar plasma by adopting
the Helmholtz equation of state (\citealt{2000ApJS..126..501T}),
which includes contributions from blackbody radiation, ideal Boltzmann
gases of a defined set of fully ionized nuclei, and degenerate or
relativistic electrons and positrons.

The SN model also traces the products of explosive nucleosynthesis
that took place during the first seconds of the explosion (thus in
model W15-2-cw), for which purpose a small $\alpha$-network was
employed (see \citealt{2013A&A...552A.126W, 2015A&A...577A..48W}).
This nuclear reaction network includes 11 species: protons ($^{1}$H),
$^{4}$He, $^{12}$C, $^{16}$O, $^{20}$Ne, $^{24}$Mg, $^{28}$Si,
$^{40}$Ca, $^{44}$Ti, $^{56}$Ni, and an additional ``tracer nucleus''
$^{56}$X. The latter represents Fe-group species synthesized in
neutron-rich environments; such conditions are found in neutrino-heated
ejecta (see \citealt{2017ApJ...842...13W} for details). As discussed
in \cite{2013A&A...552A.126W}, this network is very useful in
providing rough information on the nucleosynthesis products obtained
in the earlier phases of SN evolution, but it gives inaccurate
estimates for the yields of individual nuclear species; for instance,
it leads to a significant overestimation of the $^{44}$Ti production.
A more accurate calculation would require a much larger network.
However, for the purpose of the paper, we are mainly interested in
the spatial distributions of different chemical elements relative
to each other and do not put very much weight on the absolute amounts
of the nucleosynthesized masses. For this reason the small network
serves our needs sufficiently well. We tested this by comparing the
yields and their distributions with the results of big network
calculations (see \citealt{2017ApJ...842...13W}). Although the
relative production of $^{44}$Ti and $^{56}$Ni depends on local
conditions of temperature, density, and electron fraction in a
complex way, which can be captured in detail only by large nuclear
reaction networks (see \citealt{pllumbi2015, 2020ApJ...895...82V}),
turbulent mass motions and multi-dimensional mixing processes
(through RT and Kelvin-Helmholtz - KH - instabilities) during the
explosion led to satisfactory overall agreement of the final spatial
distributions of the two nuclear species, i.e., the dominant,
large-scale structures of their distributions revealed little
differences when their production was following by the small network
or by the large network in a post-processing step.

\begin{figure}[!t]
  \begin{center}
    \leavevmode
        \epsfig{file=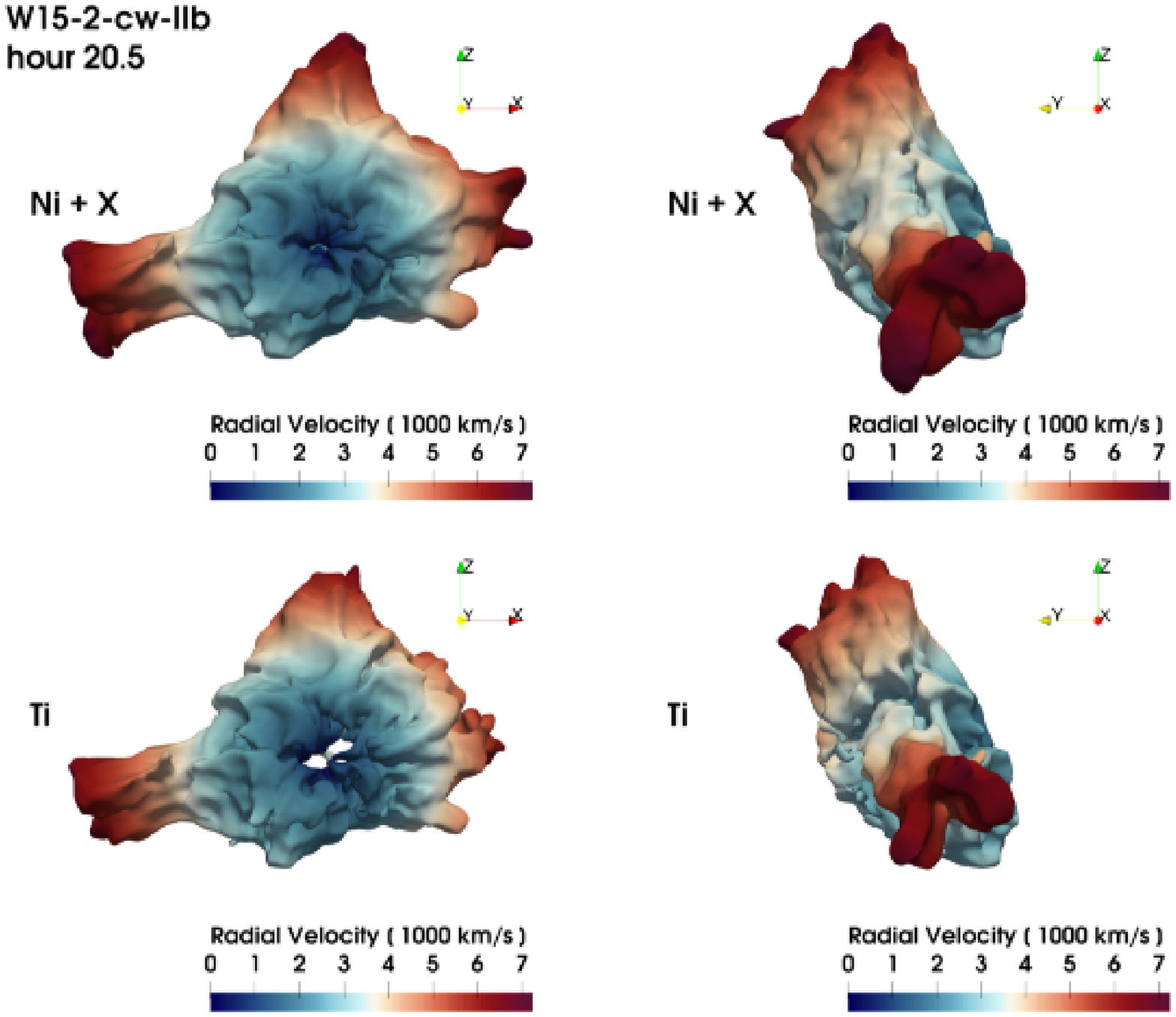, width=9cm}
	\caption{Isosurfaces of mass fractions of $(^{56}{\rm
	Ni} + 0.5\, ^{56}{\rm X})$ (upper panels) and $^{44}$Ti
	(lower panels) from different viewing angles for model
	W15-2-cw-IIb at $t = 20.5$ hours after core bounce (from
	\citealt{2017ApJ...842...13W}). The isosurfaces correspond
	to a value of mass fraction which is at 30\% of the peak
	mass fraction of each species. The colors give the radial
	velocity in units of 1000 km s$^{-1}$ on the isosurface; the
	color coding is defined at the bottom of each panel.}
  \label{ini_cond}
\end{center} \end{figure}

The output of model W15-2-cw-IIb at $t = 73940$~s ($\approx 20.5$~hr)
after core bounce was used as initial condition for the structure
and chemical composition of the ejecta in our 3D MHD/HD simulations of
the SNR (see Sect.~\ref{sec:snr_model}). To produce results which
can be easily compared with observations of \casa\, we rotated the
original system about the three axes to roughly point the modeled
Ni-rich fingers toward the extended Fe-rich regions observed
in \casa, namely $i\rs{x} = -30^{\rm o}$, $i\rs{y} = 70^{\rm o}$,
$i\rs{z} = 10^{\rm o}$. We assumed this orientation in the whole
paper; the Earth vantage point lies on the negative $y$-axis.
Fig.~\ref{ini_cond} shows the resulting distributions of $^{44}$Ti
and the Fe-group elements $(^{56}{\rm Ni} + 0.5\, ^{56}{\rm X})$
(hereafter [Ni + X] for brevity). As discussed in
\cite{2017ApJ...842...13W}, Ti and Ni are both mostly
concentrated in the northern hemisphere, opposite to the direction
of the CCO kick velocity pointing southward toward the observer
(see also \citealt{2010ApJ...725L.106W}). The distributions
of the two species are closely linked to each other with most of
their masses concentrated in widely distributed clumps and knots
of different sizes.

\begin{figure}[!t]
  \begin{center}
    \leavevmode
        \epsfig{file=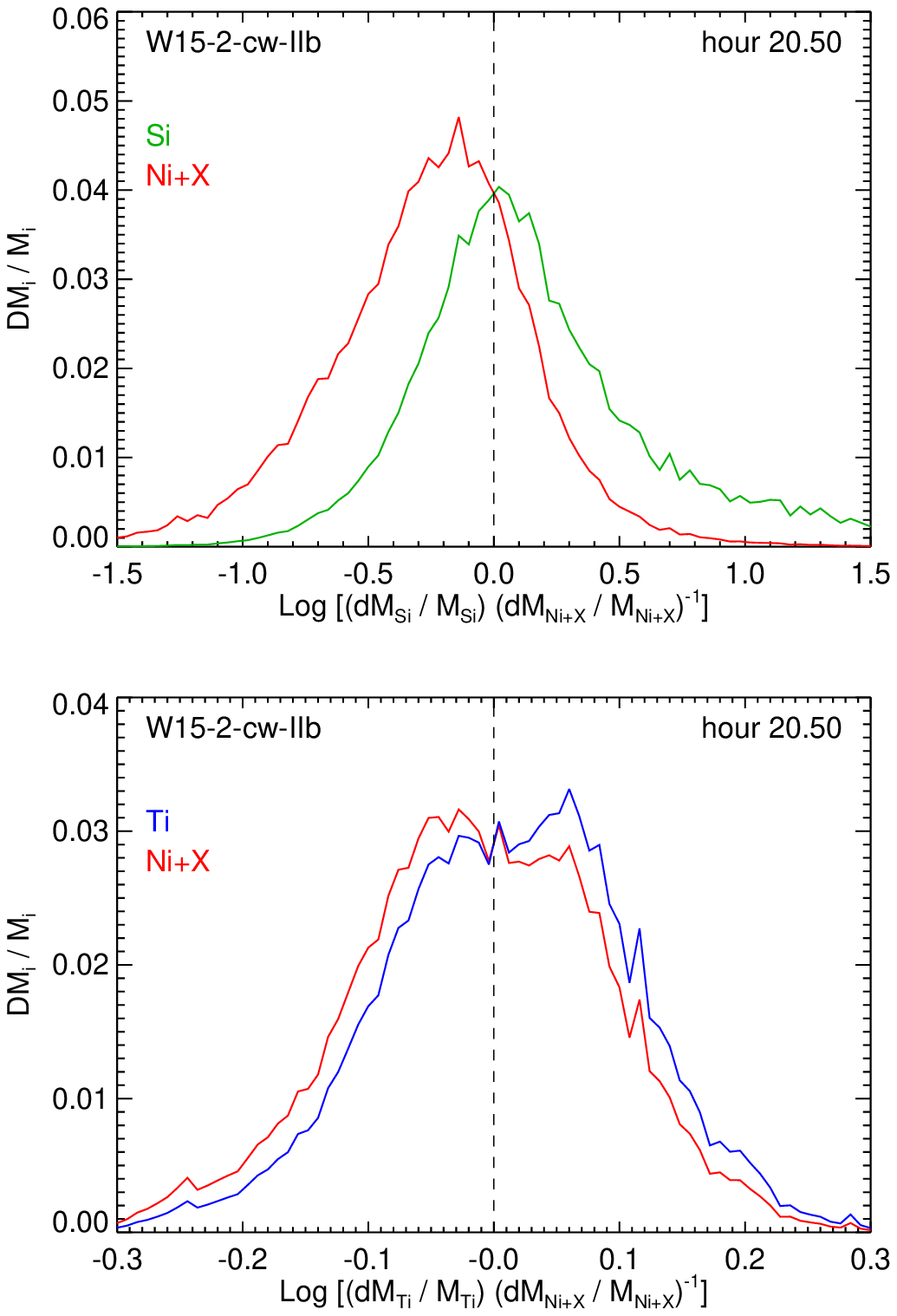, width=8.5cm}
	\caption{Mass distribution of selected species versus the
	ratio $R\rs{i,Ni+X} = \log [(dM\rs{i} / M\rs{i})(dM\rs{Ni
	+ X} / M\rs{Ni + X})^{-1}]$ (where $i$ stands for Si or Ti)
	at $t \approx 20.5$~hr after core bounce for model W15-2-cw-IIb
	(\citealt{2017ApJ...842...13W}). The quantity
	[Ni + X] denotes $(^{56}{\rm Ni} + 0.5\, ^{56}{\rm X})$.}
  \label{fe_vs}
\end{center} \end{figure}

The correlation in the spatial distributions of two elements $i$
and $j$ can be further investigated by considering the mass
distribution of the two species, $DM\rs{i} / M\rs{i}$ and $DM\rs{j}
/ M\rs{j}$, versus their abundance ratio, in log scale, formulated
as $R\rs{i,j} = \log [(dM\rs{i} / M\rs{i})(dM\rs{j} / M\rs{j})^{-1}]$,
where $DM\rs{i}$ ($DM\rs{j}$) is the mass of the $i$-th ($j$-th)
element in the range $[R\rs{i,j}; R\rs{i,j} + dR\rs{i,j}]$,
$dM\rs{i}$ ($dM\rs{j}$) is the mass of the $i$-th ($j$-th) element
in each grid cell, and $M\rs{i}$ ($M\rs{j}$) is the total mass of
the $i$-th ($j$-th) element. Fig.~\ref{fe_vs} shows the mass
distributions of Si, Ti and [Ni + X] versus $R\rs{i,Ni+X}$,
considering 300 bins in the selected range of $R\rs{i,Ni+X}$.
Negative values of $R\rs{i,Ni+X}$ indicate relatively higher
concentrations of [Ni + X], while positive values mean higher
concentrations of the $i$-th element. These distributions are useful
to investigate the relative abundances of plasma after interaction
with the reverse shock (see Sect.~\ref{sec:results}). The figure
shows that the distributions of Ti and [Ni + X] are very
similar to each other, narrow and symmetric with respect to
$R\rs{Ti,Ni+X} = 0$ (lower panel in Fig.~\ref{fe_vs}), thus confirming
that most of the Ti and [Ni + X] coexist in the mass-filled
volume. However, the spread of values in the range $-0.3 < R\rs{Ti,Ni+X}
< 0.3$ reflects some local differences of the abundance ratios of
these species with ejecta clumps which are either Ti- or Ni-rich
(as also evident from Fig~\ref{ini_cond}). In the case of Si,
the mass distributions are much broader (in the range $-2 <
R\rs{Si,Ni+X} < 2$) and significantly asymmetric with respect to
$R\rs{Si,Ni+X} = 0$ (upper panel in Fig.~\ref{fe_vs}).  This reflects
quite large differences of the abundance ratios of these species
as expected.

\begin{figure*}[!t]
  \begin{center}
    \leavevmode
        \epsfig{file=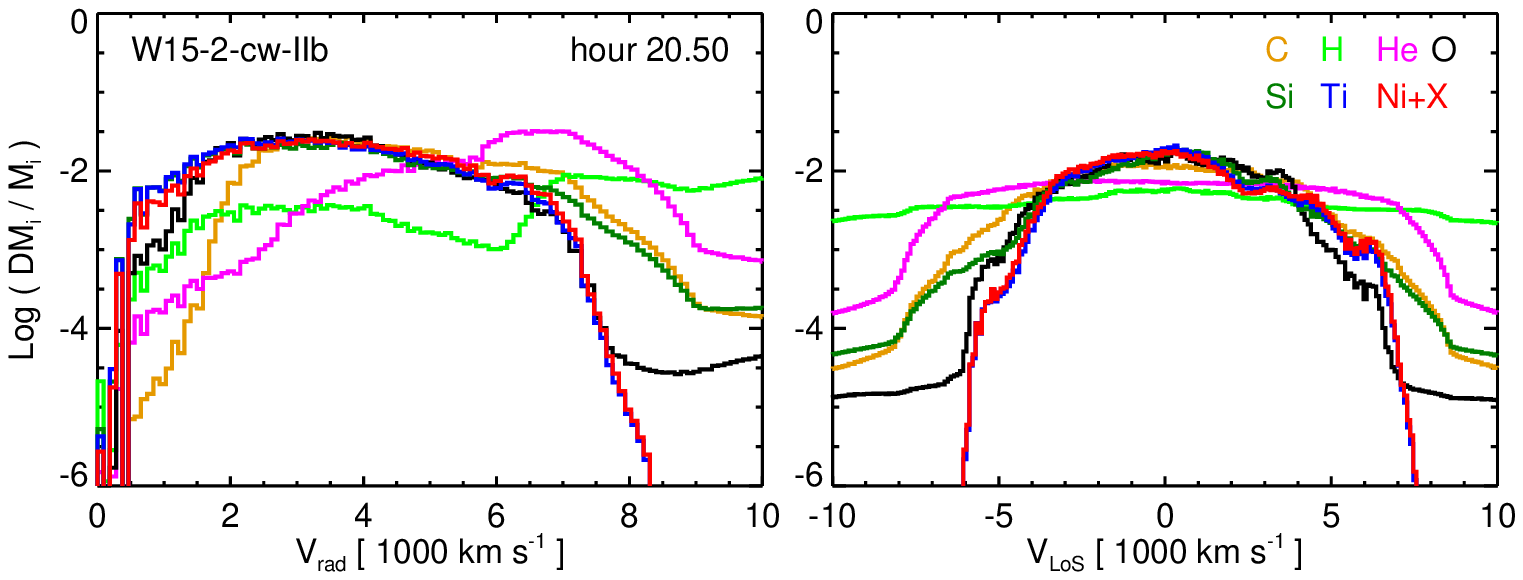, width=17cm}
	\caption{Mass distributions of $^{1}$H, $^{4}$He, $^{12}$C,
	$^{16}$O, $^{28}$Si, $^{44}$Ti, and the quantity $(^{56}{\rm
	Ni} + 0.5\, ^{56}{\rm X})$ (indicated as $[Ni+X]$) versus
	radial velocity, $v\rs{rad}$ (left panel), and velocity
	component along the LoS, $v\rs{LoS}$ (right panel) at $t
	\approx 20.5$~hr after core bounce for model W15-2-cw-IIb
	(\citealt{2017ApJ...842...13W}).}
  \label{fig1}
\end{center}
\end{figure*}
 
Figure~\ref{fig1} shows the mass distributions of selected elements
versus the radial velocity, $v\rs{rad}$ (left panel), and the
velocity component along the line-of-sight (LoS), $v\rs{LoS}$ (right
panel), at $t \approx 20.5$~hr, namely at the beginning of our SNR
simulations. In the figure, $DM\rs{i}$ is the mass of the
$i$-th element in the velocity range $[v; v + dv]$ and the
velocity is binned with $dv = 100$~km~s$^{-1}$. Again the
velocity $v\rs{LoS}$ is derived assuming the Earth vantage point
lying on the negative $y$-axis.  As discussed in
\cite{2017ApJ...842...13W}, the removal of the H envelope
allows Fe-group elements, Ti, Si, and O to be
distributed in a broad maximum below $\approx 7000$~km~s$^{-1}$
(for comparison, in model W15-2-cw - including the H envelope
- the broad maximum is below $\approx 5000$~km~s$^{-1}$; see Fig.~5
in \citealt{2017ApJ...842...13W}), with a minor fraction of elements
populating the high-velocity tail up to $\approx 9000$~km~s$^{-1}$
(in model W15-2-cw, $^{44}$Ti and $^{56}$Ni reach velocities not
larger than $\approx 5000$~km~s$^{-1}$).


\subsection{Modeling the evolution of the supernova remnant}
\label{sec:snr_model}

Model W15-2-cw-IIb was used to initiate 3D MHD/HD simulations which
describe the long-term evolution of the blast wave and ejecta, from
the shock breakout to the interaction of the remnant with the wind
of the progenitor star, covering $\approx 2000$ years of evolution.
We adopted the numerical setup of the SNR model described in
\cite{2019A&A...622A..73O, 2020A&A...636A..22O}. In particular, the
evolution of the blast wave and ejecta were modeled by numerically
solving the time-dependent MHD equations of mass, momentum, energy,
and magnetic flux conservation in a 3D Cartesian coordinate system
$(x,y,z)$ and assuming an ideal gas law, $P = (\gamma
-1)\rho\epsilon$, where $P$ is the pressure, $\gamma = 5/3$ the
adiabatic index, $\rho$ the mass density and $\epsilon$ the specific
internal energy. The effects of self-gravity were neglected during
the remnant evolution (starting $\approx 20$~hours after the
core-collapse), because the ejecta are in free expansion\footnote{We
compared the gas potential energy per unit of mass of each grid
cell with its kinetic energy per unit of mass and found that the
ratio of kinetic energy to potential energy is much higher than one
for all the expanding ejecta. This holds also at the time of the
mapping from the SN to the SNR simulation, except for a small number
of cells near the very center of the spatial domain (namely some
cells close to the explosion center), which, however, contain only
very little mass.}. The simulations include all physical processes
relevant to describe the interaction of the stellar debris with the
CSM: the deviations from equilibrium of ionization and from
electron-proton temperature equilibration; the effects of energy
deposition from radioactive decay of $^{56}$Ni and $^{56}$Co; the
effects of an ambient magnetic field.

The simulations of the expanding SNR were performed using PLUTO
v4.3 (\citealt{2007ApJS..170..228M, 2012ApJS..198....7M}), a modular
Godunov-type code intended mainly for astrophysical applications
and high Mach number flows in multiple spatial dimensions. The code
is designed to make efficient use of massively parallel computers
using the message-passing interface (MPI) for interprocessor
communications. We used the MHD/HD module available in PLUTO,
configured to compute intercell fluxes with a two-shock Riemann
solver: the linearized Roe Riemann solver based on characteristic
decomposition of the Roe matrix in the case of HD simulations, and
the Harten-Lax-van Leer discontinuities (HLLD) approximate Riemann
solver in the case of MHD simulations. As for the time-marching
algorithm, we adopted either the Runge-Kutta scheme at third order
(RK3) or the characteristic tracing (ChrTr) scheme. The combination
of a two-shock Riemann solver (either Roe or HLLD) with one
of the algorithms used to advance the solution to the next time
level (either RK3 or ChrTr) in a spatially unsplit fashion
yields the corner-transport upwind method (\citealt{1990JCoPh..87..171C,
2005ApJS..160..199M, 2005JCoPh.205..509G}), one of the most
sophisticated (and least diffusive) algorithms available in PLUTO.
A monotonized central difference flux, limiter (MC\_LIM,
the least diffusive limiter available in PLUTO) for the primitive
variables is used to prevent spurious oscillations that would
otherwise occur in the presence of strong shocks. In the case of
MHD simulations, the solenoidal constraint of the magnetic field
is controlled by adopting a hyperbolic/parabolic divergence cleaning
technique (\citealt{2002JCoPh.175..645D, 2010JCoPh.229.5896M}).

The code was extended by additional computational modules to calculate
the deviations from electron-proton temperature equilibration and
from equilibrium of ionization, and the energy deposition from
radioactive decay. The former are calculated by assuming an almost
instantaneous heating of electrons at shock fronts up to $kT = 0.3$
keV by lower hybrid waves (see \citealt{2007ApJ...654L..69G}) and
by implementing the effects of Coulomb collisions for the calculation
of ion and electron temperatures in the post-shock plasma (see
\citealt{2015ApJ...810..168O} for further details). The deviations
from equilibrium of ionization of the most abundant ions are
calculated through the maximum ionization age in each cell of the
spatial domain (\citealt{2015ApJ...810..168O}). The energy deposition
from radioactive decay is implemented following the approach
described in \cite{2019ApJ...877..136F} which is based on the general
formalism of \cite{1999astro.ph..7015J} and \cite{1994ApJS...92..527N}.
In particular we considered the dominant decay chain in which
$^{56}$Ni (half-life 6.077 days) decays in $^{56}$Co (half-life
77.27 days) and the latter decays in stable $^{56}$Fe.

The chemical evolution of the ejecta is followed by adopting a
multiple fluids approach as in \cite{2016ApJ...822...22O}. The
fluids correspond to the species calculated in model W15-2-cw-IIb
and are initialized with the abundances in the output of the SN
model. The continuity equations of the fluids are solved in addition
to our set of MHD equations. The different fluids mix together
during the remnant evolution and, in particular when the ejecta
interact with the reverse shock that develops during the expansion
of the remnant. The density of a specific element in a fluid cell
at time $t$ is calculated as $\rho_{\rm i} = \rho C_{\rm i}$,
where $C_{\rm i}$ is the mass fraction of each element and the
index ``i'' refers to the considered element. This allows mapping of the
spatial distribution of heavy elements both inside and outside the
reverse shock at different epochs during the evolution.

At the beginning of the SNR simulations, the ejecta are distributed
within a sphere with radius $\approx 10^{14}$~cm. Observations
suggest that the morphology and expansion rate of \casa\ are both
consistent with a blast wave still expanding through the wind of
the progenitor star\footnote{Some authors considered the wind
as that of a red supergiant progenitor (e.g. \citealt{2003ApJ...593L..23C,
2014ApJ...789....7L, 2020ApJ...891..116W}). However, after the outer
H envelope of the progenitor star was stripped away, we do not
expect the remaining star to be a red supergiant. In fact,
\cite{2020NatAs...4..584K} recently suggested a scenario in which
\casa\ resulted from a type IIb supernova explosion of a blue
supergiant with a thin H envelope.} Thus, for the CSM, we assumed
a spherically symmetric wind with gas density proportional to
$r^{-2}$ (where $r$ is the radial distance from the center of
explosion).  Following
\cite{2016ApJ...822...22O}, we fixed the wind density $n\rs{w} =
0.8$~cm$^{-3}$ at $r\rs{fs} = 2.5$~pc (a rough estimate of the
current outer radius of the remnant, assuming a distance of $\approx
3.4$~kpc), which is slightly smaller than the best-fit value inferred
by X-ray observations of \casa\ ($n\rs{w} = 0.9 \pm 0.3$~cm$^{-3}$;
\citealt{2014ApJ...789....7L}) but well within the range of values
constrained. We note that the
$r^{-2}$ wind profile is appropriate to describe the past evolution
of \casa\ till the current epoch. However, at later times, the
remnant will expand through an ambient medium of which we ignore
the structure and density distribution. The $r^{-2}$ wind profile
seems unlikely to extend at radii much larger than the current
radius of the remnant because, there, it predicts unrealistic low
values of the wind density. Since we ignore the structure of the
still unshocked CSM, we assumed a progressive flattening of the
wind profile to a uniform density $n\rs{c}= 0.1$~cm$^{-3}$; in this
way, our wind density profile was described as $n = n\rs{w}
(r\rs{fs}/r)^2 + n\rs{c}$.

\begin{table}
\caption{Setup for the Simulated Models}
\label{tabmod}
\begin{center}
\begin{tabular}{lcc}
\hline
\hline
Parameter  &  \multicolumn{2}{c}{Value}  \\
\hline
Explosion energy               &  \multicolumn{2}{c}{$1.5\times 10^{51}$~erg ($=1.5$~B)}   \\
Ejecta mass                    &  \multicolumn{2}{c}{$3.3\,M_{\odot}$}   \\
$E\rs{exp}/M\rs{ej}$           &  \multicolumn{2}{c}{$0.45$~B$/M_{\odot}$} \\
Wind density at $r = 2.5$~pc   &  \multicolumn{2}{c}{$0.8$~cm$^{-3}$}   \\ \\
Model                          &  radioactive  & magnetic  \\
                               &  decay        & field     \\
\hline
W15-2-cw-IIb-HD                &  no  &  no  \\
W15-2-cw-IIb-HD+dec            &  yes &  no  \\
W15-2-cw-IIb-MHD+dec           &  yes &  yes \\
\hline
\end{tabular}
\end{center}
\end{table}

We investigated the effects of energy deposition from radioactive
decay and the effects of an ambient magnetic field by performing
three long-term simulations with the above effects switched either
on or off. The models are summarized in Table~\ref{tabmod}. The
first is a pure HD simulation that does not include the effects of
radioactive decay (model W15-2-cw-IIb-HD); this simulation allowed
us to study how the pristine structures and large-scale asymmetries
originating from the neutrino-driven SN explosion contribute to
shaping the remnant morphology at different epochs. Then we evaluated
the effects of heating due to radioactive decay of $^{56}$Ni and
$^{56}$Co by performing a HD simulation with these effects included
(model W15-2-cw-IIb-HD+dec). Finally, we performed a MHD simulation
(that also accounts for radioactive decay heating) to investigate
the effects of an ambient magnetic field on the evolution of ejecta
clumps and on the development of HD instabilities (responsible for
the clump fragmentation) as the reverse shock interacts with the
ejecta (model W15-2-cw-IIb-MHD+dec). Previous studies have shown
that the magnetic field can envelope the expanding clumps of ejecta,
thus limiting the growth of HD instabilities because of the continuous
increase of the magnetic pressure and field tension at the clump
border (e.g. \citealt{2012ApJ...749..156O}). As a consequence, the
clumps can survive for a longer time.

In model W15-2-cw-IIb-MHD+dec, we assumed that the magnetic field
is the relic of the field of the progenitor star. In this case,
the simplest field configuration resulting from the rotation of the
star and from the expanding stellar wind is a spiral-shaped magnetic
field known as ``Parker spiral'' (\citealt{1958ApJ...128..664P}).
In our case, we considered a pre-SN magnetic field characterized
by an average strength at the stellar surface\footnote{After the
removal of the H envelope, the pre-SN star of our simulations
has a radius of $\approx 21.4\,R_{\odot}$ at collapse (see model
W15-IIb in \citealt{2017ApJ...842...13W}).} $B\rs{0} \approx 500$~G,
which is a value well within the range observed for magnetic
massive stars (mean surface field intensities inferred from Zeeman
splitting in the range from about 2 to 30 kG;
\citealt{2009ARA&A..47..333D}). We note that the simulations
of the progenitor star as well as that of the SN explosion adopted
here did not consider any magnetic field. So, following
\cite{2019A&A...622A..73O}, we described the magnetic field in the
initial remnant interior to be the same as in the medium outside
the progenitor star (namely the ``Parker spiral''). Although this
is a crude approximation\footnote{A more realistic field in the
remnant interior (especially in the immediate surroundings of the
remnant compact object) would be much more complex than that adopted
here and it should reflect the field of the stellar interior before
the collapse of the progenitor star.}, the magnetic field is expected
to play a role only locally in preserving clumps of ejecta from
complete fragmentation after interaction with the reverse shock and
not to influence the overall expansion and evolution of the remnant,
which is characterized by a high plasma $\beta$ (defined as the
ratio of thermal pressure to magnetic pressure). Thus, for our
purposes (namely to investigate the effects of the magnetic field in
determining the structure of the mixing region between the forward
and reverse shock), it was enough to introduce the Parker spiral
in our MHD simulation.

Our mesh configuration is that described in \cite{2019A&A...622A..73O,
2020A&A...636A..22O} which allows us to follow the large physical
scales spanned during the remnant expansion. The initial computational
domain is a Cartesian box that extends between $-1.2\times 10^{14}$~cm
and $1.2\times 10^{14}$~cm in all directions, thus including the
spatial domain of the output of model W15-2-cw-IIb. The box is
covered by a uniform grid of $1024^3$ zones, leading to a spatial
resolution of $\approx 2.3\times 10^{11}$~cm. The center of explosion
in model W15-2-cw-IIb is assumed to sit at the origin of the 3D
Cartesian coordinate system $(x_0, y_0, z_0) = (0, 0, 0)$. During
the evolution, the computational domain was gradually extended
following the expansion of the remnant through the CSM and the
physical quantities were remapped in the new domains. The domain
is extended by a factor of 1.2 in all directions when the forward
shock reaches one of the boundaries of the Cartesian box. The number
of mesh points is kept the same at each remapping, so that the
spatial resolution gradually decreases following the remnant
expansion. All the physical quantities in the extended region are
set to the values of the pre-SN CSM. In previous works, we
estimated the errors on conservation of mass, momentum and energy
introduced during the successive remapping to larger and larger
grids; this approach did not introduce errors larger than 0.1\%
after 40 remaps (\citealt{2013ApJ...773..161O, 2019A&A...622A..73O}).
In the present case, we found that 69 remappings were necessary
to follow the interaction of the blast wave with the CSM during
2000 years of evolution and we found that the errors introduced
are not larger than 1\% (with the largest errors in the conservation
of momentum). The final domain extends between $-9.4$~pc and
$9.4$~pc in all directions, with a spatial resolution of $\approx
0.018$~pc. All physical quantities were fixed to the values of the
pre-SN CSM at all boundaries.


\section{Results}
\label{sec:results}

\subsection{The remnant expansion through the stellar wind}
\label{rem_exp}

The ejecta distribution soon after the shock breakout is already
characterized by large-scale asymmetries (see Fig.~\ref{ini_cond}).
These reflect anisotropies of the 3D SN explosion developed
stochastically\footnote{Apart from the explosion mechanism, the
properties of these asymmetries also depend on the density structure
of the progenitor star (e.g.~\citealt{2015A&A...577A..48W}).} mainly
by convective overturn in the neutrino-heating layer and by the
standing accretion shock instability (SASI)
(\citealt{2017ApJ...842...13W}).

After the shock breakout, the interaction of the ejecta with the
wind of the progenitor star drives a reverse shock that moves
backwards through the ejecta.  During the early phases of propagation
of the blast wave through the wind, the metal-rich ejecta expand
almost homologously, thus carrying the fingerprints of the asymmetric
explosion. However, about 30 years after the SN, when almost all
$^{56}$Ni and $^{56}$Co have already decayed in stable $^{56}$Fe,
the Fe-group elements start to interact with the reverse
shock.  The deceleration of the swept-up ejecta passing through the
reverse shock leads to the development of HD instabilities (RT,
Richtmyer-Meshkov, and KH shear instability; \citealt{1973MNRAS.161...47G,
1991ApJ...367..619F, 1992ApJ...392..118C}), which fragment the
dominant high-entropy plumes of Fe- and Ti-rich ejecta into numerous
small fingers. The RT growth is triggered by a crossing of the
density and pressure gradients in the shocked ejecta: $\partial
P/\partial r > 0$ and $\partial \rho/\partial r < 0$. This determines
a large-scale spatial mixing of metal-rich ejecta (see
Sect.~\ref{sec:mix}).

\begin{figure}[!t]
  \begin{center}
    \leavevmode
        \epsfig{file=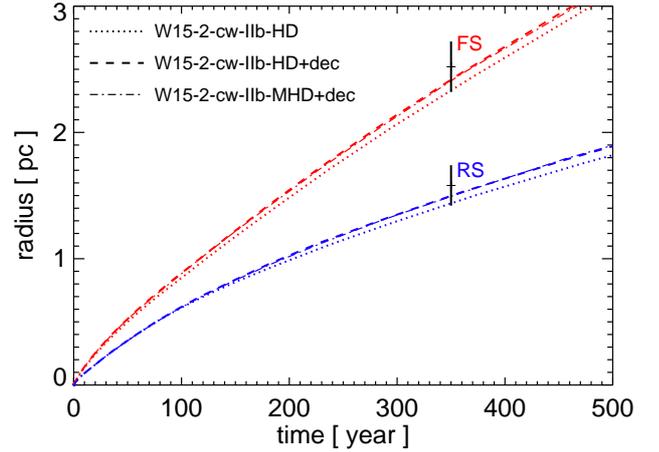, width=9cm}
	\caption{Angle-averaged radii of the forward (red lines)
	and reverse (blue lines) shocks versus time for the three
	models investigated. The black crosses show the corresponding
	observational values at the current age of \casa\
	(\citealt{2001ApJ...552L..39G}); the vertical lines of the
	crosses show the observational uncertainty.}
  \label{pos_shock}
\end{center}
\end{figure}

Figure~\ref{pos_shock} compares the angle-averaged radii of the
forward and reverse shocks resulting from our models at the age of
$\approx 350$~years with those currently observed in \casa\
(\citealt{2001ApJ...552L..39G, 2008ApJ...686.1094H}). All SNR models
are able to reproduce the observations within the error bars,
although they slightly underestimate both shock radii. This is not
surprising because the explosion energy of model W15-2-cw-IIb
($E\rs{exp} = 1.5$~B; see Table~\ref{tabmod}) is smaller than the
value inferred from the observations, $E\rs{exp} \approx 2$~B
(e.g.~\citealt{2003ApJ...597..347L, 2003ApJ...597..362H,
2020ApJ...893...49S}), although the ejecta mass of our models
($M\rs{ej} = 3.3\,M_{\odot}$; see Table~\ref{tabmod}) is within the
range of values discussed in the literature, $M\rs{ej} = [2-4]\,M_{\odot}$
(e.g. \citealt{2003ApJ...597..347L, 2003ApJ...597..362H,
2006ApJ...640..891Y}).  A slightly higher explosion energy for the
same ejecta mass would increase the radii for forward and reverse
shock at the present age of the remnant. Once again we note that
the SN model adopted here (W15-2-cw-IIb) was not tuned to describe
the SN that produced the SNR \casa, but was selected because it
roughly reproduces post-explosion anisotropies which resemble the
structure of \casa\ (\citealt{2017ApJ...842...13W}). So, the fact
that the SNR models roughly reproduce the forward and reverse shock
radii at the age of \casa\ encourages us to consider the adopted
SN model appropriate for a comparison with \casa.

Figure~\ref{pos_shock} also shows that the models including the
radioactive decay have results in slightly larger radii of the
reverse and forward shocks than in model W15-2-cw-IIb-HD. In fact
the energy
deposition due to radioactive decay provides an additional pressure
to the plasma which inflates instability-driven structures that
possess a high mass fraction of decaying elements against their
surroundings, thus powering the expansion of the ejecta
(\citealt{2020arXiv200801763G}). As for the ambient magnetic field, as
expected, it does not influence the overall expansion and evolution
of the blast wave, which is characterized by a high plasma $\beta$:
the evolution of reverse and forward shock radii is basically the
same in models W15-2-cw-IIb-HD+dec and W15-2-cw-IIb-MHD+dec.
Nevertheless, we expect that the magnetic
field plays a role in preserving inhomogeneous structures (clumps)
of ejecta from complete fragmentation by limiting the growth of HD
instabilities at their borders (e.g.~\citealt{2008ApJ...678..274O,
2012ApJ...749..156O}).

The effects of radioactive decay on the ejecta dynamics are also
visible in the amount of Fe, Ti and Si that is heated
by the reverse shock. Figure~\ref{shocked_mass} shows the fraction
of shocked masses of these elements versus time. When the heating
deposition by radioactive decay is not included in the calculation
(model W15-2-cw-IIb-HD), the shocked masses are systematically lower
than the corresponding shocked masses when the heating deposition
is included (models W15-2-cw-IIb-HD+dec and W15-2-cw-IIb-MHD+dec).
At the age of \casa, the models predict that $\approx 28-32$\% of
Ti and $\approx 30-34$\% of Fe are shocked. Considering the
tracer-particle-based post-processing with the large nuclear network,
\cite{2017ApJ...842...13W} estimated for model W15-2-cw-IIb a mass
of $1.57\times 10^{-4}\,M_{\odot}$ of $^{44}$Ti and $9.57\times
10^{-2}\,M_{\odot}$ of $^{56}$Ni. The initial mass of Ti is
consistent with the analysis of NuSTAR observations, which suggests
a total initial mass of $^{44}$Ti of $1.54\pm 0.21 \times
10^{-4}\,M_{\odot}$ (\citealt{2017ApJ...834...19G}). Thus, at the
age of \casa, the amount of shocked Ti (not considering its
decay in $^{44}$Ca) and Fe is in the ranges $[4.4-5.0]\times
10^{-5}\,M_{\odot}$ and $[2.9-3.3]\times 10^{-2}\,M_{\odot}$,
respectively. Some words of caution are needed here. In the present
paper, we considered only $^{56}$Fe from $^{56}$Ni decay. However,
there is additional Fe (other isotopes, e.g. $^{52}$Fe), which
increase the total mass by some 10\% (see \citealt{2017ApJ...842...13W},
footnote~9). All our estimates of Fe mass should be considered as
a lower limit to the total Fe mass. As for the $^{44}$Ti, for an
e-folding time of 90~years (the half-life of $^{44}$Ti is 63~years),
its mass at the age of \casa\ (350~years) would be in the range
$[0.94-1.1]\times 10^{-6}\,M_{\odot}$.

Finally, we estimated the fraction of shocked Ti and Fe if
the model had an explosion energy of $E\rs{exp} \approx 2$~B, as
inferred from observations (e.g.~\citealt{2003ApJ...597..347L,
2003ApJ...597..362H, 2020ApJ...893...49S}). In this case, the ratio
$E\rs{exp}/M\rs{ej}$ would have been $\approx 0.6$~B$/M_{\odot}$
(whereas, in our simulations, $E\rs{exp}/M\rs{ej} \approx
0.45$~B$/M_{\odot}$) and the average velocity of the ejecta soon
after the shock breakout (approximately given by $\overline{v}\rs{ej}
\approx \sqrt{2E\rs{exp} / M\rs{ej}}$) would have been a factor
1.15 higher than in our simulations. The slightly higher velocities
with $E\rs{exp} = 2$~B would have advanced the interaction with the
reverse shock, thus leading to a fraction of shocked Ti\footnote{For
an e-folding time of 90~years, the mass of $^{44}$Ti is in the range
of masses is $[1.1-1.2]\times 10^{-6}\,M_{\odot}$}  and Fe at the
age of \casa\ in the ranges $[5.0-5.6]\times 10^{-5}\,M_{\odot}$
($\approx 32-36$\%) and $[3.2-3.6]\times 10^{-2}\,M_{\odot}$ ($\approx
34-38$\%), respectively.

\subsection{Mass distribution in velocity space}
\label{sec:mix}

HD instabilities that develop during the interaction of the ejecta
with the reverse shock determine the structure and mixing of
shock-heated ejecta in the region between the reverse and the forward
shocks. Apart from the asymmetries already present in the
ejecta at the shock breakout (see Fig.~\ref{ini_cond}), the velocity
distributions of elements at the various SNR ages also reflect the
mixing between layers of different chemical composition, driven by
the interaction of the ejecta with the reverse shock.
Figure~\ref{prof_vel_hd_dec} shows the mass distributions of selected
elements versus the radial velocity, $v\rs{rad}$ (left panels), and
the LoS velocity, $v\rs{LoS}$ (right panels), for model
W15-2-cw-IIb-HD+dec. The other two models (either without the
radioactive decay, W15-2-cw-IIb-HD, or including the effects of an
ambient magnetic field, W15-2-cw-IIb-MHD+dec) produce similar results
and are presented in Appendix \ref{app:vel_distr}. As in Fig.~\ref{fig1},
$v\rs{LoS}$ is derived assuming the Earth vantage point lying on
the negative $y$-axis and the remnant oriented as in the left
panels in Fig.~\ref{ini_cond}. Fig.~\ref{prof_vel_hd_dec} reports
$^{56}$Fe instead of [Ni + X] (as in Fig.~\ref{fig1}) because almost
all $^{56}$Ni and $^{56}$Co already decayed in $^{56}$Fe at the
epochs considered.

\begin{figure}[!t]
  \begin{center}
    \leavevmode
        \epsfig{file=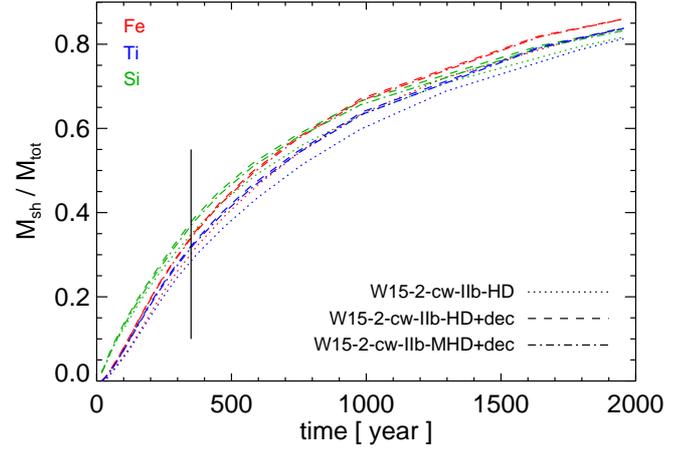, width=9cm}
	\caption{Mass of shocked Si (green lines), Ti (blue),
	and Fe (red) normalized to their total mass versus time
	for the three SNR models analyzed. The vertical black line
	marks the age of \casa.}
  \label{shocked_mass}
\end{center} \end{figure}

\begin{figure*}[!t]
  \begin{center}
    \leavevmode
        \epsfig{file=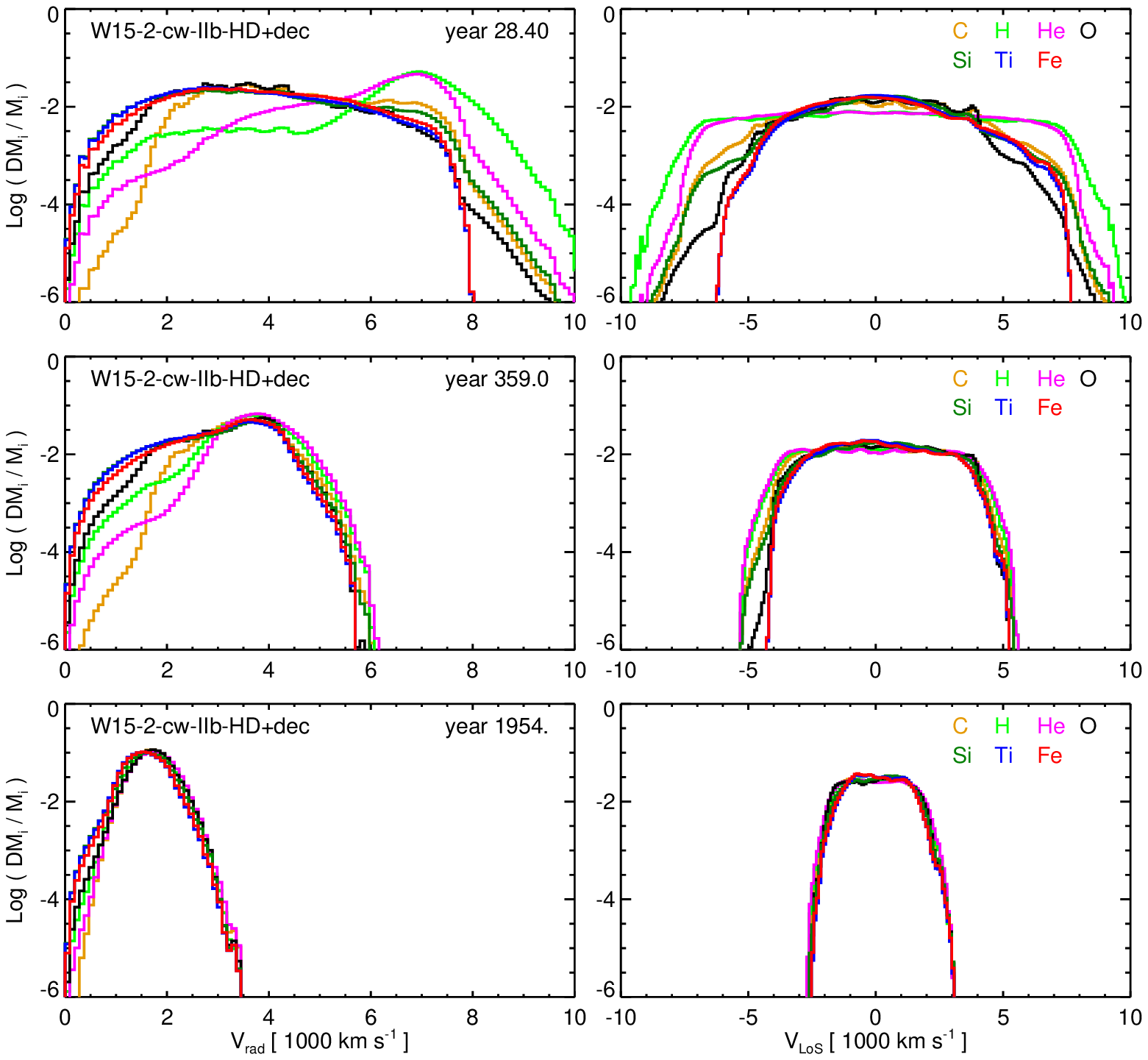, width=17cm}
	\caption{Mass distributions of $^{1}$H, $^{4}$He, $^{12}$C,
	$^{16}$O, $^{28}$Si, $^{44}$Ti, and $^{56}$Fe versus radial
	velocity, $v\rs{rad}$ (left panels), and velocity component
	along the LoS, $v\rs{LoS}$ (right panels) at the labeled
	times for model W15-2-cw-IIb-HD+dec. The upper panels
	correspond to the time when Fe and Ti start to
	interact with the reverse shock, the middle panels to the
	time corresponding to the age of \casa, and the lower panels
	to $t\approx 2000$~years.}
  \label{prof_vel_hd_dec}
\end{center}
\end{figure*}

The upper panels in Fig.~\ref{prof_vel_hd_dec} show the mass distribution
of elements at the time when the reverse shock hits the Fe- and
Ti-rich ejecta. By comparing these panels with Fig.~\ref{fig1}, at
this stage the reverse shock has already slowed down the expanding
outer layers of the ejecta. The distributions of intermediate-mass and
light elements at velocities larger than $\approx 8000$~km~s$^{-1}$
have similar shapes with a slope much steeper than in the initial
condition (soon after the shock breakout). This is a sign of efficient
mixing in the region between the reverse and forward shocks.
Nevertheless the initial order of the elements is roughly preserved,
with light elements (H and He) residing at larger radii with higher
velocities, and intermediate-mass elements (C, Si, O) at smaller
radii with smaller velocities. At this time, the mass distributions
of Fe and Ti are similar to those of the initial condition, suggesting
an almost homologous expansion of these species before interaction
with the reverse shock, although their broad maxima now extend up
to $\approx 7500$~km~s$^{-1}$ ($\approx 7000$~km~s$^{-1}$ in the
initial condition). This is mainly due to radioactive decay that
powers the ejecta residing in regions of heating deposition (compare
upper panels in Fig.~\ref{prof_vel_hd_dec} with upper panels in
Fig.~\ref{prof_vel_hd}). The mass distributions versus the LoS
velocity show large asymmetries in heavy and intermediate-mass
elements, thus reflecting the large-scale asymmetries developed in
the 3D SN explosion: Fe and Ti traveling toward (away from)
the observer reach peak velocities up to $\approx 6000$~km~s$^{-1}$
($\approx 7500$~km~s$^{-1}$).

The middle panels in Fig.~\ref{prof_vel_hd_dec} show the mass
distributions at the current age of \casa. At this stage, the reverse
shock has moved inward through the Fe- and Ti-rich plumes of the
ejecta, heating a significant fraction of their masses ($\approx
30$\% in model W15-2-cw-IIb-HD and $\approx 34$\% in models with
radioactive decay included; see Fig.~\ref{shocked_mass}). The reverse
shock has considerably slowed down the shocked ejecta and made the
overall shapes of the high-velocity parts of the mass distributions
of all elements more similar to each other. The broad maximum in
the distributions now extends up to $\approx 5000$~km~s$^{-1}$.
The maxima and the positive slopes below the maxima are more or
less the same as in the initial condition, thus indicating again
that the unshocked ejecta continue to roughly expand homologously
(although some effect due to radioactive decay is present in models
W15-2-cw-IIb-HD+dec and W15-2-cw-IIb-MHD+dec; see Appendix
\ref{app:vel_distr}). The tails of the fastest ejecta above the
distribution maxima are very similar to each other and characterized
by steep slopes that stretch out the distributions up to $\approx
6000$~km~s$^{-1}$, suggesting a considerable mixing between layers
of different chemical composition driven by HD instabilities.

The mass distributions versus $v\rs{LoS}$ still show some asymmetries
between red shifted and blue shifted profiles, although more reduced
than at earlier times. The mass distributions of metal-rich ejecta
extend in the velocity range $-4000\, {\rm km\, s}^{-1} < v\rs{LoS}
< 5500\, {\rm km\, s}^{-1}$ with the redshifted part more prominent
than the blueshifted one, consistent with the fact the the CCO
approaches the observer. These results are in very nice agreement
with the range of values inferred from observations of \casa\ in
the optical, infrared, and X-ray bands, which have revealed an overall
velocity asymmetry of $-4000$ to $+6000$~km~s$^{-1}$ in the ejecta
distribution (e.g.
\citealt{1995AJ....109.2635L, 1995ApJ...440..706R, 2002A&A...381.1039W,
2010ApJ...725.2038D}), and prominent redshifted LoS velocities of
$1000 - 6000\, {\rm km\, s}^{-1}$ for Ti (\citealt{2014Natur.506..339G,
2017ApJ...834...19G}).

The lower panels in Fig.~\ref{prof_vel_hd_dec} show the mass distributions
at $t\approx 2000$~years, when most of Fe- and Ti-rich ejecta have
been shocked ($\approx 80$\%; see Fig~\ref{shocked_mass}). Now the
reverse shock has reached the deepest layers of the ejecta and the
mass distributions significantly differ from those at the initial
condition. The shapes of the distributions of all elements are very
similar to each other due to very efficient mixing by HD instabilities.
The maxima of the distributions versus $v\rs{rad}$ now extend between
1000 and 3000 km s$^{-1}$ with peak velocities of 3500 km s$^{-1}$.
The asymmetries in the mass distributions versus $v\rs{LoS}$ are
much more reduced with respect to previous epochs, but still showing
a prominent redshift with LoS velocities up to 3000 km s$^{-1}$
($2000 - 2500$ km s$^{-1}$ in the blueshifted part). In principle,
the fingerprints of the asymmetric explosion might still be visible
after 2000 years of evolution. However, at this epoch, the remnant
is likely to have interacted with an inhomogeneous ambient environment
characterized by molecular or atomic clouds. In this case, its shape
and the distribution of the ejecta might be largely influenced by this
interaction, washing out the fingerprints of the explosion.

\subsection{$^{44}$Ti and $^{56}$Fe distributions in the early phase
of SNR evolution}
\label{sec:distrib}

\begin{figure*}[!t]
  \begin{center}
    \leavevmode
        \epsfig{file=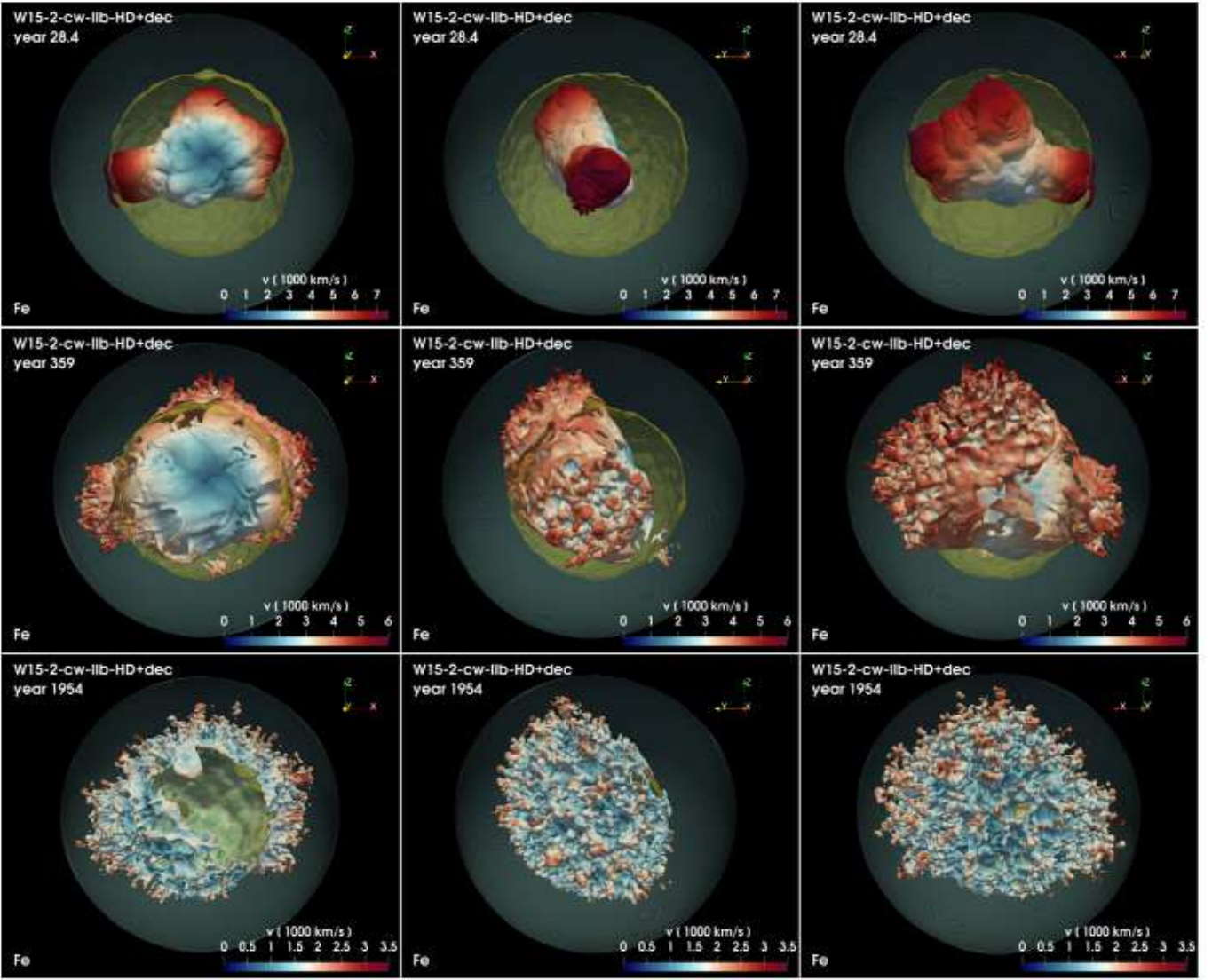, width=18cm}
	\caption{Isosurfaces of the distribution of Fe at the
	time when the reverse shock starts to interact with the Fe-
	and Ti-rich plumes of ejecta (upper panels), at the age of
	\casa\ (middle panels), and at $t=2000$~years (lower panels)
	for different viewing angles (from left to right) for model
	W15-2-cw-IIb-HD+dec. The opaque irregular isosurfaces
	correspond to a value of Fe density which is at 5\% of the
	peak density; their colors give the radial velocity in units
	of 1000 km s$^{-1}$ on the isosurface (the color coding is
	defined at the bottom of each panel). The semi-transparent
	clipped quasi-spherical surfaces indicate the forward (green)
	and reverse (yellow) shocks. The Earth vantage point lies
	on the negative $y$-axis. See online Movie 1 for an animation
	of these data; a navigable 3D graphic of the Fe spatial
	distribution at the age of \casa\ is available at https://skfb.ly/6TKRK.}
  \label{distr_fe_time}
\end{center} \end{figure*}

The supersonic expansion of the ejecta through the wind environment
leads to the development of an almost spherical forward shock and
a more corrugated reverse shock due to propagation through ejecta
inhomogeneities. As expected, HD instabilities (RT, Richtmyer-Meshkov,
and KH shear instability; \citealt{1973MNRAS.161...47G,
1991ApJ...367..619F, 1992ApJ...392..118C}) develop and grow up
at the contact discontinuity between shocked ejecta and shocked
wind (see online Movie 1). The energy deposition due to the radioactive
decay chain $^{56}{\rm Ni} \rightarrow\ ^{56}{\rm Co} \rightarrow\
^{56}{\rm Fe}$ is effective during the first year of evolution. In
this phase, regions rich in $^{56}{\rm Ni}$ and $^{56}{\rm Co}$ can
be significantly heated to temperatures up to a few millions degrees.

As mentioned in Sect.~\ref{sec:mix}, the reverse shock reaches
regions of Fe- and Ti-rich ejecta about 30 years after the SN event.
The upper panels in Fig.~\ref{distr_fe_time} show the spatial
distribution of Fe during this phase in model W15-2-cw-IIb-HD+dec.
The other two models show similar distributions (see Appendix
\ref{app:ej_distr}). Assuming that the remnant is oriented in such
a way that the Fe-rich fingers point toward the same direction as
the extended Fe-rich regions observed in \casa, the figure shows
different viewing angles: with the perspective of the remnant in
the plane of the sky (i.e.  the vantage point is at Earth; left
panel), with the perspective rotated by $90^{\rm o}$ about the
$z$-axis (center panel), and with the perspective rotated by $180^{\rm
o}$ about the $z$-axis (namely the vantage point is from behind
\casa; right panel). At this time, a significant fraction of
intermediate-mass elements (as Si, O, and C) have already passed
through the reverse shock (see upper panels in Fig.~\ref{prof_vel_hd_dec}).

At the age of $\approx 30$~years, the Fe and Ti distributions are
similar to those at the shock breakout, confirming that their
evolution was almost homologous till the interaction with the reverse
shock (see online Movie 1). The only significant change in their
spatial distributions is due to heating by radioactive decay that
inflated regions dominated by the decaying elements against their
surroundings and provided an additional boost to the ejecta expansion
(compare Fig.~\ref{distr_fe_time} with Fig.~\ref{distr_fe_early}).
As a result, a dense shell forms at the interface between regions
dominated by the decaying elements and the surrounding ejecta.
During the evolution, the Ti has a spatial distribution very
similar to that of Fe and the two species trace a similar 3D
geometry. This is not surprising because, as discussed in
\cite{2017ApJ...842...13W}, the two nuclei were basically synthesized
in the same regions of (incomplete) Si burning and of processes
like $\alpha$-particle-rich freezeout (see also
\citealt{1997ApJ...486.1026N, 2010ApJS..191...66M, 2020ApJ...895...82V}),
and there is no physical process during their expansion able to
decouple or decompose them.

As for the other species, after passing through the reverse shock,
Fe and Ti become affected by HD instabilities which gradually grow
as the species approach to the contact discontinuity. As discussed
in detail in the literature, the growth conditions for such
instabilities originate from the deceleration of the denser ejecta
at their collision with the compressed but less dense shell of
shocked CSM material, and they grow as long as a reverse shock moves
backward into the slower ejecta (e.g. \citealt{1992ApJ...392..118C,
1996ApJ...465..800J, 2001ApJ...549.1119W, 2001ApJ...560..244B}).
The development of HD instabilities is enhanced by the presence of
shock deformation and pre-shock small- and large-scales perturbations
of the ejecta (e.g.~\citealt{2012ApJ...749..156O, 2013MNRAS.430.2864M,
2020A&A...642A..67T}).
These conditions are present in our case, in which the SN ejecta,
before colliding with the reverse shock, are characterized by
small-scale clumping and large-scale asymmetries that originated
stochastically from the convective overturn in the neutrino-heating
layer and SASI activity (e.g.~\citealt{2015A&A...577A..48W,
2017ApJ...842...13W}).

\subsection{Structure of shocked ejecta at the age of \casa}
\label{sh_ejecta}

\begin{figure*}[!t]
  \begin{center}
    \leavevmode
        \epsfig{file=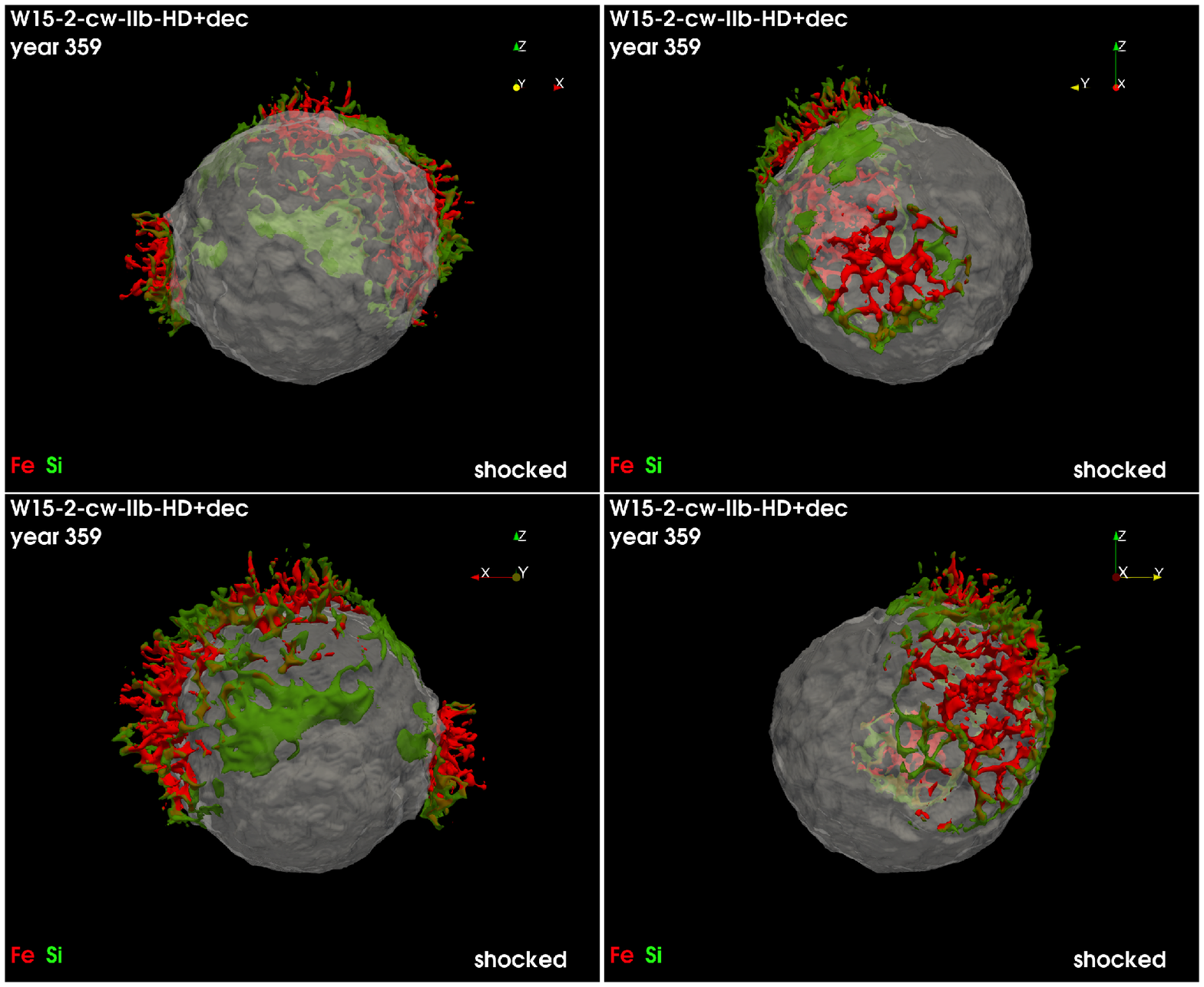, width=18cm}
	\caption{Isosurfaces of the distributions of shocked Fe
	(red) and Si (green) at the age of \casa\ for different
	viewing angles for model W15-2-cw-IIb-HD+dec. The upper
	left panel corresponds to the vantage point at Earth for
	\casa.  The opaque irregular isosurfaces correspond to a
	value of Fe and Si density which is at 25\% of the respective
	peak density. The semi-transparent white surface indicates
	the reverse shock. See online Movie 2 for an animation of
	these data; a navigable 3D graphic of these distributions which
	includes also unshocked Fe and Si is available at
	https://skfb.ly/6UJYu.}
  \label{sh_fe_si_casa}
\end{center} \end{figure*}

At the age of \casa\ (middle panels in Fig.~\ref{distr_fe_time}),
about $2.45\,M_{\odot}$ of ejecta were shocked. The HD instabilities
are already well developed and have determined the progressive
fragmentation of the initially large, metal-rich plumes of ejecta
(rich in Fe and Ti) to smaller fingers of dense ejecta gas that
protrude into the shocked wind material. The instabilities also
enhance the mixing between layers of different chemical composition
as evident from the middle panels in Fig.~\ref{prof_vel_hd_dec}.
At this age, large regions of shocked Fe-rich ejecta form in
coincidence with the original large-scale fingers of Fe-group
elements. There, several Fe- and Ti-rich shocked filamentary
structures extend from the reverse toward the forward shock. As in
\casa, shocked Fe, Ti, and Si are arranged in a torus-like geometry,
which is tilted approximately $30^{\rm o}$ with respect to the plane
of the sky (see Fig.~\ref{sh_fe_si_casa} and online Movie 2). This
structure reflects the large-scale asymmetry of the Fe-group
ejecta at the shock breakout, consisting mainly of three extended
Fe-rich fingers. The latter define a plane in which metal-rich
ejecta extend much more than perpendicularly to it. As a result,
the ejecta lying close to this plane reach the reverse shock well
before those in other directions, producing the observed torus-like
structure. The plane is oriented with an $\approx -30^{\rm o}$
rotation about the $x$-axis (the east-west\footnote{Throughout the
paper, we define that ``east'' is left and ``west'' is right on
images, according to observers.} axis in the plane of the sky) and
an $\approx 10^{\rm o}$ rotation about the $z$-axis (the north-south
axis in the plane of the sky).  Interestingly, this orientation is
very similar to that found by \cite{2016ApJ...822...22O} for the
plane in which the three Fe bullets lie that reproduce the observed
distribution of shocked Fe and Si/S (namely $i\rs{x} \approx -30^{\rm
o}$ and $i\rs{z}\approx 25^{\rm o}$, respectively).

\begin{figure*}[!t]
  \begin{center}
    \leavevmode
        \epsfig{file=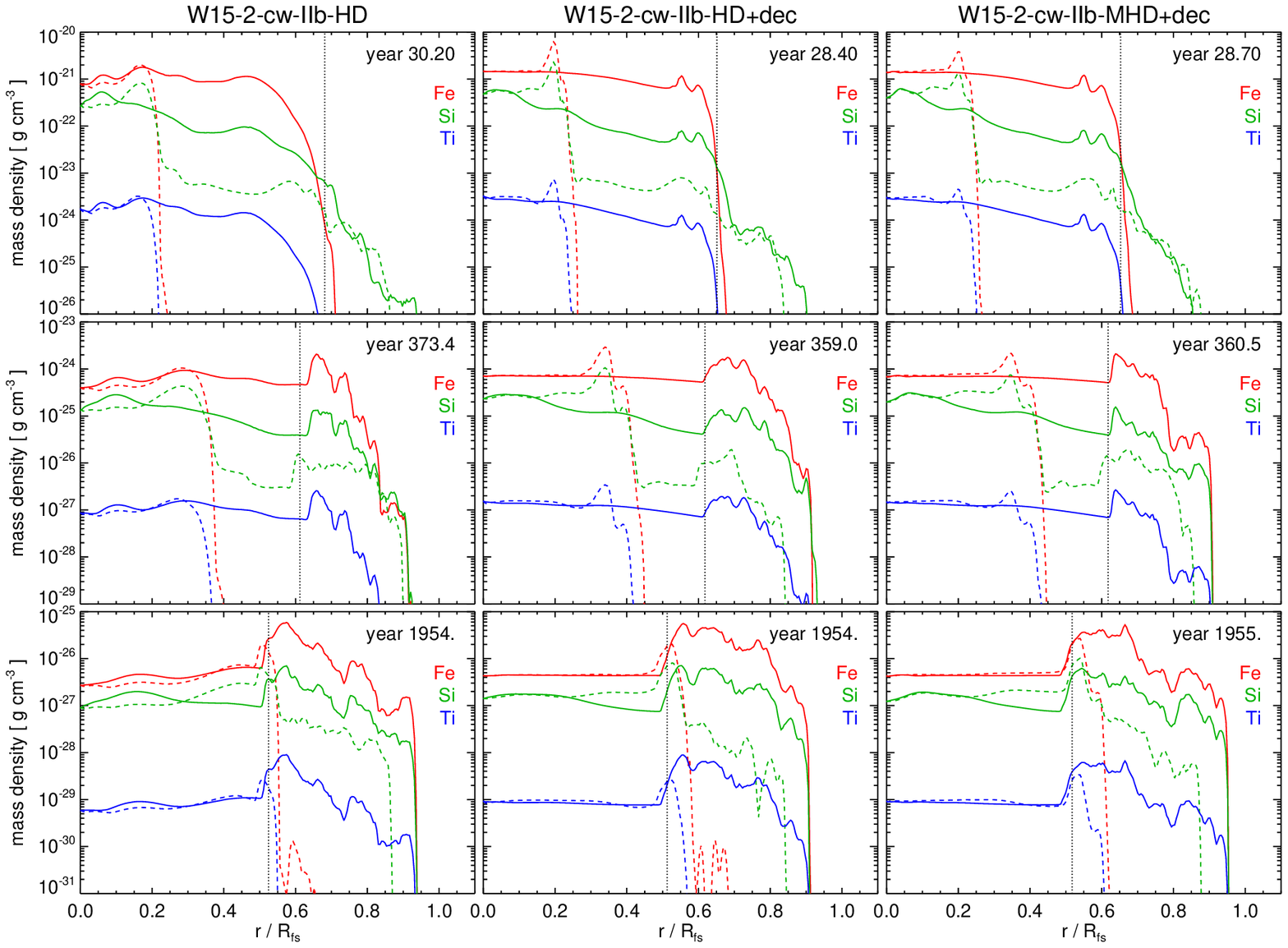, width=18.5cm}
	\caption{Radial profiles of Fe (red), Ti (blue), and Si
	(green) averaged over a solid angle of $30^{\rm o}$ in the
	direction of the east Fe-rich finger (roughly along the
	negative $x$-axis in the left panels of
	Fig.~\ref{distr_fe_time}; solid lines) and of the negative
	$z$-axis (i.e. where no Fe-rich finger is present; dashed
	lines) for our three models, at three different epochs. The
	vertical dotted line in each panel marks the average position
	of the reverse shock. The density of Ti was calculated by
	considering the tracer-particle-based post-processing with
	a large nuclear network (\citealt{2017ApJ...842...13W}), but
	not considering its decay in $^{44}$Ca (for an e-folding
	time of 90~years, the Ti density is scaled by a factor of
	0.027).}
  \label{inversion}
\end{center} \end{figure*}

The ejecta in the torus are dominated by morphological structures
that resemble the cellular structure (appearing as rings on the
surface of a sphere) of [Ar II], [Ne II], and Si XIII observed in
\casa\ (e.g.~\citealt{2010ApJ...725.2038D, 2013ApJ...772..134M}).
In some cases, the ring structures have RT fingers which extend
outward, giving them the appearance of crowns as also observed in
\casa\ (e.g.~\citealt{2013ApJ...772..134M}). Ring- and crown-like
structures originate from a combined action of small-scale structures
in the initial Fe-rich fingers and the development of HD instabilities
after the passage of the reverse shock. In fact, soon after the
breakout of the SN shock from the star, half of the mass of Ni-rich
ejecta is concentrated in relatively small, highly enriched clumps
and knots (see Fig.~10 in \citealt{2017ApJ...842...13W}). Later on,
this ejecta structure is kept in the distribution of Fe (the final
product of Ni decay). A significant fraction of these small-scale
clumps is located in the outermost tips of the largest Ni/Fe-rich
fingers (see Fig.~10 in \citealt{2017ApJ...842...13W}), so that
they start to interact with the reverse shock at relatively early
times ($\approx 30$~years). The clumpy structure of the ejecta
enhances the development of HD instabilities and leads to the
formation of a filamentary pattern of shocked ejecta with ring-like
features (e.g.~\citealt{2012ApJ...749..156O, 2016ApJ...822...22O}).

The comparison between models W15-2-cw-IIb-HD and W15-2-cw-IIb-HD+dec
shows that the radioactive decay affects the structure of shocked
ejecta by further boosting Fe and Ti. Fig.~\ref{inversion} shows
radial profiles of Fe, Ti, and Si averaged over a solid angle of
$30^{\rm o}$ in the direction of the east Fe-rich finger (roughly
along the negative x-axis for an observer on Earth) and of the
negative $z$-axis (i.e.  where no Fe-rich fingers are present) for
our three models. We note that a forward shock radius of $\approx
2.5$~pc is reached earlier in models W15-2-cw-IIb-HD+dec and
W15-2-cw-IIb-MHD+dec ($\approx 360$~years) than in model W15-2-cw-IIb-HD
($\approx 370$~years). The inflated ejecta produce a dense shell
at the edge of regions with a high concentration of the decaying
elements, which becomes evident at the age $\approx 30$~years as
bumps in the radial profiles of Fe, Ti, and Si immediately before
the reverse shock (see upper panels in Fig.~\ref{inversion}; see
also discussion in Sect.~\ref{sec:distrib}).  Indeed, in models
including the decay heating, the density at the outermost tip of
the Fe-rich plumes is about a factor of $\approx 7$ higher than in
model W15-2-cw-IIb-HD (compare the red profiles at $r/R\rs{fs} =
0.6$, in the upper panels in Fig.~\ref{inversion}), thus increasing
the density contrast of the Fe-rich clumps with respect to the
surrounding ejecta. As the
Fe-rich plumes hit the reverse shock, the inflated dense ejecta in
models W15-2-cw-IIb-HD+dec and W15-2-cw-IIb-MHD+dec penetrate more
efficiently in the mixing region, leading to a more effective
development of the ring- and crown-like structures (compare the
middle panels in Fig.~\ref{inversion} at the age of \casa, and
Fig.~\ref{sh_fe_si_casa} with Fig.~\ref{sh_fe_si_HD_casa}).

\begin{figure*}[!t]
  \begin{center}
    \leavevmode
        \epsfig{file=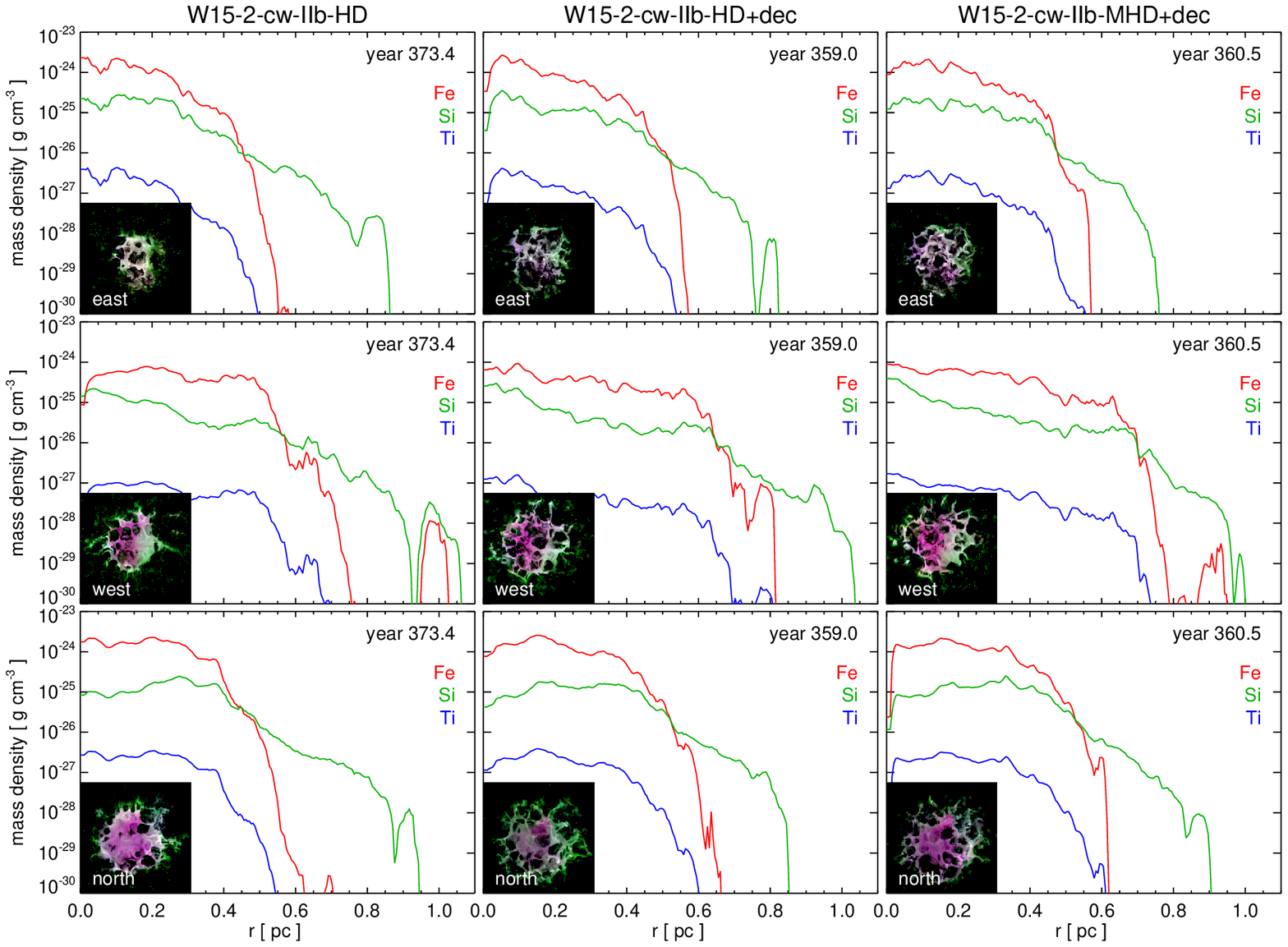, width=18.5cm}
	\caption{Profiles of Fe (red), Ti (blue), and Si (green)
	as a function of the distance from the center of the Fe
	finger that roughly points to the east (top row), to the
	west (middle row) and to the north (botton row), at the age
	of \casa, for our three models. The profiles are averaged
	over all directions in a plane tangential to the sphere at
	$r\approx 0.7\,R\rs{fs}$ from the center of the explosion
	(namely immediately above the reverse shock). The density
	of Ti was calculated by considering the tracer-particle-based
	post-processing with a large nuclear network
	(\citealt{2017ApJ...842...13W}), but not considering its
	decay in $^{44}$Ca (for an e-folding time of 90~years, the
	Ti density is scaled by a factor of 0.027). The inset in
	the lower left corner of each panel shows a three-color
	composite image of the mass density distributions of Fe
	(red), Ti (blue), and Si (green) in the plane at $r\approx
	0.7\,R\rs{fs}$.}
  \label{rings}
\end{center} \end{figure*}

We note that the Fe-rich regions are circled by rings of Si-rich
ejecta, another striking resemblance with the \casa\ structure.
This is evident in Fig.~\ref{sh_fe_si_casa} (see also online Movie
2) where it is possible to clearly identify an almost complete ring
of Si-rich ejecta around the north-west Fe-rich region (lower right
panel), and an incomplete ring around the east region (upper
right panel). These rings reflect the structure of the ejecta at
the shock breakout and the dynamics during the subsequent remnant
evolution. This can be seen from Fig.~\ref{inversion} showing that,
before the Fe-rich finger starts to interact with the reverse shock
(upper panels), the radial distribution of post-shock Si is almost
the same along the two directions analyzed. In fact, in this early
phase, the distribution of shocked Si is almost spherical. Fe-rich
layers (also rich in Si) are enveloped by a Si-rich layer of the ejecta
(compare red and green lines in the figure). As a result, as soon
as Fe-rich fingers interact with the reverse shock, the Si density
strongly increases in shocked ejecta located in the immediate
surroundings of regions of shocked Fe. This is clearly evident in
Fig.~\ref{rings} that shows, for the three Fe fingers (pointing
roughly to the east, to the west and to the north), the profiles
of Fe, Ti and Si as a function of the distance from the center of
the Fe fingers; the profiles are averaged over all directions in
a plane tangential to the sphere at $r\approx 0.7\,R\rs{fs}$
(immediately above the reverse shock). In all the models, a shell
rich in Si envelopes each Fe-rich region.

\begin{figure*}[!t]
  \begin{center}
    \leavevmode
        \epsfig{file=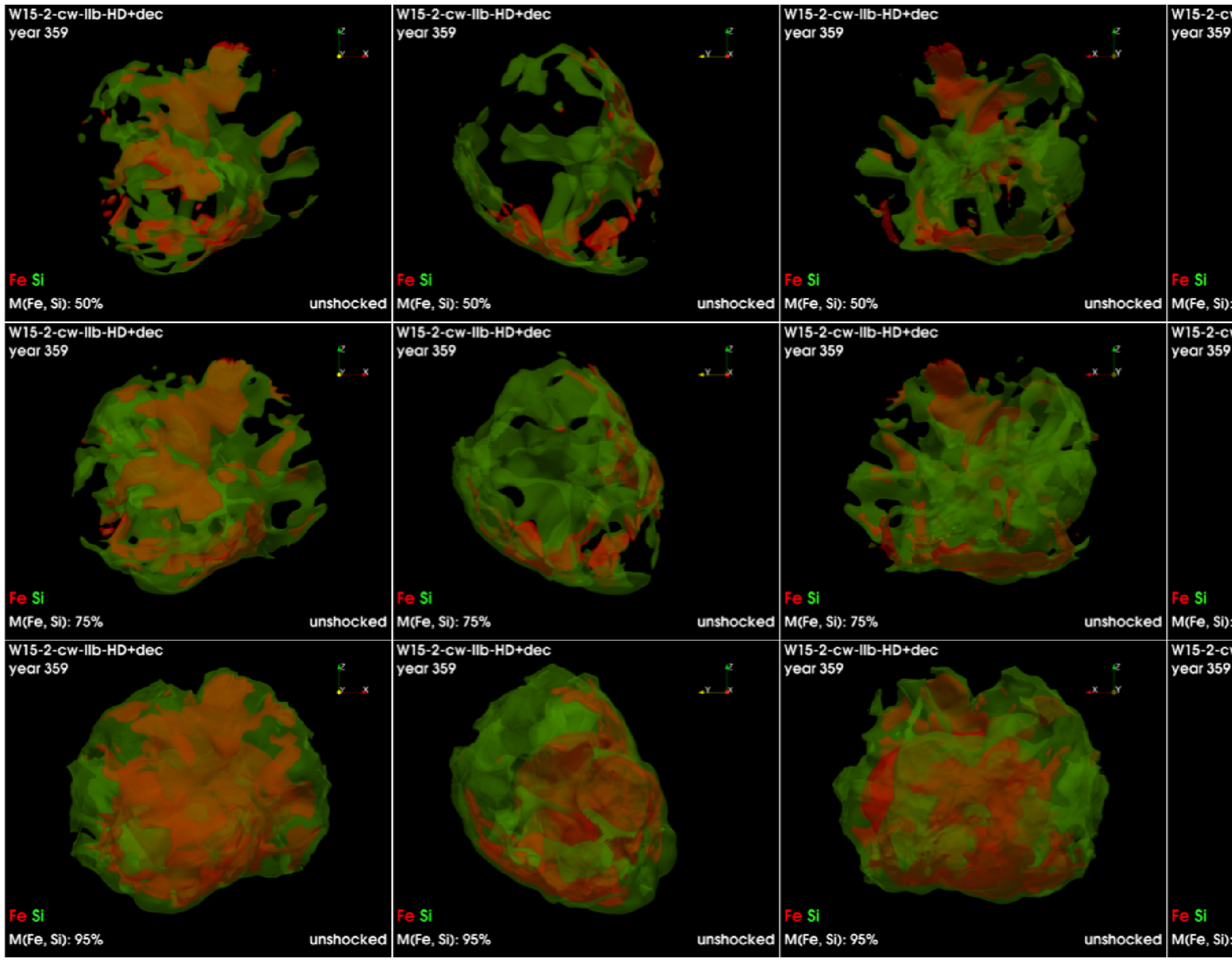, width=18cm}
	\caption{Isosurfaces of the distributions of unshocked Fe
	(red) and Si (green) at the age of \casa\ for different
	viewing angles (from left to right) for model W15-2-cw-IIb-HD+dec.
	The left panels correspond to the observer direction for
	\casa. The isosurfaces enclose 50\% (upper row), 75\% (center
	row), and 95\% (lower row) of the total mass of unshocked
	Fe and Si and correspond to a value of Fe (Si)
	density, in units of $10^{-26}$~g~cm$^{-3}$, which is 81
	(25), 70 (16), and 55 (6.0), respectively.}
  \label{unsh_fe_si}
\end{center} \end{figure*}

Our models also predict that the initial large-scale asymmetries
resulting from the SN explosion produce a spatial inversion of
ejecta layers at the age of \casa, leading locally to Fe-rich ejecta
placed at a greater radius than Si-rich ejecta. This feature forms
as a result of the high velocity and dense Fe-rich plumes of ejecta
in a way similar to that described in \cite{2016ApJ...822...22O}
for Fe-rich pistons. Our models show that, soon after the shock
breakout, the large-scale Fe-rich fingers have already protruded
through the chemically distinct layers above and are enveloped by
a less dense Si-rich layer. After the interaction with the reverse
shock, the dense Fe-rich fingers push out the less dense material
above (including Si), breaking through some of it, and leading
to the spatial inversion of the ejecta layers. This is evident from the
middle panels in Fig.~\ref{inversion} by comparing the average
radial distribution of Fe along the east Fe-rich finger (red
solid line) with the distribution of Si along the negative $z$-axis
(where no Fe-rich fingers are present; dashed green lines). The
inversion of the ejecta layers is much more evident in models including
the radioactive decay, which contributes boosting the Fe-rich fingers
out (see also \citealt{2020arXiv200801763G}). In these cases, the
shocked Fe is denser than in model W15-2-cw-IIb-HD and, already
at the age of \casa, it extends to a greater radius than Si-rich
ejecta outside Fe-rich fingers (compare red solid with green dashed
lines), whereas in model W15-2-cw-IIb-HD the inversion is evident
only at later times (see lower panels in Fig.~\ref{inversion}). The
inversion also causes the material of the outer layers (in particular
Si) to be swept out by the Fe-rich fingers and to accumulate to the
side of the shocked Fe regions (see Fig.~\ref{rings}).

\subsection{Structure of unshocked ejecta at the age of \casa}
\label{sec:unshocked}

The 3D spatial distributions of unshocked Fe and Si at the age of
\casa\ for model W15-2-cw-IIb-HD+dec are displayed in
Fig.~\ref{unsh_fe_si}. An animation shows the 3D distributions
rotating about the north-south axis (Movie 2). At this epoch, the
mass of unshocked ejecta from the models is $\approx 0.85\,M_{\odot}$
which is in the range of values inferred from observations, namely
between $< 0.4\,M_{\odot}$ (\citealt{2012ApJ...746..130H,
2014ApJ...785....7D}) and $\approx 3\,M_{\odot}$
(\citealt{2018A&A...612A.110A}), and in excellent agreement with
recent findings inferred from infrared observations
($0.47^{+0.47}_{-0.24}\,M_{\odot}$; \citealt{2020arXiv201007718L}).
This wide range of values inferred from observations is basically
due to several sources of uncertainties, including the temperature
and the clumpy structure of the unshocked ejecta (see also
\citealt{2018ApJ...866..128R, 2018ApJ...866..139K}).

Our models show that the unshocked ejecta are highly structured.
The amount of unshocked $^{56}$Fe is $\approx 0.062\,M_{\odot}$;
additional mass of Fe ($\approx 10$\%; \citealt{2017ApJ...842...13W})
comes from other Fe-group species, leading to a total of $\approx
0.068\,M_{\odot}$ of unshocked Fe.  This estimate is consistent
with the $\approx 0.07\,M_{\odot}$ of Fe that may be present
in diffuse gas in the inner ejecta, as suggested by
\cite{2020arXiv201007718L}. About 50\% of the unshocked Fe and Si
are distributed in an irregular shell located in the proximity of
the reverse shock (see upper row in Fig.~\ref{unsh_fe_si}).  The
two distributions are characterized by two large cavities (evident
in the central columns of Fig.~\ref{unsh_fe_si}) corresponding to
the directions of propagation of the large-scale Fe-rich plumes of
ejecta (the initial asymmetry). In fact, the regions of shocked
Fe in the mixing region are located exactly above these cavities
(compare Figs.~\ref{sh_fe_si_casa} and \ref{unsh_fe_si} and see
Movie~2). Since the cavities are present in all simulations either
including or not the Ni decay, we conclude that they form mainly
because of the fast expansion of Fe in the directions of the
large-scale plumes. The initial asymmetry in the SN explosion is
mainly responsible for the largest cavities in our simulations. On
the other hand, we note that the cavities are more extended in
models including the Ni decay, thus reflecting the inflation of the
Fe-rich regions driven by the radioactive decay heating of the
initial Ni. This effect enhances the formation of the large-scale
Si-rich rings visible in the shocked ejecta and encircling the
Fe-rich regions. As a result, these rings are physically
connected with the large cavities, a characteristic feature also
present in the ejecta morphology of \casa\ (e.g.
\citealt{2015Sci...347..526M}).  This is particularly evident for
the north-west Fe-rich region (compare Figs.~\ref{sh_fe_si_casa}
and \ref{unsh_fe_si}; see online Movie~2).

The overall structure of the unshocked Si is similar to that found
by \cite{2016ApJ...822...22O}, who considered parameterized initial
post-explosion anisotropies in the ejecta, and it very closely
resembles the bubble-like morphology of the \casa\ interior
(\citealt{2015Sci...347..526M}). The analysis of near-infrared
spectra of \casa, including the [S III] 906.9 and 953.1 nm lines,
shows a distribution of unshocked sulfur being very structured in
the remnant interior (\citealt{2015Sci...347..526M}), at odds with
the results of \cite{2016ApJ...822...22O}, but in nice agreement
with the Si distribution of our simulations. This is basically due
to the fact that the SN model in \cite{2016ApJ...822...22O} was
one-dimensional, thus missing the mixing processes between different
chemically inhomogeneous layers of the ejecta. In our simulations,
the morphology of unshocked Fe and Si is enriched by several
smaller-scale cavities that, especially in the case of Si, extend
toward the remnant interior. Silicon was present in the Si
shell of the progenitor and some of it formed from explosive oxygen
burning behind the outgoing shock. Naturally, some of it was left
in the remnant interior and some of it was mixed outward by the
radial instabilities. Fe, Ti and Si were formed in neighboring
regions, and all of them were affected by the same mixing processes,
starting with the convective overturn triggered by neutrino heating,
during the first second of the explosion, and followed by subsequent
RT instability as the SN shock crosses the composition shell
interfaces of the progenitor.  These detailed dynamical effects
during the early supernova stages
were missing in the simulations of \cite{2016ApJ...822...22O}.

Another characteristic feature of the modeled ejecta structure which
strikingly resembles that inferred from \casa\ is the evidence of a
``thick-disk'' geometry for ejecta rich in Si, Ti, and Fe and which is
tilted by an angle of $\approx -30^{\rm o}$ with respect to the plane
of the sky (e.g~\citealt{2010ApJ...725.2038D, 2017ApJ...834...19G}).
This slanted thick-disk originates from the initial large-scale
explosion asymmetry (see also \citealt{2017ApJ...842...13W} and
Fig.~11 there) and, in fact, is oriented as the plane in which the
three dominant high-entropy plumes lie at the time of shock
breakout\footnote{According to simulations of core-collapse
SNe, the occurrence of three large plumes is quite common in the case
of highly asymmetrical explosions, though in some cases a single
plume (large dipole asymmetry of the explosion) or two large plumes
(bipolar structure) may develop (e.g. \citealt{2013A&A...552A.126W,
2015A&A...577A..48W, 2019A&A...624A.116U}, Utrobin et al. in
preparation).}. The thick-disk appearance is due to the fast
expansion of the three plumes which makes the Si-, Ti-, and Fe-rich
ejecta expanded more parallel to this plane than perpendicular to
it.

\begin{figure*}[!t]
  \begin{center}
    \leavevmode
        \epsfig{file=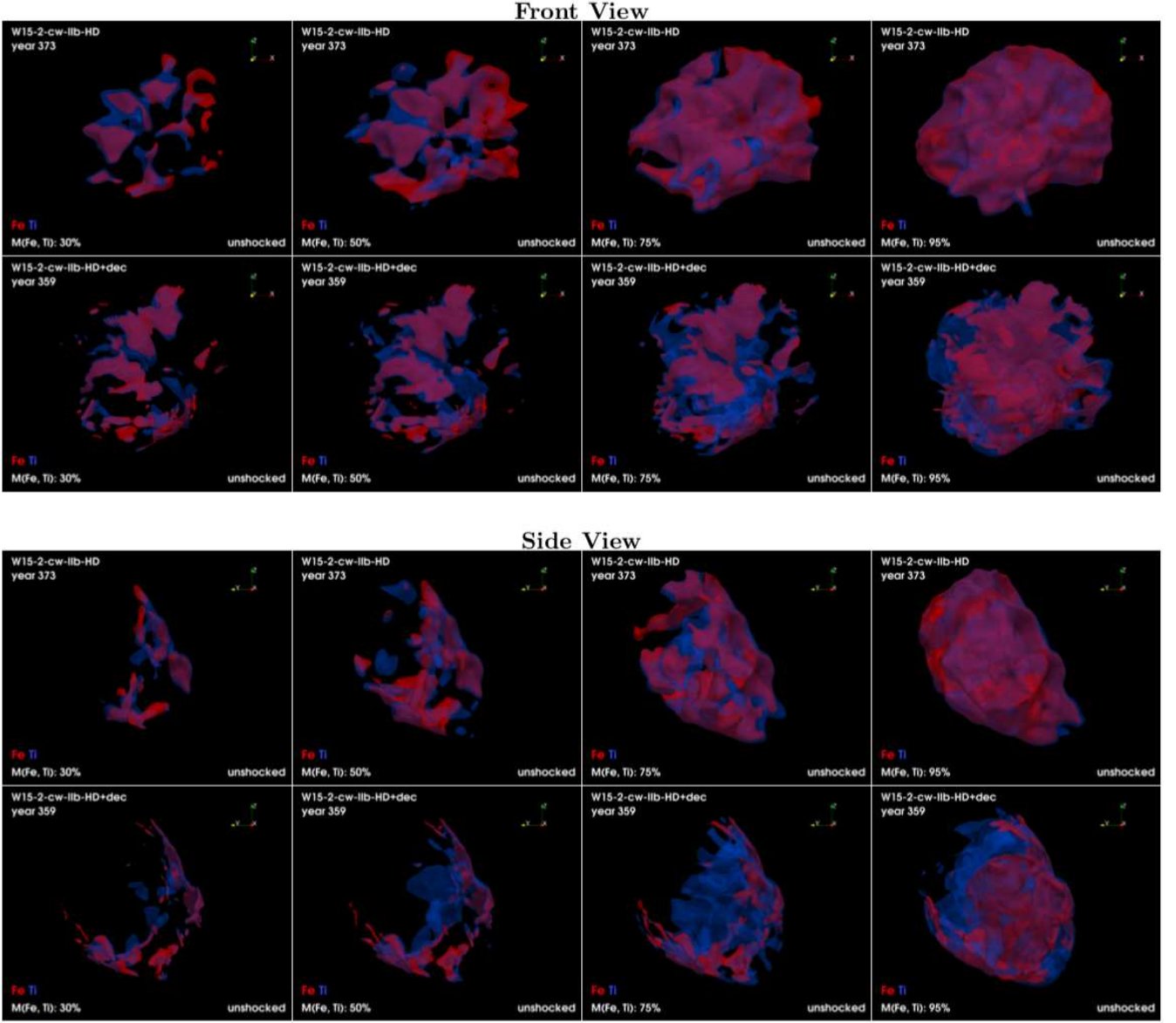, width=18cm}
        \caption{Isosurfaces of the distributions of unshocked Fe
	(red) and Ti (blue) at the age of \casa\ for different
	values of the mass fractions (decreasing from left to right;
	see the values reported at the end of the caption) and
	different viewing angles (front and side views) for models
	W15-2-cw-IIb-HD and W15-2-cw-IIb-HD+dec. The values of the
	mass fractions are selected so that the corresponding
	isosurfaces enclose, from left to right, 30\%, 50\%, 75\%,
	and 95\% of the total mass of the respective unshocked
	species. The red isosurfaces correspond to a value of Fe
	density in model W15-2-cw-IIb-HD (W15-2-cw-IIb-HD+dec) which
	is, in units of $10^{-26}$~g~cm$^{-3}$, 119 (89), 95 (81),
	71 (70), 40 (55), respectively; the blue isosurfaces
	correspond to a value of Ti density in model W15-2-cw-IIb-HD
	(W15-2-cw-IIb-HD+dec) which is, in units of $10^{-27}$~g~cm$^{-3}$
	(considering the tracer-particle-based post-processing with
	a large nuclear network, \citealt{2017ApJ...842...13W}, but
	not considering its decay in $^{44}$Ca), 2.1 (1.7), 1.7
	(1.4), 1.2 (1.1), 0.7 (0.8), respectively (considering the
	decay of $^{44}$Ti in $^{44}$Ca, these values are scaled
	by a factor of 0.027, for an e-folding time of 90~years).
	See online Movie~3 for an animation of these data; a navigable
	3D graphic of these distributions is available at
	https://skfb.ly/6UMxI.}
  \label{unsh_fe_ti}
\end{center} \end{figure*}

The structure of the unshocked ejecta can be significantly affected
by the radioactive decay of $^{56}$Ni and $^{56}$Co during the first
year of the evolution (see also \citealt{2020arXiv200801763G}).
Fig.~\ref{unsh_fe_ti} shows the distributions of Fe and Ti at the
age of \casa\ for models W15-2-cw-IIb-HD and W15-2-cw-IIb-HD+dec
(see also Movie~3). The distributions of Fe and Ti are very similar
to each other in model W15-2-cw-IIb-HD, although there are some
regions with dominant concentration of Fe and other with dominant
concentration of Ti; in the presence of decay heating the differences
are much more evident (see model W15-2-cw-IIb-HD+dec).

The distributions of unshocked Fe and Ti (especially in model
W15-2-cw-IIb-HD) have properties similar to those witnessed after
the shock breakout. In fact, they result from an almost homologous
expansion of the ejecta. The innermost, still unshocked, Fe- and
Ti-rich ejecta are more diluted than the outer shocked ones,
consistent with the distributions of Ni and Ti obtained soon after
the shock breakout (see Fig.~10 in \citealt{2017ApJ...842...13W}).
This is also consistent with the fact that the unshocked ejecta
expanded homologously, whereas the shocked ejecta were strongly
compressed by the reverse shock. A mass fraction of about 50\% of
unshocked Fe and Ti is enclosed in highly enriched clumps (second
column in Fig.~\ref{unsh_fe_ti}), which roughly form an irregular
shell enclosing regions of more diluted Fe and Ti (in Fig.~\ref{unsh_fe_ti}
compare the different isosurfaces enclosing 30\%, 50\%, 75\%, and
95\% of the total unshocked ejecta mass of the respective element).

When the effects of radioactive decay of $^{56}$Ni and $^{56}$Co
are taken into account, soon after the shock breakout the heating
by the energy release inflates the ejecta in regions with a high
concentration of the decaying elements, pushing them against their
surroundings (see also \citealt{2020arXiv200801763G}). As a result,
the ejecta in these regions expand faster than in regions with a
low concentration of Ni and Co. As in model W15-2-cw-IIb-HD, at the
age of \casa, the bulk of unshocked Fe (the decay product of Co)
is enclosed in enriched clumps of ejecta forming an irregular shell.
However, in model W15-2-cw-IIb-HD+dec, the shell is more developed
and expanded (inflated) than in W15-2-cw-IIb-HD due to the additional
pressure provided by the decay heating: about 75\% of the total
ejecta mass of unshocked Fe is concentrated in the shell (see 3rd
column in Fig.~\ref{unsh_fe_ti} for model W15-2-cw-IIb-HD+dec; see
also Movie~3). The tilted thick-disk appearance of Fe- and Ti-rich
ejecta is slightly less evident if decay heating is effective.
Regions with a high concentration of Ti (and a low concentration
of Fe) do not expand as fast as Fe-rich regions and, in fact, a
significant fraction of Ti is distributed in the innermost part of
the remnant (see 3rd column in Fig.~\ref{unsh_fe_ti} for model
W15-2-cw-IIb-HD+dec).  With decay heating, the unshocked ejecta
become more structured with sharper contours, and Ti-enriched regions
appear more clearly offset from volumes with high Fe abundance.
Without decay heating the Fe and Ti structures seem to be more
interwoven (see model W15-2-cw-IIb-HD in Fig.~\ref{unsh_fe_ti}).

\begin{figure*}[!t]
  \begin{center}
    \leavevmode
        \epsfig{file=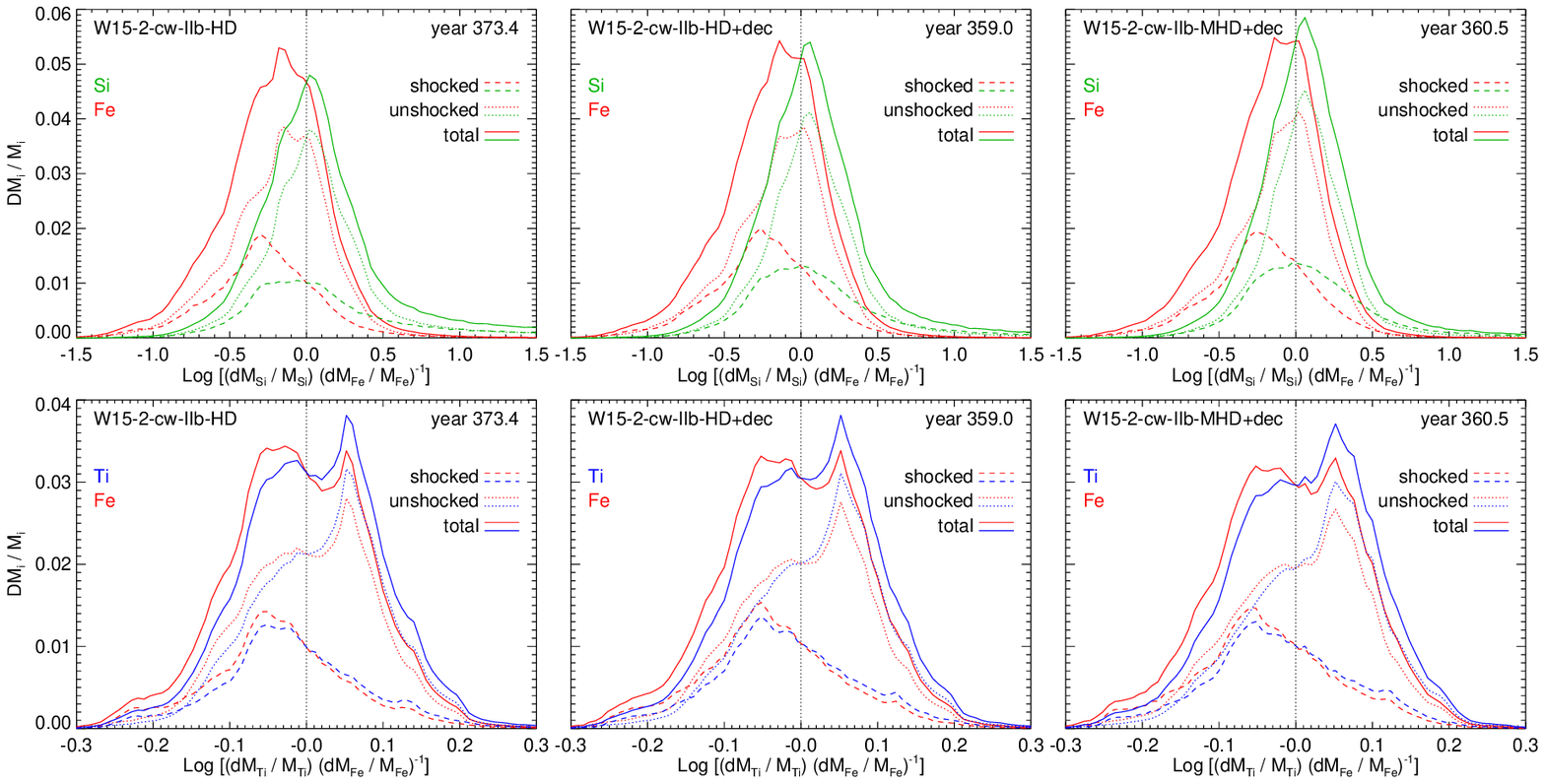, width=18cm}
	\caption{Mass distribution of Si (green), Ti (blue), and
	Fe (red) versus the ratio $R\rs{i,Fe} = \log [(dM\rs{i}
	/ M\rs{i})(dM\rs{Fe} / M\rs{Fe})^{-1}]$ (where $i$
	stands for Si or Ti) at the age of \casa\ for the three
	models. The solid lines correspond to the total distributions; the
	dotted (dashed) lines display the distributions of unshocked
	(shocked) ejecta. The vertical dotted black lines indicate
	an equal concentration of two species ($R\rs{i,Fe} = 0$).}
  \label{fe_vs_ti-si_snr}
\end{center} \end{figure*}

We note that our models do not contain any mechanisms able to
decouple Fe from Ti. Fig.~\ref{fe_vs_ti-si_snr} shows the mass
distributions of Si, Ti, and Fe versus the abundance ratio, $R\rs{i,Fe}$
(where $i$ stands for Si or Ti; see Sect.~\ref{sec:sn_model}), at
the age of \casa\ for our three models. Negative values of $R\rs{i,Fe}$
indicate relatively higher concentrations of Fe, and vice versa.
The mass distributions of the selected elements versus $R\rs{i,Fe}$
do not change significantly during the whole remnant evolution,
thus indicating that the relative abundances of these elements in
each computational cell do not change appreciably during the
interaction of the ejecta with the reverse shock or because of the
effects of Ni decay. This is plausible
because neither of these effects can segregate the nuclear species
on macroscopic scales. In fact the mass distributions obtained with
the models including the Ni decay (models W15-2-cw-IIb-HD+dec,
W15-2-cw-IIb-MHD+dec) are almost the same as those of the model
neglecting the Ni decay (model W15-2-cw-IIb-HD).

The distributions of Ti and Fe versus $R\rs{Ti,Fe}$ are similar to
each other and have almost the same narrow and symmetric shape
observed for Ti and [Ni + X] soon after the shock breakout (compare
lower panels in Figs.~\ref{fe_vs} and \ref{fe_vs_ti-si_snr}). This
result confirms that, at the age of \casa, the bulk of Ti and Fe
almost coexists in the mass-filled volume. This is not surprising
because $^{44}$Ti and $^{56}$Ni were both synthesized in regions
of Si burning and of processes like $\alpha$-particle-rich
freezeout and, after the shock breakout, the $^{56}$Ni decayed in
$^{56}$Co and the latter in $^{56}$Fe. The spread of values in the
range $-0.3 < R\rs{Ti,Fe} < 0.3$ reflects some local differences
of the abundance ratios of these species with clumps of ejecta which
are with a higher concentration of either Ti or Fe.  The mass
distributions of shocked (dashed lines in Fig.~\ref{fe_vs_ti-si_snr})
and unshocked (dotted lines) ejecta indicate that the former have
a higher concentration of Fe (distributions peaking at negative
values of $R\rs{Ti,Fe}$), whereas the latter have a higher concentration
of Ti (distributions peaking at positive values). For instance, in
model W15-2-cw-IIb-HD+dec, about 35\% of Fe against 33\% of Ti are
shocked at the age of \casa. Nevertheless, our models predict that
a significant amount of Ti has already interacted with the reverse
shock at this time. This is basically due to the fact that the
outermost tip of the Fe- and Ti-rich plumes (the first to interact
with the reverse shock) have a higher concentration of Fe than Ti,
but carry also considerable amounts of Ti.

In the case of Si and Fe, the mass distributions are broad and
asymmetric like those derived soon after the shock breakout (compare
upper panels in Figs.~\ref{fe_vs} and \ref{fe_vs_ti-si_snr}), thus
reflecting large differences in the abundance ratios of these two species.
At the age of \casa, however, the distributions of Si and Fe versus
$R\rs{Si,Fe}$ appear sharper than those derived soon after the shock
breakout, suggesting a significant mixing between the two species
during the interaction with the reverse shock. As found for Ti, the
mass distributions of shocked and unshocked ejecta suggest that the
former have, on average, a higher concentration of Fe, whereas the
latter have a higher concentration of Si. This reflects the
spatial inversion of ejecta layers discussed in Sect.~\ref{sh_ejecta}:
Fe seems to be more concentrated at larger radii than Si, opposite
to how they are formed before and during the explosion.

\subsection{$^{44}$Ti and $^{56}$Fe distributions at the age
of \casa: comparison with NuSTAR observations}
\label{sec:nustar}

\begin{figure*}[!t]
  \begin{center}
    \leavevmode
        \epsfig{file=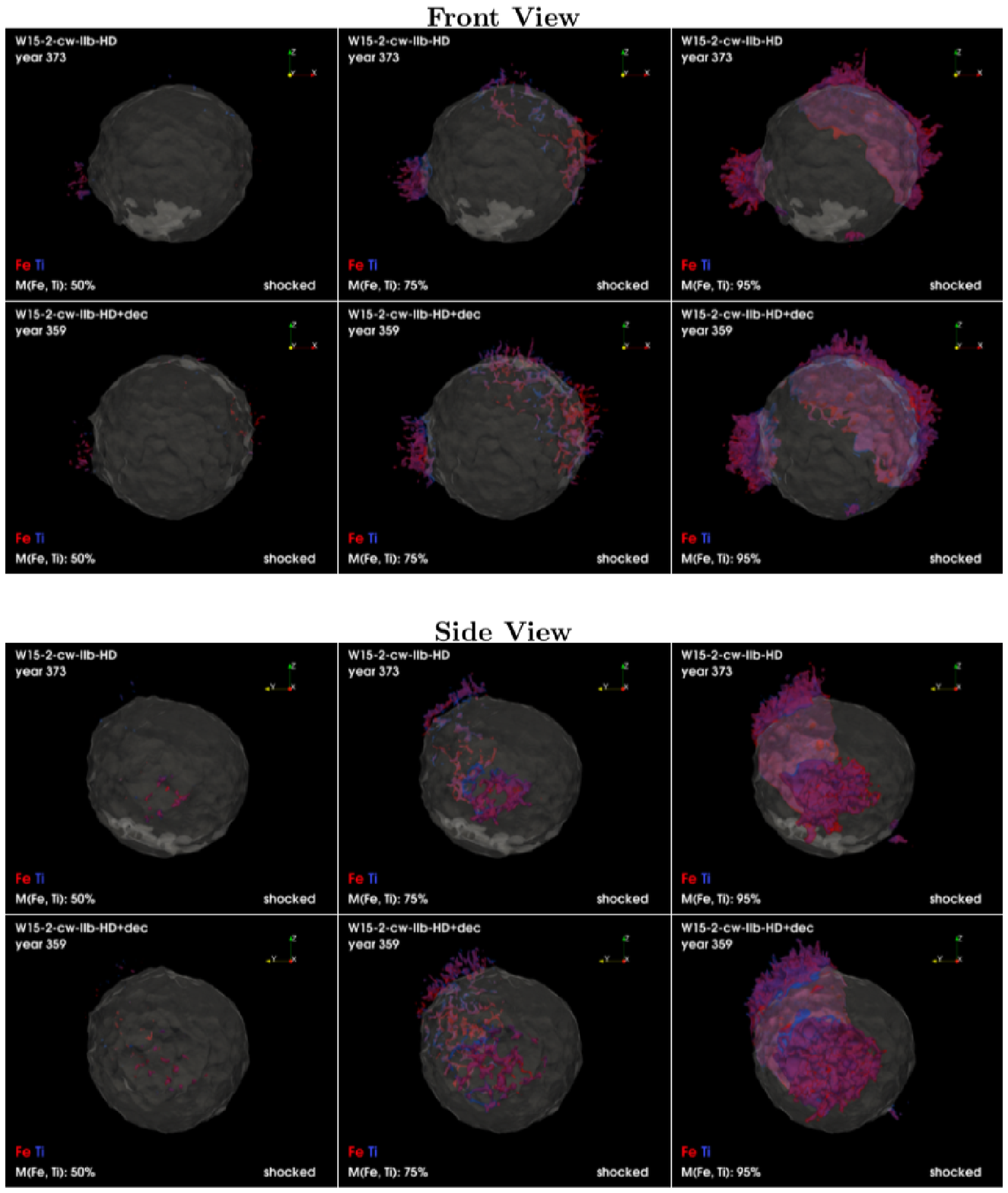, width=17cm}
	\caption{As in Fig.~\ref{unsh_fe_ti} but for the distributions
	of shocked Fe (red) and Ti (blue). The values of
	the mass fractions are selected so that the corresponding
	isosurfaces enclose, from left to right, 50\%, 75\%, and
	95\% of the total mass of the respective shocked species.
	The red isosurfaces correspond to a value of Fe density in
	model W15-2-cw-IIb-HD (W15-2-cw-IIb-HD+dec) which is, in
	units of $10^{-25}$~g~cm$^{-3}$, 72 (50), 35 (26), 7.5
	(7.9), respectively; the blue isosurfaces correspond to a
	value of Ti density in model W15-2-cw-IIb-HD (W15-2-cw-IIb-HD+dec)
	which is, in units of $10^{-26}$~g~cm$^{-3}$ (considering
	the tracer-particle-based post-processing with a large
	nuclear network, \citealt{2017ApJ...842...13W}, but not
	considering its decay in $^{44}$Ca), 21 (15), 10 (7.9), 2.8
	(2.5), respectively (considering the decay of $^{44}$Ti in
	$^{44}$Ca, these values are scaled by a factor of 0.027,
	for an e-folding time of 90~years). A navigable 3D graphic
	of these distributions is available at https://skfb.ly/6UMxI.}
  \label{sh_fe_ti}
\end{center}
\end{figure*}

To further investigate the distribution of Ti versus Fe and to
compare our model results with the findings from the analysis of
NuSTAR observations (\citealt{2014Natur.506..339G, 2017ApJ...834...19G}),
we derived the distributions of shocked Ti and Fe (see Fig.~\ref{sh_fe_ti})
in addition to the distributions of unshocked Ti and Fe displayed
in Fig.~\ref{unsh_fe_ti} (see also Sect.~\ref{sec:unshocked}). In
fact, present NuSTAR observations compared the observed distribution
of Ti only with the distribution of the shocked Fe. Fig.~\ref{sh_fe_ti}
shows that regions of shocked Fe are also rich in Ti. As expected,
these regions are more extended in model W15-2-cw-IIb-HD+dec than
in W15-2-cw-IIb-HD due to the inflation of the Fe-rich regions
driven by the radioactive decay heating of the initial Ni. For
the same reason, the small-scale RT fingers developing from the
regions of shocked Fe and Ti are more developed in the presence of
decay heating. We also note that 50\% of the mass of the shocked
Fe and Ti is enclosed in a few small clumps of the ejecta (see left
panels in Fig.~\ref{sh_fe_ti}). These structures are the relic of
the small-scale clumps present in the tips of the Fe-rich plumes
of the ejecta soon after the shock breakout at the stellar surface
(see Fig.~10 in \citealt{2017ApJ...842...13W}).

\begin{figure*}[!t]
  \begin{center}
    \leavevmode
        \epsfig{file=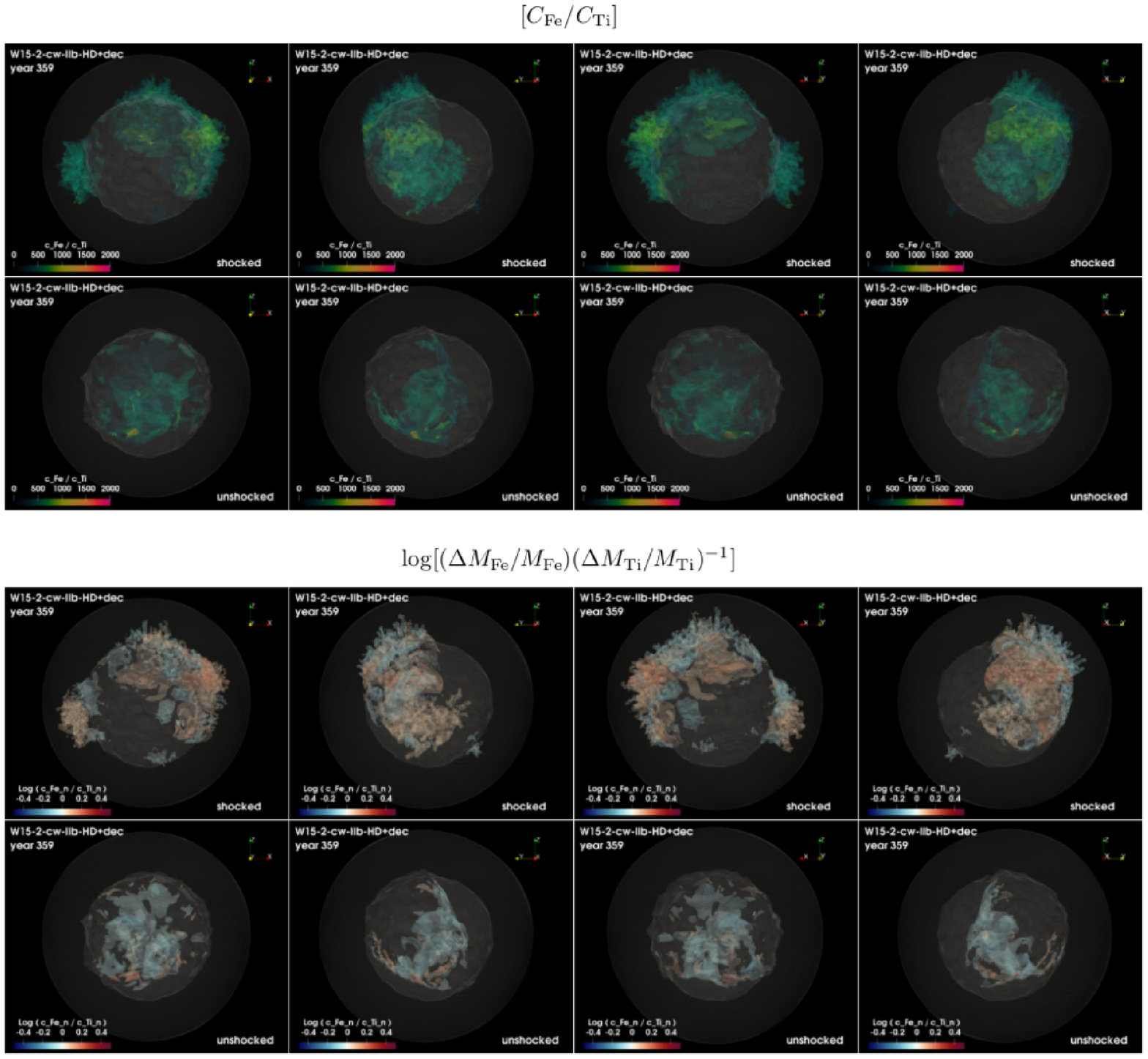, width=17cm}
	\caption{3D distributions of mass fraction ratio $[C\rs{Fe}
	/ C\rs{Ti}]$ (upper 2 rows) and log of abundance ratio,
	$\log [(dM_{\rm Fe} / M_{\rm Fe})\times (dM_{\rm
	Ti} / M_{\rm Ti})^{-1}]$ (lower 2 rows), for different
	viewing angles (from left to right) for model W15-2-cw-IIb-HD+dec
	at the age of \casa. The results for both shocked and
	unshocked ejecta are shown. The two semi-transparent
	quasi-spherical gray isosurfaces indicate the forward
	(outermost surface) and reverse (inner surface) shocks.
	The ratios are calculated in cells with density of Fe and
	Ti greater than 5\% of the maximum density of the respective
	species.}
	\label{fe_vs_ti}
\end{center} \end{figure*}

Fig.~\ref{fe_vs_ti} shows the mass fraction ratio $[C\rs{Fe} /
C\rs{Ti}]$ (upper 2 rows) and the logarithm of the abundance ratio,
$\log [(dM_{\rm Fe} / M_{\rm Fe})\times (dM_{\rm Ti} /
M_{\rm Ti})^{-1}]$ (lower 2 rows, see also online Movie 4), where
$dM_{\rm Fe}$ and $dM_{\rm Ti}$ are the fractions of
Fe and Ti in the domain cell. The mass fraction ratio can vary by
3 orders of magnitude in the 3D domain (in agreement with
\citealt{2020ApJ...895...82V}; see also \citealt{1998ApJ...492L..45N}).
Assuming that the masses of Fe (total mass\footnote{As mentioned
in Sect.~\ref{rem_exp}, here we considered only $^{56}$Fe from
$^{56}$Ni decay. The Fe mass, therefore, is underestimated by about
10\% (e.g.  \citealt{2017ApJ...842...13W}). Additional mass comes
from other isotopes as $^{52}$Fe.} of $9.57\times 10^{-2}\,M_{\odot}$)
and Ti (total mass\footnote{The mass was estimated by considering
the results of the tracer-particle-based post-processing with the
large nuclear network (\citealt{2017ApJ...842...13W}), but not
considering the decay of $^{44}$Ti in $^{44}$Ca. At the age of
\casa, the density of $^{44}$Ti is scaled by a factor of 0.027, for
an e-folding time of 90~years.} of $1.57\times 10^{-4}\,M_{\odot}$)
are diluted in a similar mass-filled volume, we expect an average
mass fraction ratio of $\approx 600$.
Fig.~\ref{fe_vs_ti} shows that the variations of the mass fraction
ratio around this average value can be even larger than a factor
of 10: regions with a ratio $> 600$ (in yellow and red in the figure)
have the highest concentration of Fe relative to Ti, whereas regions
with ratio below $\approx 600$ (in dark blue) have the lowest
concentration of Fe with respect to Ti. The regions with dominant
concentration of Fe (in red) and those with dominant concentration
of Ti (in blue) are clearly evident in the 3D distribution of the
abundance ratio (lower 2 rows of Fig.~\ref{fe_vs_ti}).

From Figs. \ref{sh_fe_ti} and \ref{fe_vs_ti},\, we conclude that our
model is in line with the observation of (shocked) regions with
high concentrations of Fe but low abundance of $^{44}$Ti as
reported by \cite{2017ApJ...834...19G} on grounds of NuSTAR results.
In particular, these authors found Ti-rich ejecta located both
exterior and interior to the reverse shock, as predicted in our
simulations (see Figs.~\ref{unsh_fe_ti}, \ref{sh_fe_ti}, \ref{fe_vs_ti}
and online Movie 4). As in \casa, in our models the bulk of
shock-heated Ti is located in regions of shock-heated Fe,
but there are some regions of shock-heated Fe (marked yellow in
the distribution of mass-fraction ratio and red in the distribution
of abundance ratio; see shocked ejecta in Fig.~\ref{fe_vs_ti}) with
a low concentration of Ti that can even be considerably
lower than a factor of 2 (see upper panels in Fig.~\ref{fe_vs_ti};
see also \citealt{2017ApJ...842...13W}). These variations of the
mass-fraction ratio of Fe and Ti are the relic of the structure of
the two nuclei when they were synthesized. This is also consistent
with the evidence that, in general, the concentration of Fe in
shocked ejecta is slightly higher than average (mass distributions
of shocked ejecta peaking at negative values of $R\rs{Ti,Fe}$) in
all our three models (see Fig.~\ref{fe_vs_ti-si_snr}). Our models
also predict that a significant amount of unshocked Fe is interior
to the reverse shock, preferentially in regions with a relatively
high concentration of Ti ($R\rs{Ti,Fe} \approx 0.05$). However
unshocked Fe cannot be detected in \casa\ until it is heated by the
reverse shock.

\subsection{Properties of X-ray emitting ejecta at the age of \casa}

\begin{figure*}[!t]
  \begin{center}
    \leavevmode
        \epsfig{file=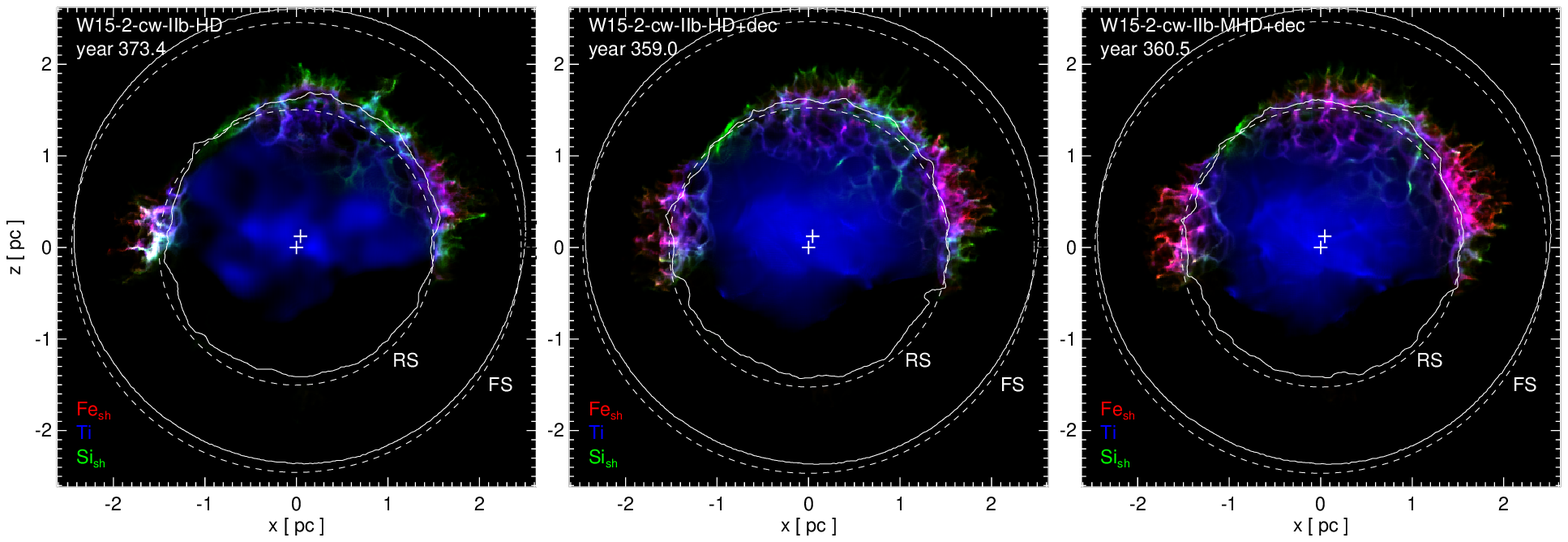, width=18cm}
	\caption{Emission measure, EM, of shocked Fe (red) and Si
	(green) integrated along the $y$-axis (i.e. the LoS assuming
	the vantage point at Earth), and column density, $n\rs{Ti}$,
	of Ti (both shocked and unshocked; blue) at the age
	of \casa\ for our three models. The solid contours mark
	cuts of the forward and reverse shocks in the plane of the
	sky passing through the center of the explosion; the dashed
	lines mark the same cuts but for spheres with the centers
	coincident with the center of explosion and radii corresponding
	to the average radii of the forward and reverse shocks. The
	two crosses in each panel mark the center of the explosion
	and the geometric center of the forward and reverse shocks
	offset to the north-west.}
  \label{maps_obs}
\end{center} \end{figure*}

\casa\ observations show X-rays by thermal emission arising from
optically thin shocked plasma and by nuclear transitions of unstable
elements (in particular $^{44}$Ti). In the first case, the contribution
of the $i$-th atomic species is proportional to the Emission Measure
(EM), defined as EM\,$= \int n\rs{e} n\rs{i} dV$ (where $n\rs{e}$
is the electron density, $n\rs{i}$ the density of species $i$, and
$dV$ the volume element); in the second case, the emission is
proportional to the density of the unstable element. For a better
comparison of the model results with X-ray observations of \casa,
Fig.~\ref{maps_obs} shows the EM of shocked Fe (red) and Si (green)
integrated along the $y$-axis (i.e. the LoS assuming the vantage
point at Earth) at the age of \casa\ for our three models; the
figure also shows the column density, $n\rs{Ti}$, of shocked and
unshocked Ti (both contributing to X-ray emission; blue) at
the same epoch. In Appendix \ref{app:nh_em}, we compare the EM
distributions of shocked Fe and Ti integrated along the $y$-axis
at the age of \casa\ for our three models and the corresponding
column density maps of unshocked Fe and Ti.

The resemblance of our models with the morphology of \casa\ is
impressive. Shocked Fe and Si are characterized by a highly
inhomogeneous and filamentary structure with Fe-rich regions
surrounded by Si-rich ejecta as observed in \casa. These characteristics
are more evident in models including the radioactive decay, because
the Fe-rich ejecta penetrating the reverse shock are faster and
denser (or, in other words, have a higher contrast of velocity and
density with respect to the surrounding ejecta) than in model
W15-2-cw-IIb-HD (see Sect.~\ref{sh_ejecta}). Once more, this
underlines the role played by the decay heating in shaping the final
remnant morphology.

The fingers of shocked Fe are more evident (and well defined) in
model W15-2-cw-IIb-MHD+dec than in the others. In this case, the magnetic
field lines in the mixing region follow the plasma structures formed
during the growth of HD instabilities at the contact discontinuity.
This leads to the development of a preferentially radial component
around the RT fingers (e.g. \citealt{1994ApJ...433..757M,
1996ApJ...472..245J}). The tension of the field lines determines
a more laminar flow around the fingers and limits the growth
of HD instabilities which would destroy and fragment the Fe-rich
structures (see also \citealt{2005ApJ...619..327F, 2008ApJ...678..274O,
2012ApJ...749..156O}). As a result, the fragmentation of the
crown-like structures in the mixing region is less efficient and
the structures survive for a longer time. The effect of magnetic
field is also evident in Fig.~\ref{inversion}, where the density
structure of Fe, Ti, and Si is different when comparing models
W15-2-cw-IIb-HD+dec and W15-2-cw-IIb-MHD+dec (see also Appendices
\ref{app:vel_distr} and \ref{app:ej_distr} for the effects of an
ambient magnetic field on the structure of the ejecta).

Fig.~\ref{maps_obs} also shows the column density, $n\rs{Ti}$, of
$^{44}$Ti at the age of \casa, a proxy of the X-ray emission expected
from nuclear transitions of unstable Ti. The bulk of this
species is concentrated in the northern hemisphere, consistent with
the initial asymmetry, suggesting that its momentum direction points
to the north and is tilted by $\approx 30^{\rm o}$ out of the plane
of the sky away from the observer. The momentum direction of Ti is
roughly opposite to the direction of the CCO kick velocity pointing
southward toward the observer (see also \citealt{2017ApJ...842...13W}).
This is in rough agreement with NuSTAR observations of \casa\ that
have shown an average momentum direction of Ti to the north, tilted
by $58^{\rm o} \pm 20^{\rm o}$ out of the plane of the sky away
from the observer, opposite to the inferred direction of motion of
the CCO (\citealt{2017ApJ...834...19G}).  The figure also shows
that Ti-rich ejecta would emit from regions located both interior
and exterior to the reverse shock, in agreement with the Ti
distribution discussed in Sect.~\ref{sec:nustar} and, again, with
NuSTAR observations ((\citealt{2017ApJ...834...19G}). In the case
of shocked ejecta, most of the emission from Ti would originate
from regions with significant emission from shocked Fe (purple areas
in Fig.~\ref{maps_obs}), although there are also regions with
prominent Fe emission and low contribution from shocked Ti (more
reddish areas in Fig.~\ref{maps_obs}).

The map of $n\rs{Ti}$ in model W15-2-cw-IIb-HD is more clumpy than
in the other two cases, suggesting that the distribution of Ti was
smoothed by Ni radioactive heating occurring at early times in
models W15-2-cw-IIb-HD+dec and W15-2-cw-IIb-MHD+dec. Interestingly,
the NuSTAR images show a clumpy distribution of $^{44}$Ti which is
remarkably similar to that produced by model W15-2-cw-IIb-HD. We
note that, in models with decay heating, all of the decay energy
is assumed to be deposited in the ejecta instead of allowing for
some $\gamma$-ray leakage from the inner part of the remnant. Thus,
these models most likely overestimate the smoothing effect by decay
heating. In the light of this, the models W15-2-cw-IIb-HD and
W15-2-cw-IIb-HD+dec represent the two extreme cases of either no
or full energy deposition, and the case of \casa\ should be
between the two.

Fig.~\ref{maps_obs} also shows that both the forward and reverse
shocks deviate from a spherically symmetric expansion around the
center of the explosion (compare solid and dashed contours in
Fig.~\ref{maps_obs}). They expand faster to the north-west than to
the southeast. This is consistent with the initial large-scale
asymmetry in which Ni and Ti were both mostly concentrated in the
northern hemisphere propagating away from the observer, whereas the
CCO travels in the southern hemisphere toward the observer with a
kick velocity of $\approx 600$~km~s$^{-1}$ (\citealt{2010ApJ...725L.106W,
2013A&A...552A.126W, 2017ApJ...842...13W}). We note that observations
of \casa\ in different bands also show that the forward and reverse
shocks deviate from a spherically symmetric expansion; in particular,
the reverse shock is offset to the north-west by $\approx 0.22$~arcmin
($\approx 0.2$~pc at the distance of $3.4$~kpc) from the geometric
center of the forward shock (\citealt{2001ApJ...552L..39G}). However,
at odds with the observations, in our simulations the geometric
centers of the two shocks coincide and both are offset to the
north-west by $\approx 0.13$~pc from the center of the explosion.

\begin{figure}[!t]
  \begin{center}
    \leavevmode
        \epsfig{file=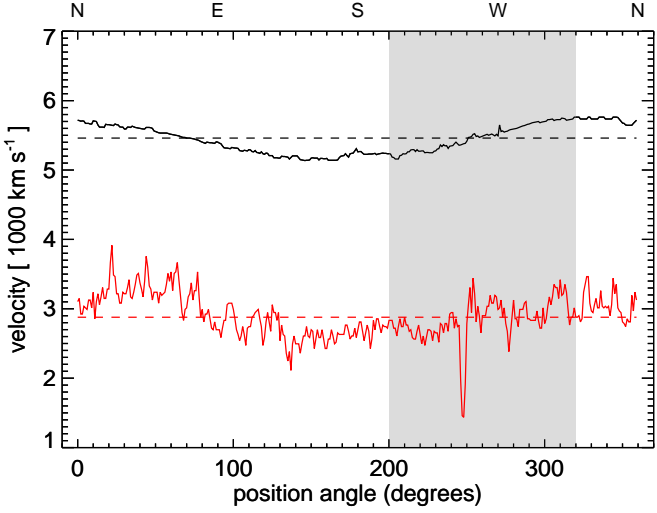, width=9cm}
	\caption{Forward (black) and reverse (red) shock velocities
	versus the position angle in the plane of the sky for model
	W15-2-cw-IIb-HD+dec at the age of \casa. The dashed horizontal
	lines mark the median values of the respective velocities.
	The shaded area marks the position angle where the reverse
	shock appears to be stationary or even moving inward in
	observations of \casa\ (e.g.~\citealt{2008ApJ...686.1094H,
	2018ApJ...853...46S}).}
  \label{shock_vel}
\end{center} \end{figure}

Fig.~\ref{shock_vel} displays the profiles of the forward and reverse
shock velocities versus the position angle in the plane of the sky
for model W15-2-cw-IIb-HD+dec at the age of \casa\ (again, we found
similar results for the other two models). These values can be
compared with the forward and reverse shock velocities measured in
the observer frame from the analysis of observations. The expansion
of the forward shock shows velocities around $\approx 5500$~km~s$^{-1}$
consistent with observations in different wavelength bands
(e.g.~\citealt{2003ApJ...589..818D, 2009ApJ...697..535P,
2019sros.confE..32F}). As for the reverse
shock, we found that our models predict an average expansion with
velocities ranging between $2000$~km~s$^{-1}$ and $4000$~km~s$^{-1}$.
This is consistent with the velocities inferred in the eastern and
northern hemisphere of the remnant (e.g.~\citealt{2018ApJ...853...46S,
2019sros.confE..32F}), but it is in clear contrast to the values
inferred from observations in the western hemisphere, where the
reverse shock appears to be stationary or even moving inward in the
observer frame (see shaded area in Fig.~\ref{shock_vel};
e.g.~\citealt{2008ApJ...686.1094H, 2018ApJ...853...46S}).

The origin of this asymmetry in the reverse shock properties is
vividly debated in the literature. Our ejecta morphology and many
other model features show great resemblance to \casa. However, the
fact that our simulations do not reproduce the displacement of the
center of the reverse shock relative to the center of the forward
shock points to something missing. Our simulations assume an isotropic
wind; but the interaction of the forward shock with an anisotropic
CSM might play a role. For instance, an initial spherically symmetric
blast wave expanding through an ambient environment characterized
by global anisotropy can produce an offset between the geomeric
centers of the forward and reverse shocks (see, for instance, upper
left panel in Fig.~2 in \citealt{2007A&A...470..927O}).  So, our
results seem to suggest that environmental asymmetries might play
a role in displacing the center of the forward shock from the
explosion center and probably also in shifting it relative to the
center of the reverse shock (Orlando et al. in preparation). 

\begin{figure}[!t]
  \begin{center}
    \leavevmode
        \epsfig{file=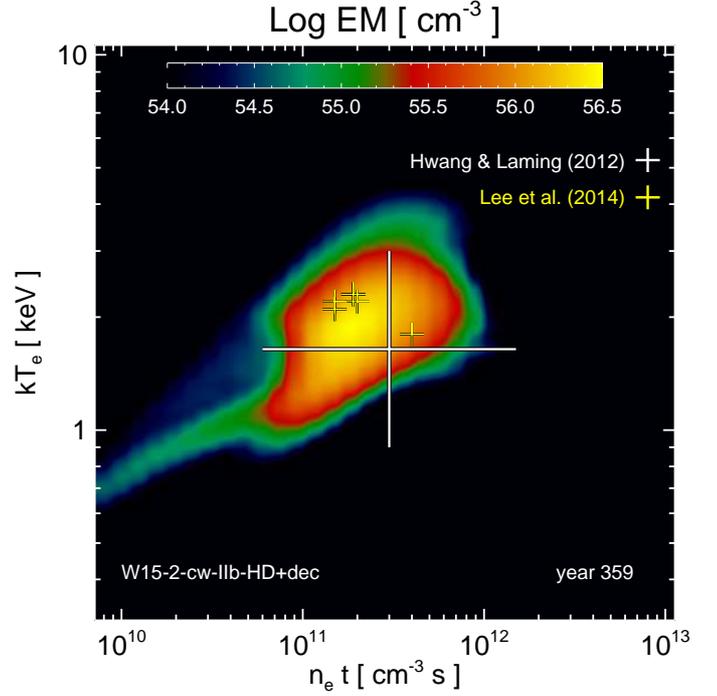, width=9cm}
	\caption{Distribution of emission measure vs. electron
	temperature ($kT\rs{e}$) and ionization age ($n\rs{e}t$)
	at the epoch of \casa\ derived from model W15-2-cw-IIb-HD+dec
	compared with the results of {\em Chandra} X-ray observations.
	The large white cross marks the ranges of $kT\rs{e}$ and
	$n\rs{e}t$ derived from an X-ray survey of ejecta in \casa\
	(\citealt{2012ApJ...746..130H}); the small yellow crosses
	mark the values inferred from the analysis of regions dominated
	by thermal emission of shocked ambient gas
	(\citealt{2014ApJ...789....7L}).}
  \label{map_EM}
\end{center} \end{figure}

From the models, we derived the distribution of the EM as a function
of electron temperature, $kT\rs{e}$, and ionization age, $n\rs{e}t$
(a measure of the deviations from equilibrium of ionization; see
Sect.~\ref{sec:snr_model} and \citealt{2016ApJ...822...22O} for a
description of how $kT\rs{e}$ and $n\rs{e}t$ are calculated), for
the shocked plasma. These distributions are very useful for a
quantitative comparison of the model results with X-ray observations.
Fig.~\ref{map_EM} shows the distributions derived from model
W15-2-cw-IIb-HD+dec at the age of \casa. The distributions derived
from the other two models give analogous results. The figure shows
that the X-ray emitting plasma is largely out of equilibrium of
ionization with the EM distribution peaking at $kT\rs{e}\approx
2$~keV and $n\rs{e}t\approx 2\times 10^{11}$~cm$^{-3}$~s. This
result is similar to that found by \cite{2016ApJ...822...22O} who
designed a model with artificial ad-hoc initial conditions to
reproduce the structure of \casa.  Indeed, our models (in which the
initial large-scale asymmetry originates from stochastic anisotropies
that developed through HD instabilities triggered by neutrino heating
soon after core bounce) predict an EM distribution at the age of
\casa\ that is in excellent agreement with the results from the
analysis of X-ray observations of \casa\
(e.g.~\citealt{2012ApJ...746..130H, 2014ApJ...789....7L,
2020A&A...638A.101G}).  In particular, we found that the distributions
of $kT\rs{e}$ and $n\rs{e}t$ derived from the models span ranges
of values similar to those observed (the observed ranges are indicated
by the white cross in Fig.~\ref{map_EM}), and they are highly peaked
in agreement with the findings of \cite{2012ApJ...746..130H}. As
suggested by these authors, the peaked distributions reflect multiple
secondary shocks following reverse shock interaction with ejecta
inhomogeneities.

\begin{figure}[!t]
  \begin{center}
    \leavevmode
        \epsfig{file=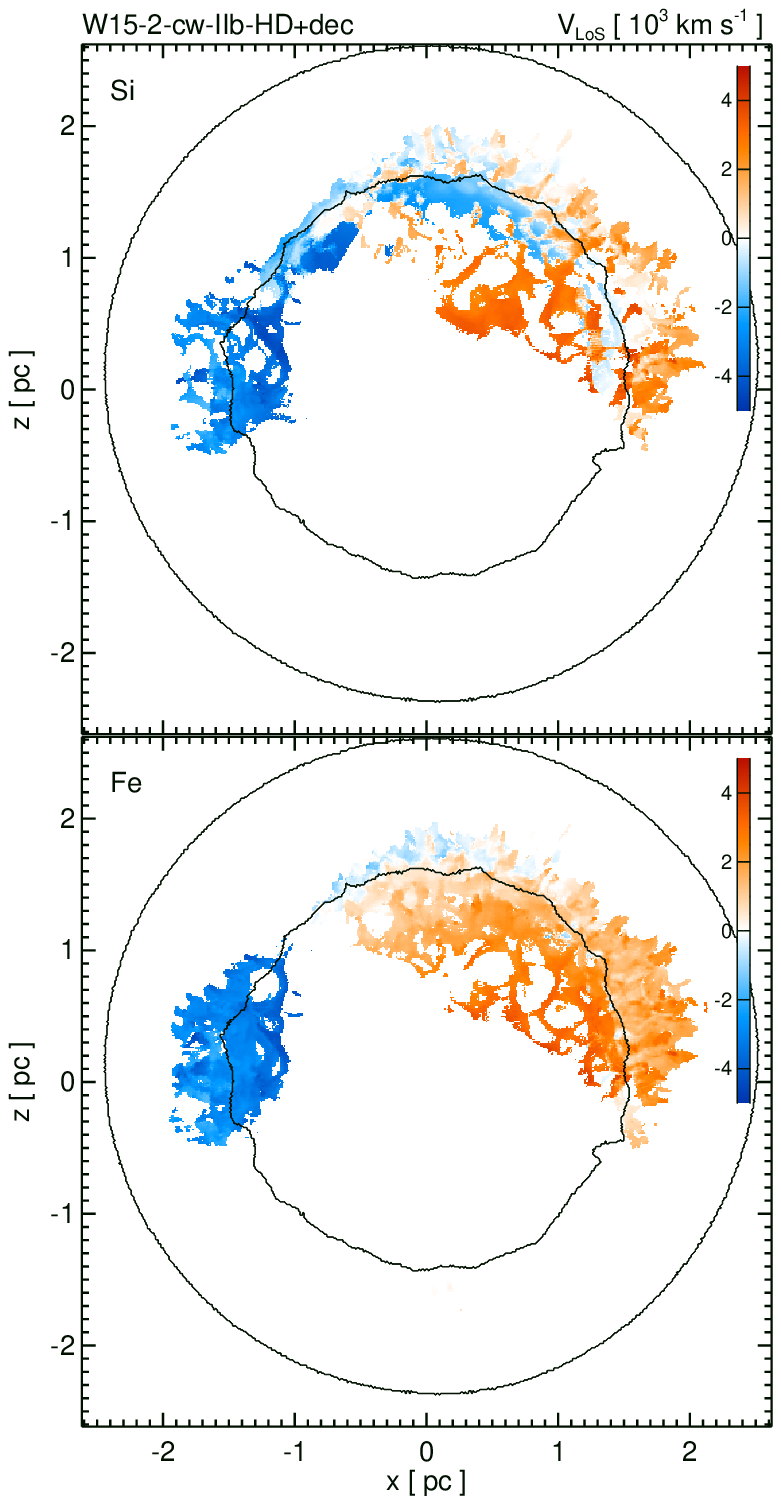, width=9cm}
	\caption{Maps of average EM-weighted velocity along the LoS
	for shocked Si (upper panel) and Fe (lower panel) at the
	age of \casa\ for model W15-2-cw-IIb-HD+dec. The approximate
	velocity range is $\pm 5000$~km~s$^{-1}$ (see color bar on
	the right in units of $1000$~km~s$^{-1}$). The solid contours mark
	the cuts of the forward and reverse shocks in the plane of
	the sky passing through the center of the explosion. These
	images correspond to the effective Doppler maps derived from
	observations.}
  \label{map_doppler}
\end{center} \end{figure}

Fig.~\ref{map_doppler} shows the average EM-weighted velocity of
shocked Fe and Si along the LoS derived from model W15-2-cw-IIb-HD+dec
(again the other two models yield similar results). As expected,
the bulk of high-velocity shocked ejecta that are rich in Si and Fe is
concentrated in regions where the large-scale metal-rich ejecta plumes
interact with the reverse shock. The highest redshifted
Si-rich ejecta ($v\rs{LoS} \approx +5000$~km~s$^{-1}$) form a large
ring-like feature to the north-west; this is the result of the Fe-rich
plume which breaks through the Si layer and determines a large
Fe-rich region. The highest blueshifted Si-rich ejecta ($v\rs{LoS}
\approx -5000$~km~s$^{-1}$) are concentrated in the east region.
These two features have the highest absolute values of the velocity
along the LoS. The shocked Fe is also concentrated in the large
redshifted region to the north-west and the blueshifted region to
the east. The velocity pattern resembles the Doppler images
derived from observations of \casa\ (e.g.  \citealt{2002A&A...381.1039W,
2010ApJ...725.2038D}), and matches remarkably well the approximate
velocity range inferred from observations ($-4000 < v\rs{LoS} <
+5000$~km~s$^{-1}$).

\begin{figure*}[!t]
  \begin{center}
    \leavevmode
        \epsfig{file=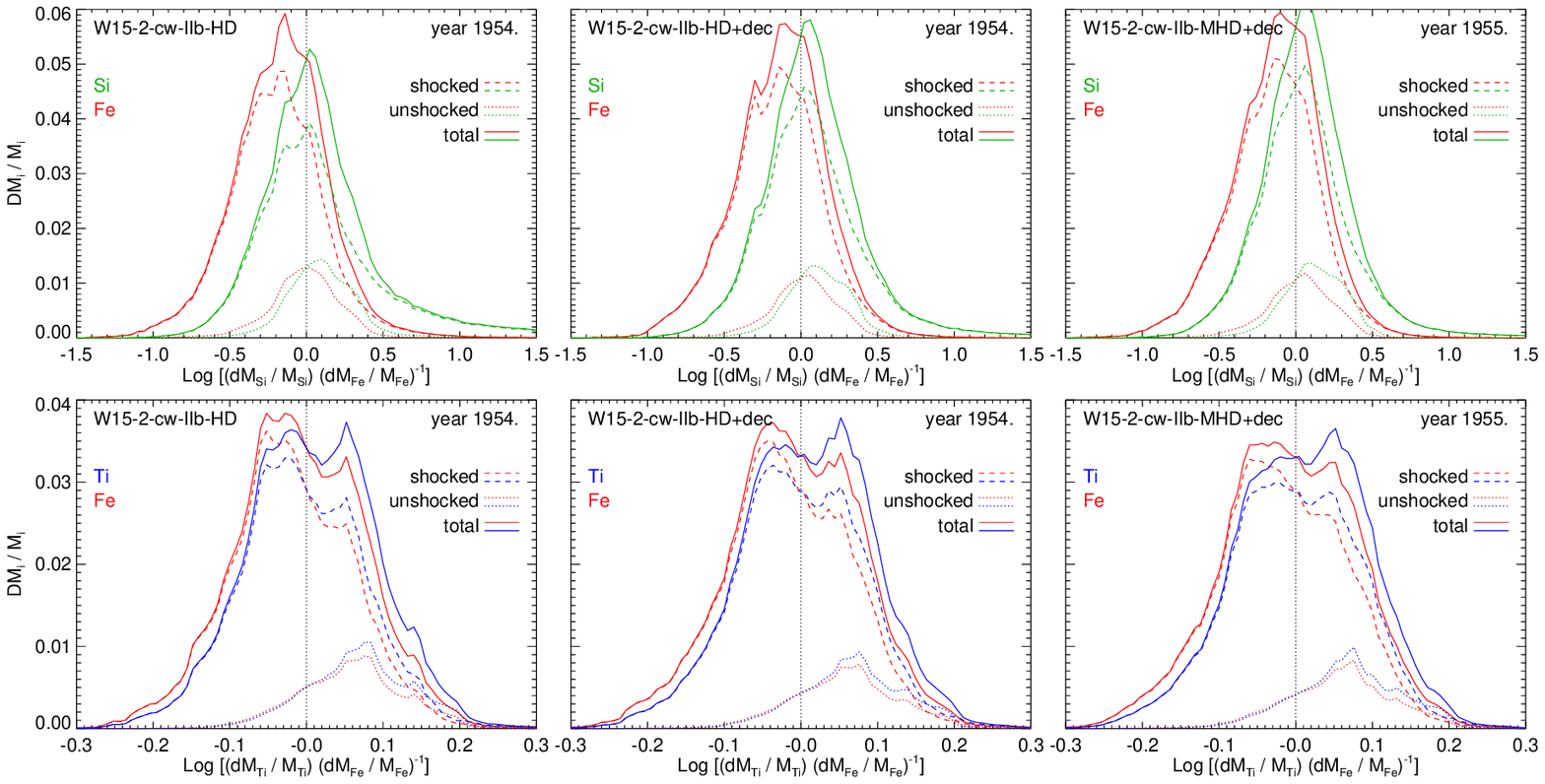, width=18cm}
	\caption{As in Fig.~\ref{fe_vs_ti-si_snr} but at the age of
	$\approx 2000$~years.}
  \label{fe_vs_ti-si_snr_69}
\end{center}
\end{figure*}

It is worth to note that the model presented in \cite{2016ApJ...822...22O}
reproduces the distribution of the ejecta in \casa\ in many details.
In that case, the model adopted artificial initial conditions with
the aim to identify a parameterized initial large-scale asymmetry
and to constrain the masses and energies of the post-explosion
anisotropies that are able to reproduce the observed distributions
of Fe and Si/S. Conversely, in the models presented here, the
initial large-scale asymmetry originates from the stochastic growth
of anisotropies due to HD instabilities triggered by neutrino heating
soon after core bounce. No fine-tuning was performed to match the
morphology of \casa. These models therefore demonstrate that the
basic properties of the ejecta structure in \casa\ can be naturally
explained as originating from asymmetries arising from post-shock
convective instability developing in the first second of the
explosion. The observed features are the fingerprints of a
neutrino-driven SN explosion. Nevertheless, our new simulations
fail in reproducing another crucial aspect of the \casa\ morphology,
namely the formation of the two Si-rich wide-angle ``jets'' (also
called Si-rich ``sprays'' in the literature). These features may
result from post-explosion phenomena (as, for instance, disk accretion
of fallback matter by the new-born neutron star;
\citealt{1989ApJ...346..847C, 2011ApJ...736..108P}) not considered
in our simulations.

\subsection{Distribution of ejecta at later times}

At later times, the remnant expands further through the wind of the
progenitor star. At the end of our simulations, at the age of $\approx
2000$~years, most of the metal-rich ejecta ($\approx 80$\%) have
been shocked (see Fig.~\ref{shocked_mass}). HD instabilities have led to
well developed large-scale fingers and dominate the structuring
and considerable mixing of shocked ejecta. The efficient mixing is
evident from the high-velocity tails of the mass distributions of elements
versus radial velocity that are very similar to each other (see
lower right panel in Figs.~\ref{prof_vel_hd_dec}, \ref{prof_vel_hd},
and \ref{prof_vel_mhd_dec}). The remnant interior is characterized
by material with density much lower than that of the shock-heated ejecta
due to the adiabatic expansion of the unshocked plasma and the high
compression of shock-heated ejecta by the reverse shock; however,
voids and cavities remain, even when the unshocked ejecta are very
diluted. The spatial distribution of shocked Fe and Ti (or its
decay product $^{44}$Ca) suggests that most of their masses are
concentrated in the remnant hemisphere that propagates away from
the observer, taking on the appearance of the spiny outer
shell\footnote{The hole in the shell evident in the lower panels
in Fig.~\ref{distr_fe_time} forms because the bulk of Fe-rich ejecta,
denser than 5\% of the peak density of Fe, has not reached yet the
reverse shock.} of a chestnut (see lower panels in
Fig.~\ref{distr_fe_time}). This is consistent with the initial
large-scale asymmetry that launched most of Fe and Ti to the north,
away from the observer and opposite to the CCO kick velocity (see
Fig.~\ref{ini_cond} and \citealt{2017ApJ...842...13W}).

Figure~\ref{fe_vs_ti-si_snr_69} shows the mass distributions of Si,
Ti (or, most likely, its decay product $^{44}$Ca), and Fe versus
the abundance ratio, $R\rs{i,Fe}$ (where $i$ stands for Si or Ti),
at $t\approx 2000$~years for our three models.  These distributions
appear more peaked than those derived at previous times, although
the changes are small. Again this indicates that, on average, the
relative abundances of these elements in each computational cell
do not change appreciably after the passage through the reverse
shock. At this epoch, the distribution of Si vs. $R\rs{Si,Fe}$
slightly differs among the three models, i.e. the distribution being
more peaked in models including the radioactive decay (compare upper
panels in Fig.~\ref{fe_vs_ti-si_snr_69}). The distributions of Ti
($^{44}$Ca) and Fe versus $R\rs{Ti,Fe}$ are still similar to each
other in the three models indicating that, after 2000 years of
evolution, the bulk of Ti ($^{44}$Ca) and Fe almost coincides. As
during previous epochs, the figure shows that, on average, shocked
ejecta (dashed lines) have a higher concentration of Fe, whereas
unshocked ejecta (dotted lines) have a higher concentration of Ti
($^{44}$Ca), which is again a relic of the efficient outward mixing
of highly Fe-enriched ejecta.

Our simulations at the age of $\approx 2000$~years can be compared
with other O-rich SNRs that are more evolved than \casa, although some word
of caution is needed here: the progenitor model adopted for our
study was a stripped star that had lost most of its H envelope,
its He-core mass was specific to be in the ballpark of estimates
for \casa, and the explosion energy is also not generally valid for
all SNe. Nevertheless,
an interesting candidate for a comparison with our models is an
older cousin of \casa, the Galactic core-collapse SNR G292.0+1.8
(in the following G292), with an estimated age of $\approx 3000$~years
(e.g.~\citealt{2005ApJ...635..365G, 2009ApJ...692.1489W,
2003ApJ...583L..91G}) and average forward and reverse shock radii
of $\approx 7.7$~pc and $\approx 3.8$~pc, respectively (at a distance
of $\approx 6$~kpc; \citealt{2007ApJ...670L.121P, 2003ApJ...594..326G,
2015ApJ...800...65B}), similar to those derived at the end of our
simulations ($\approx 8.2$~pc and $\approx 4.2$~pc, respectively,
at the age of $\approx 2000$~years).  The progenitor was most likely
a star with a zero-age-main-sequence mass in the range between 13
and $30\,M_{\odot}$ (\citealt{2019ApJ...872...31B}). The total
ejecta mass inferred from radio and X-ray observations is in the
range of $6-8\,M_{\odot}$ (\citealt{2003ApJ...594..326G,
2015ApJ...800...65B}), whereas the estimated explosion energy is
$E\rs{exp}\leq 1$~B (\citealt{2003ApJ...583L..91G}), although the
latter value has large uncertainties. We note that
the ejecta mass, $M\rs{ej}$, in G292 is higher than in our model (and
in \casa) and the explosion energy, $E\rs{exp}$, is lower (see
Table~\ref{tabmod}). This means that the ratio $E\rs{exp}/M\rs{ej}
\approx 0.14$~B$/M_{\odot}$ is considerably lower than in our
simulations ($E\rs{exp}/M\rs{ej} \approx 0.45$~B$/M_{\odot}$), thus
suggesting an average velocity of the ejecta soon after the shock
breakout at the stellar surface (approximately given by
$\overline{v}\rs{ej} \approx \sqrt{2E\rs{exp} / M\rs{ej}}$) of
$\approx 3700$~km~s$^{-1}$ versus the $\overline{v}\rs{ej} \approx
6600$~km~s$^{-1}$ of our models. So, we advise the reader to be
cautious when comparing our models with G292. In particular, our
models with their higher ejecta velocities should evolve on a shorter
time scale than G292. Having said that, we note that, as \casa\
(and as predicted by our models), G292 is characterized by a knotty
and filamentary structure.  At odds with our models, the morphology
of G292 also shows evidence of an equatorial belt, probably originating
from shocked dense circumstellar material (not modeled in our
simulations), and thin circumferential filaments
(e.g.~\citealt{2004ApJ...602L..33P, 2007ApJ...670L.121P}).

\begin{figure}[!t]
  \begin{center}
    \leavevmode
        \epsfig{file=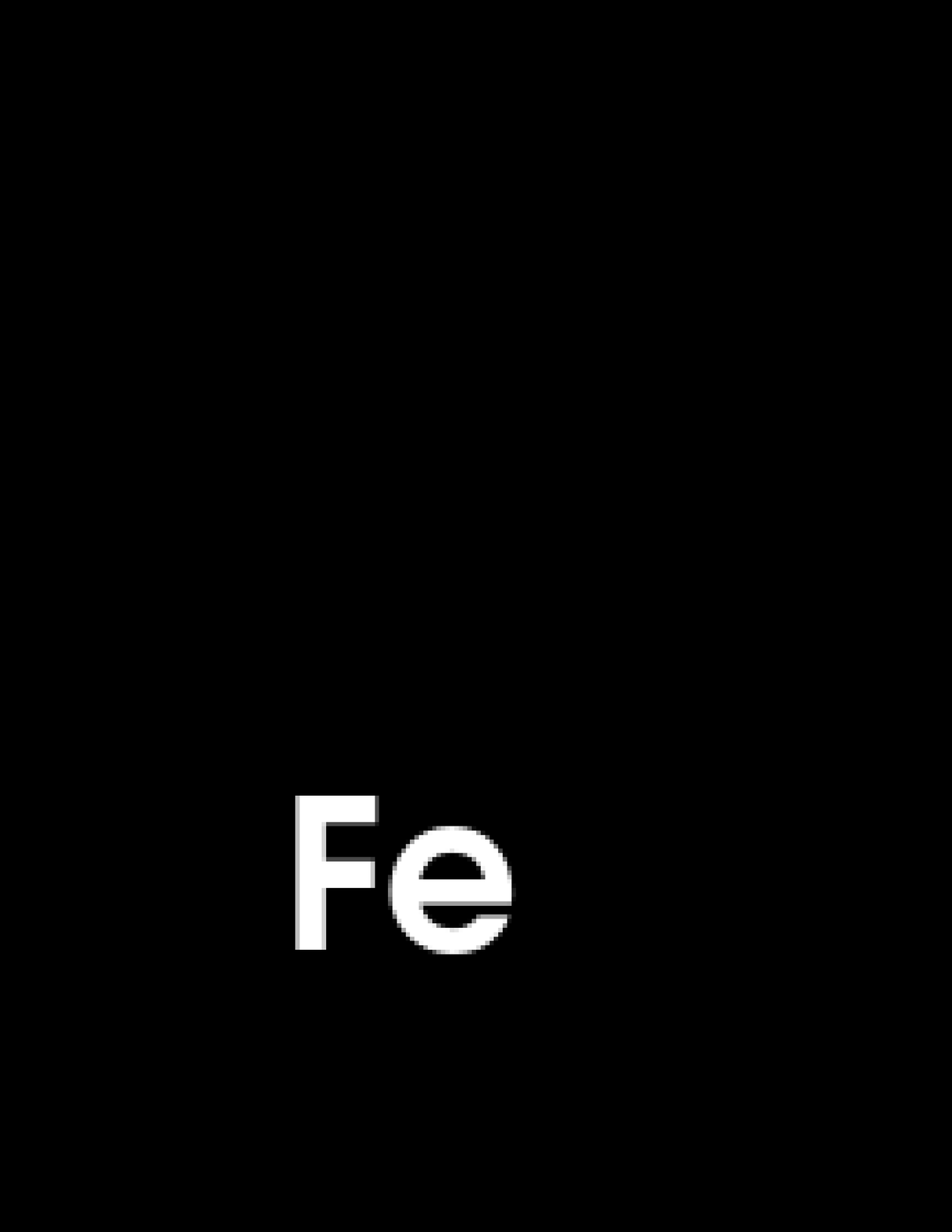, width=9cm}
	\caption{Isosurface of the distribution of Fe at
	$t=2000$~years for model W15-2-cw-IIb-HD+dec. The model is
	oriented to roughly match the direction of the neutron star
	kick and the remnant asymmetries observed in G292. The
	opaque corrugated isosurface corresponds to a value of the Fe
	density which is at 5\% of the peak density; the colors
	give the radial velocity in units of 1000 km s$^{-1}$ on
	the isosurface (the color coding is defined at the bottom
	of the panel). The semi-transparent quasi-spherical surface
	indicates the forward shock. The Earth vantage point lies
	on the negative $y$-axis.}
  \label{sh_fe_g292}
\end{center}
\end{figure}

\begin{figure}[!t]
  \begin{center}
    \leavevmode
        \epsfig{file=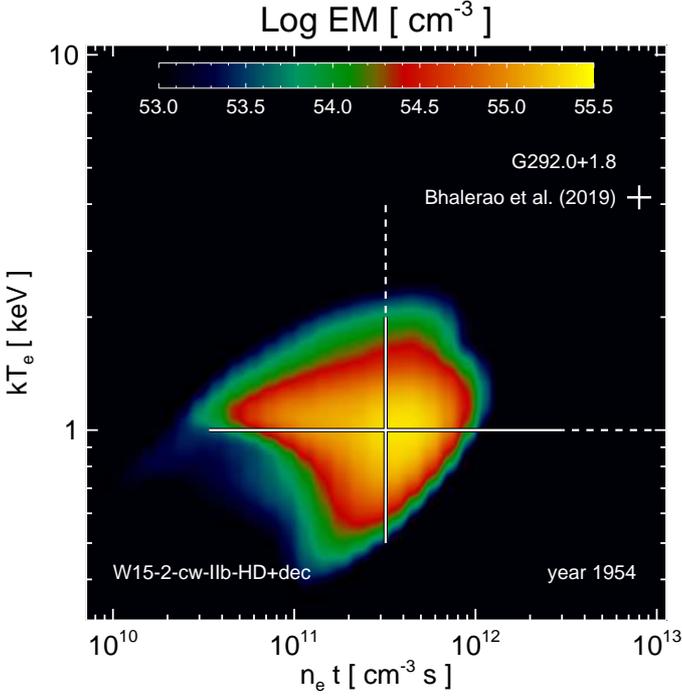, width=9cm}
	\caption{As in Fig.~\ref{map_EM} but at the age of $\approx
	2000$~years for model W15-2-cw-IIb-HD+dec. The solid white
	cross marks the ranges of $kT\rs{e}$ and $n\rs{e}t$ of the
	bulk of the ejecta derived from {\em Chandra} X-ray observations
	of the SNR G292 (\citealt{2019ApJ...872...31B}); the dashed
	lines mark the extreme values of $kT\rs{e}$ and $n\rs{e}t$
	inferred for isolated knots in the north-west hemisphere of
	the remnant.}
  \label{map_EM_69}
\end{center} \end{figure}

For a better comparison between our models and G292, we rotated the
system about the three axes to roughly match the orientation of the
neutron star kick inferred for G292 and the observed remnant
asymmetries, namely $i\rs{x} = -30^{\rm o}$, $i\rs{y} = 60^{\rm
o}$, $i\rs{z} = -90^{\rm o}$ (see Fig.~\ref{sh_fe_g292}). The
analysis of {\em
Chandra\,} observations of G292 has revealed electron temperatures
ranging, in general, between 0.5 and 2 keV, although there are
isolated knots in the north-west hemisphere with significantly higher
electron temperatures (up to $kT\rs{e} \approx 4$~keV;
\citealt{2019ApJ...872...31B}). The ionization time ranges, in
general, between $3.4 \times 10^{10}$~cm$^{-3}$~s and $3\times
10^{12}$~cm$^{-3}$~s, with few small regions in the north-west with
values up to $n\rs{e}t \approx 10^{13}$~cm$^{-3}$~s.  In
Fig.~\ref{map_EM_69} we compare the inferred ranges of values
mentioned above with the distribution of the EM as a function of
$kT\rs{e}$ and $n\rs{e}t$ at $t\approx 2000$~years for model
W15-2-cw-IIb-HD+dec.
At this stage of the evolution, the EM distribution is about an
order of magnitude lower than that at the age of \casa\ due to the
remnant expansion that has diluted the X-ray emitting plasma.  The
ejecta at this epoch are on average characterized by lower temperatures
and slightly higher ionization ages than found at the age of \casa,
in agreement with the adiabatic expansion of the remnant and with
the progressive approach of the shocked plasma to ionization
equilibrium. In particular, the bulk of the shocked ejecta mixed
with shocked wind material extends over the ranges $0.6$~keV$< kT\rs{e}
< 2$~keV and $10^{11}$~cm$^{-3}$~s~$<n\rs{e}t < 10^{12}$~cm$^{-3}$~s,
peaking at $kT\rs{e} \approx 1$~keV and $n\rs{e}t\approx 3.5\times
10^{11}$~cm$^{-3}$~s. These values are consistent with those inferred
for G292, although the observed remnant shows higher values of
$n\rs{e}t$, indicating that the plasma is closer to the equilibrium
of ionization. This could be a consequence of the fact that the
ratio $E\rs{exp}/M\rs{ej}$ of G292 is considerably lower than for
our models or that the remnant is expanding through an ambient
medium which is denser than that of our model. Nevertheless, we
consider it as astounding that the modeled remnant describes the
properties of the shocked plasma inferred from observations so
closely, especially because our models have not been tuned to
describe the progenitor star, the SN explosion, the CSM, and the
evolution of G292.

The observations have also shown that the ejecta in G292 are
characterized by high velocities, with the number of blueshifted
knots larger than that of redshifted ones (\citealt{2005ApJ...635..365G,
2015ApJ...800...65B}). The measurements of the radial velocities
(i.e., the velocities along the LoS) reveal ranges between
$-1800$~km~s$^{-1}$~$< v\rs{rad} < 1490$~km~s$^{-1}$ in the east-west
direction, and $-3570$~km~s$^{-1}$~$< v\rs{rad} < 2340$~km~s$^{-1}$
in the north-south direction, thus indicating that the fastest ejecta
are blueshifted and distributed along the north-south direction
(\citealt{2005ApJ...635..365G, 2009ApJ...692.1489W}).  This evident
asymmetry has been interpreted as the result of an asymmetric SN
explosion, with the stellar debris ejected fastest along an axis
oriented approximately north–south in the plane of the sky and in
direction toward the observer (\citealt{2009ApJ...692.1489W}). The
range of velocities inferred from observations is consistent with
the range of radial velocities predicted by our models at the age
of $\approx 2000$~years (see Fig.~\ref{prof_vel_hd_dec}), with the
high-velocity tails extending up to $v\rs{rad} \approx 3500$~km~s$^{-1}$.

\begin{figure}[!t]
  \begin{center}
    \leavevmode
        \epsfig{file=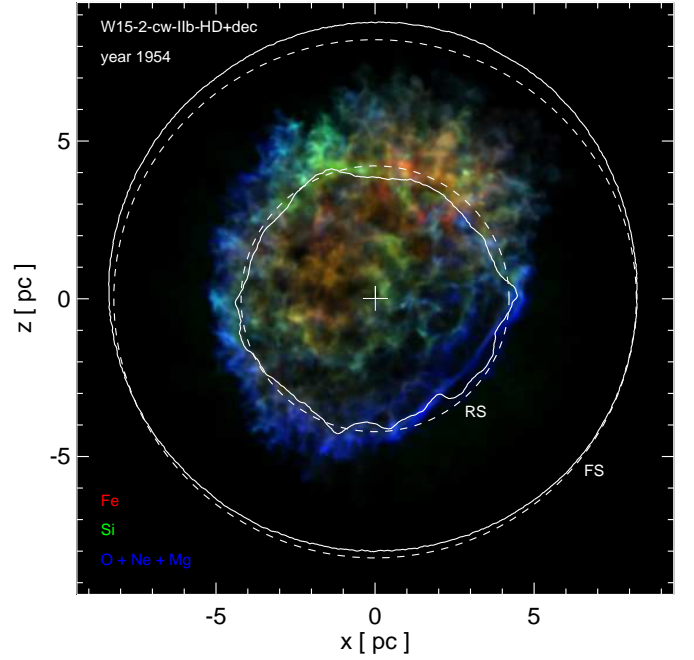, width=9cm}
	\caption{Spatial distributions of EM of shocked Fe (red),
	Si (green), and intermediate-mass elements (O, Ne, and Mg;
	blue) integrated along the LoS (assuming the same orientation
	of \casa) at the age of $\approx 2000$~years for model
	W15-2-cw-IIb-HD+dec. The solid contours mark the cuts of
	the forward and reverse shocks in the plane of the sky
	passing through the center of the explosion; the dashed
	lines mark the same cuts but for spheres with the centers
	coincident with the center of explosion (the cross) and
	radii corresponding to the average radii of the forward and
	reverse shocks.}
  \label{g292}
\end{center}
\end{figure}

The observed spatial distribution of the ejecta is highly inhomogeneous
in G292, with different morphologies for intermediate-mass elements
(as O, Ne, and Mg) on one hand, and for tracers of explosive
nucleosynthesis (most notably Si, S, and Fe) on the other hand. In
particular, O, Ne, and Mg are observed to be distributed within the whole
remnant, whereas Si-rich ejecta are concentrated mainly in the
north-west area (e.g.~\citealt{2002ApJ...564L..39P, 2014RAA....14.1279Y,
2019ApJ...872...31B}). Recently, Fe-rich ejecta have been found
concentrated in Si-rich regions to the north-west
(\citealt{2019ApJ...872...31B}). These properties of the ejecta
distributions resemble those found in our models, with Si- and Fe-rich
shocked ejecta concentrated in the north-west hemisphere where the
high-entropy plumes from the SN explosion mainly interact with the
reverse shock. Fig.~\ref{g292} shows the distributions of the EM of
shocked Fe (red), Si (green), and intermediate-mass elements (O,
Ne, and Mg; blue) integrated along the LoS for model W15-2-cw-IIb-HD+dec.
The map assumes the orientation that roughly matches the direction
of the neutron star kick inferred for G292; the LoS is along the
$y$-axis (see Fig.~\ref{sh_fe_g292}). As expected, the model shows
that most of the shocked Si and Fe are concentrated in the north-west
hemisphere, whereas intermediate-mass elements are, on average,
distributed over the whole remnant area. For asymmetric explosions
with high neutron star kicks, \cite{2013A&A...552A.126W} predicted
such an asymmetry of the spatial distributions of elements including
and heavier than Si, which are expected to be ejected predominantly
in the hemisphere opposite to the kick direction of the neutron
star. In contrast, lighter elements (C, O, Ne, Mg) were found to
exhibit little or no (anti-)correlation with the neutron star kick.

The similarities between the ejecta structure in G292 and our
simulations support the idea that the parent SN of the observed
remnant was characterized by a large-scale asymmetry with characteristics
analogous to those analyzed in our models. The direction of propagation
of the neutron star in G292 reinforces this belief as pointed out
by \cite{2017ApJ...844...84H} and \cite{2018ApJ...856...18K}. In fact, the
neutron star was discovered to travel to the southeast in direction
opposite to the bulk of Fe- and Si-rich ejecta with a kick velocity
of $\approx 440$~km~s$^{-1}$ (\citealt{2003ApJ...591L.139H,
2009ApJ...692.1489W, 2017ApJ...844...84H, 2018ApJ...856...18K,
2019ApJ...872...31B}), slightly smaller than in our simulations
(kick velocity $\approx 600$~km~s$^{-1}$; \citealt{2017ApJ...842...13W}).
Indeed, in our models the bulk of the Fe-group elements and Si are
ejected opposite to the direction in which the CCO is kicked
(\citealt{2017ApJ...842...13W}), a direct consequence of the mechanism
that produced the initial large-scale asymmetry during the SN
explosion (e.g.~\citealt{2000ApJS..127..141N, 2013A&A...552A.126W,
2017hsn..book.1095J}).

Although our models are not appropriate to describe G292 with its
detailed explosion properties, their comparison with observations
of this remnant suggests that the lack of Fe-group elements and
Si in the southeast region is the consequence of an asymmetric SN
explosion likely analogous to that which produced \casa. If true,
this might indicate that, even thousands of years after the SN, the
ejecta structure observed in SNRs might still keep memory of the
fingerprints of the processes driving a neutrino-driven SN explosion.
We note that \cite{2019ApJ...877..136F} have found that the remnants
of type Ia SNe seem to keep memory of the SN explosion for $\approx
500$~years after the explosion, but at that epoch most of the
explosion asymmetries are dominated by the HD instabilities triggered
during the interaction of the ejecta with the reverse shock. Our
simulations show that, in the remnants of core-collapse SNe, the
memory of the explosion asymmetries persist for a longer time than
in the remnants of type Ia SNe; this is consistent with the
observational evidence that the former are systematically much more
asymmetric than the latter (e.g.~\citealt{2011ApJ...732..114L}).

It is worth to mention, however, that the small amount of shocked
Fe observed in G292 (in contrast to the evidence of a large amount
of shocked Fe in our simulations) suggests that the interaction of
the reverse shock with the innermost ejecta (rich of nucleosynthesis
products such as Fe) has started only recently in this remnant
(whereas in our models it started about 30 years after the explosion).
We cannot rule out the possibility that the amount of shocked Fe
(at the age of 3000 years) can be underestimated. For instance, it
is possible that some of the Fe has already cooled too much to
emit detectable radiation, or the quality of the observations did
not allow the Fe to be fully identified yet. In fact, the low
spectral resolution of current CCD detectors strongly limits the
analysis of X-ray emission from ejecta in SNRs and may lead to a
degeneracy between the best-fit values of chemical abundances and
the plasma emission measure (e.g.~\citealt{2020A&A...638A.101G}). In
case the lack of Fe in G292 is confirmed, this may be due to a
weaker asymmetry in its parent SN explosion (with less pronounced
Fe-rich plumes) or it can be a consequence of the lower ratio of
$E\rs{exp}/M\rs{ej}$ of G292 compared to that of our simulations.
The ejecta mass of this remnant is significantly larger than in our
models (and in the case of \casa) and, possibly, its pre-SN progenitor
star had a more extended H envelope. In this latter case,
the metal-rich ejecta might not have been mixed as far outward as
in \casa\ (see, for comparison, the results for model W15-2-cw in
\citealt{2017ApJ...842...13W}) and this might be the reason for the
later interaction of Fe-rich ejecta with the reverse shock. The
comparison between our simulations and the observations of G292
confirms that our models are not suitable to describe the evolution
of this remnant in all of its morphological details but, at the
same time, it suggests that the features observed in the ejecta
structure of G292 might form through processes analogous to those
discussed in this paper.


\section{Discussion and conclusions}
\label{sec:concl}

We presented 3D HD/MHD simulations which describe the formation of
ejecta structure and asymmetries in the remnant of a neutrino-driven
SN explosion. The aim was to explore how the morphology of the
remnant and the structure of the ejecta keep memory of the asymmetries
which developed stochastically by convective overturn in the
neutrino-heating layer and by SASI activity during the first second
after core bounce. To this end, we coupled the model of a neutrino-driven
SN explosion with remarkable resemblance to basic properties of
\casa\ (\citealt{2017ApJ...842...13W}) with HD/MHD simulations that
describe the formation of the remnant (e.g.~\citealt{2016ApJ...822...22O}).
The neutrino-driven SN model was presented and discussed in detail
by \cite{2017ApJ...842...13W}. These authors showed that, in the
considered model, three pronounced Ni-rich fingers grow, naturally
but stochastically, due to HD instabilities. The initial seeds of
these structures originate from the close proximity of the collapsed
core right at the onset of the explosion (see \citealt{2013A&A...552A.126W,
2015A&A...577A..48W}), and their morphology suggests a correspondence
to the extended shock-heated Fe-rich regions observed in \casa.

Our 3D simulations start soon after the breakout of the shock wave
at the stellar surface ($\approx 20$~hours after core-collapse),
using the output of the neutrino-driven SN simulation as initial
condition. We performed SNR simulations either with or without
the heating due to radioactive decay of $^{56}$Ni and $^{56}$Co
(mainly effective during the first year of the evolution of the remnant)
and simulations either with or without an ambient magnetic field.
This specific exploration allowed us to evaluate the role played
by these effects in the evolution of the ejecta and in determining
their structure after the interaction of the reverse shock with
the large-scale asymmetries left from the earliest phases of the
explosion. The simulations covered $\approx 2000$~years of evolution.
We compared the simulation results with the observations of \casa,
at the age of $\approx 350$~years and, tentatively, to the properties
of the $\approx 3000$\,years old SNR G292; at later epochs, we
explored how and to which extent the remnant keeps memory of the
anisotropies that emerged from violent non-radial flows during the
early moments after the core-collapse.

We found that our SNR models predict radii of the forward and reverse
shocks and ejecta velocities at the age of $\approx 350$~years which
are consistent (within the error bars) with those observed in \casa\
(see Fig.~\ref{pos_shock}), assuming a density of the stellar wind
at the current position of the forward shock ($n\rs{w} = 0.8$~cm$^{-3}$)
slightly lower than the best-fit value inferred from X-ray observations,
but well within its uncertainty ($n\rs{w} = 0.9 \pm 0.3$~cm$^{-3}$;
\citealt{2014ApJ...789....7L}).  We note that our models consider
an ejecta mass ($3.3\,M_{\odot}$) in good agreement with the range
of values estimated from the analysis of observations ($M\rs{ej} =
[2-4]\,M_{\odot}$; e.g.~\citealt{2003ApJ...597..347L, 2003ApJ...597..362H,
2006ApJ...640..891Y}), but assume an explosion energy ($E\rs{exp}=1.5$~B)
lower than expected for \casa\ ($\approx 2$~B;
\citealt{2003ApJ...597..347L, 2003ApJ...597..362H, 2020ApJ...893...49S}).
A higher explosion energy would have produced a much better match
of the models with the radii of the forward and reverse shocks and
with the density of the stellar wind. Nevertheless, our models
predict a mass of unshocked ejecta ($\approx 0.85\,M_{\odot}$) and,
more specifically, unshocked Fe ($\approx 0.068\,M_{\odot}$) which
are in excellent agreement with those recently inferred from the
analysis of infrared observations (\citealt{2020arXiv201007718L}).
Furthermore, our models allowed us to address the idea that the
fundamental chemical, physical, and geometric properties observed
in \casa\ can naturally be explained in terms of the processes
associated with the asymmetric beginning of a SN explosion and to
a sequence of subsequent hydrodynamic instabilities that led to
fragmentation and mixing in the ejecta. Our main findings are
summarized as follows.

i) {\em Formation of ring- and crown-like structures.} The initial
high-entropy plumes of Fe- and Ti-rich ejecta cross the reverse
shock at the age of $\approx 30$~years. The slow-down of ejecta
passing through the reverse shock triggers the development of HD
instabilities (i.e., RT instability), which gradually fragment the
plumes into numerous small-scale fingers. In this way, extended
regions of shock-heated Fe form in the main shell of the remnant,
which resemble those observed in \casa. The pattern of small-scale
structures in these regions reflects the ejecta structure left from
the SN explosion.  In fact, core-collapse SN simulations show the
growth of fast plumes of Ni-rich ejecta into extended fingers from
which fast metal-rich clumps can detach
(e.g.~\citealt{2015A&A...577A..48W}). This leads to ejecta characterized
by pronounced clumpiness with small dense clumps and knots mainly
present at the tips of the high-entropy plumes
(\citealt{2017ApJ...842...13W}). The combination of these clumps
and knots with the HD instabilities (that develop after the interaction
of the ejecta with the reverse shock) naturally produces a filamentary
pattern of shocked ejecta with ring-like features (see also
\citealt{2012ApJ...749..156O, 2016ApJ...822...22O}). In many cases,
the ring structures have RT fingers which extend outward, giving
them the appearance of crowns. This effect is more evident in the
case of denser fingers that can survive for a longer time (thus
extending more outward) in the inter-shock region before they get
dispersed due to mixing under the effect of RT and KH instabilities.
In fact, the crown-like appearance is more evident in simulations
including the decay heating (because of a higher density contrast
of the ejecta at the tips of the plumes; see Fig.~\ref{inversion}; see
also \citealt{2012ApJ...749..156O}) and it is also more evident
when the effects of an ambient magnetic field are taken into account
(because the field limits the growth of KH shear instability at the
border of RT fingers; \citealt{1994ApJ...433..757M, 2005ApJ...619..327F,
2008ApJ...678..274O}). The resulting pattern of shocked ejecta in
our models resembles that observed in \casa\ remarkably well 
(\citealt{2010ApJ...725.2038D, 2013ApJ...772..134M}).

ii) {\em Inversion of ejecta layers.} Our models show that the initial
large-scale plumes also determine the spatial inversion of ejecta
layers, leading to Fe-rich ejecta being located at greater radii than
Si-rich ejecta. In fact, the inversion is evident only in regions
where the fast plumes of Fe-rich ejecta interact with the reverse
shock; elsewhere, the original chemical stratification is roughly
preserved (see Fig.~\ref{inversion}). The simulations show that,
already at the breakout of the shock wave from the stellar surface,
the fast plumes of Ni-rich ejecta led to a global metal asymmetry
characterized by a deep penetration of Ni fingers into the overlying
layers of the ejecta. After the decay of $^{56}$Ni to $^{56}$Co and of
the latter to $^{56}$Fe, the Fe-rich plumes have protruded into
the chemically distinct layers above and are enveloped by a less
dense medium of Si-rich material. When the plumes start to interact
with the reverse shock, the denser Fe-rich ejecta push the less
dense Si-rich layer outward, breaking through it, and leading to
the spatial inversion of the ejecta layers (see Fig.~\ref{inversion}).
The radioactive decay of Ni and Co enhances the effect of inversion
of the ejecta layers. The additional pressure due to the decay
heating inflates the Fe-rich fingers outward and sideways, making
their outermost tips denser and more efficient in piercing the Si-rich layer.
Another consequence of the dynamics of the fast and dense Fe-rich
plumes is the accumulation of swept-out material of the outer layers
(in particular Si) sideways around the shocked Fe regions (see
Fig.~\ref{rings}). As a result, Fe-rich shocked regions are circled
by rings of Si-rich shocked ejecta (see Figs.~\ref{sh_fe_si_casa},
\ref{sh_fe_si_HD_casa}, and \ref{sh_fe_si_MHD+dec_casa}). All these
modeled features, namely the inversion of ejecta layers and the
formation of Si-rich rings encircling the Fe-rich regions, nicely
match those observed in \casa\ (e.g.~\citealt{2000ApJ...528L.109H,
2010ApJ...725.2038D, 2013ApJ...772..134M}), suggesting that they
are the fingerprints of the asymmetries that developed in the earliest
phases of the SN explosion.

iii) {\em The ``thick-disk'' appearance.} Discussing the orientation
of the remnant structures compared to \casa\ observations, we define
the LoS such that the three dominant Fe-fingers coincide roughly
with the directions where shocked Fe is seen in \casa. In this
case the kick motion of the CCO points to the southern hemisphere
and toward the observer, in agreement with conclusions drawn from
the distribution of $^{44}$Ti by \cite{2017ApJ...834...19G}.  The
simulations show that, in all cases, the metal-rich ejecta are
arranged in a ``thick-disk'' geometry, with the disk tilted by an
angle of $\approx 30^{\rm o}$ with respect to the plane of the sky
(see Fig.~\ref{sh_fe_si_casa}).  This is the natural consequence
of the highly asymmetric explosion which ejected three dominant
fast high-entropy plumes of metal-rich ejecta roughly lying in a
plane tilted by $\approx 30^{\rm o}$. The fast expansion of the
three plumes makes the Si-, Ti-, and Fe-rich ejecta more expanded
parallel to this plane than perpendicular to it, giving the metal-rich
ejecta the appearance of a thick disk. Interestingly, also the
ejecta of \casa\ seem to be arranged in a thick disk which is tilted
by an angle of $\approx 30^{\rm o}$ with respect to the plane of
the sky (e.g~\citealt{2010ApJ...725.2038D, 2017ApJ...834...19G}).
Thus, the same initial large-scale asymmetry which produces the
Fe-rich regions as in \casa, the pattern of ring- and crown-like
structures of the shocked ejecta, and the spatial inversion of ejecta
layers as observed in \casa, also predicts a thick-disk distribution
of metal-rich ejecta with the same orientation as that observed in
\casa. When the decay heating is included, the tilted thick-disk
appearance of Fe- and Ti-rich ejecta is slightly less evident.
However, in this context it is worth noting that, in our models,
we assumed that all of the decay energy is deposited in the ejecta
without any $\gamma$-ray leakage from the inner part of the remnant.
Our models, most likely, overestimate the effects of decay heating.

iv) {\em Formation of cavities and voids.} The distribution of unshocked
ejecta is characterized by cavities and voids. These are particularly
evident in Si, Ti, and Fe. The largest cavities of Si correspond
to the directions of propagation of the dominant Fe-rich plumes of
ejecta. As a result, these cavities
are located immediately below the regions of shocked Fe in the main
shell of the remnant as evident by comparing Figs.~\ref{sh_fe_si_casa}
and \ref{unsh_fe_si} (see also Movie~2). Furthermore, the large
cavities are physically connected with the large-scale Si-rich rings
visible in the shocked ejecta and encircling the Fe-rich regions.
As mentioned above, Si-rich layers of ejecta envelope the
Fe-rich plumes. As one of these plumes passes through the reverse
shock, a Fe-rich region encircled by Si-rich ejecta from the layer
(both shock-heated and, therefore, visible in different bands) forms
exterior to the shock, whereas a cavity in the unshocked Si (visible
in the radio and near-infrared bands), which is filled by unshocked
Fe (not visible), forms interior to the shock. This result naturally
explains why the cavities in Si-rich ejecta observed in near-infrared
observations of \casa\ are physically connected to the bright rings
in the main-shell (e.g.~\citealt{2015Sci...347..526M}). As suggested
by \cite{2016ApJ...822...22O}, the cavities are actually filled by
low-density unshocked Fe, which is not visible until it interacts
with the reverse shock. In models including the decay heating, the
cavities are more extended due to the inflation of the Fe-rich
regions driven by the radioactive decay heating of the initial
Ni. This effect enhances the size of the cavities and the
formation of the large-scale Si-rich rings encircling the Fe-rich
regions.

v) {\em Distributions of $^{44}$Ti versus $^{56}$Fe.} Our simulations
show that, on average, Fe and Ti coexist in the mass-filled
volume, according to the fact that these two species were synthesized
in regions of Si burning and of processes like $\alpha$-particle-rich
freezeout, and later mixing cannot lead to any separation\footnote{Our
models do not consider any mechanisms able to decouple Fe from Ti.}.
Nevertheless, we found that their abundance ratio can have significant
variations (already at the initial conditions soon after the shock
breakout at the stellar surface), with regions with a high concentration
of Fe and others with a relatively high concentration of Ti.
In particular, we found that the tips of the initial high-entropy
plumes have, on average, a higher concentration of Fe. As a result,
at the age of \casa, the mass distributions of shocked and unshocked
ejecta indicate that the former have, on average, a higher concentration
of Fe, whereas the latter have, on average, a relatively higher
concentration of Ti (see Fig.~\ref{fe_vs_ti-si_snr}).  The mass
fraction ratio derived at the age of \casa\ can vary by a factor
of 10 around its average value (see Fig.~\ref{fe_vs_ti}).  These
results are fully consistent with the findings of
\cite{2017ApJ...834...19G} who found Ti-rich ejecta located both
exterior and interior to the reverse shock from the analysis of
NuSTAR observations. As in \casa, in our models the bulk of shocked
Ti is located in regions of shocked Fe, although there are also
regions of shocked Fe with a low concentration of Ti.
\cite{2017ApJ...834...19G} claimed that Ti could not be found
in Fe-forming regions if its abundance there is a factor of $\approx
2$ lower than in the main Ti clumps. In fact, our models show that,
at the center of regions of shocked Fe, the concentration of
Fe is slightly higher and the concentration of Ti lower than on average
even by factors larger than 2. According to our models, a significant
amount of unshocked Fe is interior to the reverse shock,
preferentially concentrated in highly enriched clumps, which roughly
form an irregular shell near the reverse shock, enclosing regions
of more diluted Fe and Ti. We note that this unshocked Fe cannot
be detected in \casa\ until it is heated by the reverse shock. So,
it is not astonishing that Ti is detected (emitting X-rays due to
radioactive decay) in unshocked regions but not Fe. We found that
the decay heating of Ni and Co does not affect the abundance ratio
of Fe and Ti but can modify their spatial distributions. Regions
with a high concentration of Ti (and a relatively lower concentration
of Fe) do not expand as fast as regions with a higher Ni/Fe
concentration, the latter expanding faster because of the additional
pressure of the decay heating. In such a way, most of the Fe is
pushed into the irregular shell of unshocked Fe close to the reverse
shock, whereas regions with a high concentration of Ti remain
(diluted) in the innermost part of the remnant. As a result,
Ti-enriched regions appear more clearly offset from volumes with
high Fe abundance in the case of effective decay heating.

vi) {\em Reverse shock asymmetry.} The simulations show that the
forward and reverse shocks deviate from a spherically symmetric
expansion around the center of the explosion. Both shocks appear
still roughly spherical (with small-scale deformations), but with
an offset to the north-west by $\approx 0.13$~pc from the center
of the explosion. This offset is the consequence of the initial,
asymmetric explosion, in which most of the Ni and Ti were ejected
in the northern hemisphere away from the observer. Interestingly,
observations of \casa\ show that the geometric center of the reverse
shock is offset to the north-west by $\approx 0.2$~pc (at the
distance of $3.4$~kpc) from the center of the forward shock
(\citealt{2001ApJ...552L..39G}). Our simulations cannot reproduce
the offset between the two shocks, but suggest that, in general,
the geometric center of the forward shock does not coincide with
the center of the explosion. A similar offset can be also introduced
if the remnant expands through a medium characterized by a global
anisotropy (see upper left panel in Fig. 2 in
\citealt{2007A&A...470..927O}).  A direct consequence of this offset
is that some caution should be applied when assuming the geometric
center of the forward shock as the center of the explosion to
estimate the kick velocity of a CCO. For instance, in the case of
\casa, the CCO travels in the southern hemisphere toward the observer,
opposite to the offset of the forward and reverse shocks from the
center of the explosion. Thus, assuming the geometric center of the
forward shock as the center of the explosion lead to an overestimation
of the kick velocity of the CCO. The velocities of the forward and
reverse shocks in our models are in nice agreement with those
inferred from the observations in the eastern and northern hemisphere
of the remnant (e.g.~\citealt{2018ApJ...853...46S, 2019sros.confE..32F}).
However, our simulations are not able to reproduce a reverse shock
which appears to be stationary or even inward moving in the observer
frame as suggested for \casa\ by some authors
(e.g.~\citealt{2008ApJ...686.1094H, 2018ApJ...853...46S}). The
origin of this asymmetry is still debated in the literature and it
is not clear if this is also connected with the offset between the
centers of the forward and reverse shocks. A significant role
can be played by environment asymmetries in the stellar wind (Orlando
et al. in preparation).

vii) {\em The persistence of the explosion fingerprints at later
times.} Our simulations covered $\approx 2000$~years of evolution.
During this time, most of the metal-rich ejecta have already crossed
through the reverse shock and were subject to strong mixing by the
HD instabilities. Nevertheless, the simulations show that the spatial
distributions of Fe-group elements, Ti (or its decay
product $^{44}$Ca), and Si still keep memory of the original SN
asymmetry. Cavities and voids in the unshocked ejecta are still
present even though they are much more diluted than in the epoch of
\casa. The fingerprints of the explosion are also evident in the
asymmetric distributions of the velocity along the LoS for the various
species. Even if our simulations were not tuned to the case of SNR
G292, we compared our model results with multi-wavelength observations
of this remnant and found that some of the observed asymmetries may
be interpreted in terms of large-scale asymmetries left from the
earliest phases of the explosion, analogously to what we have found
for \casa. An accurate study of G292, however, would require dedicated
simulations with ejecta mass, explosion energy, and structure of
the ambient medium appropriate for this remnant.

\bigskip
In conclusion, our study has shown that the interaction of the
reverse shock with the post-explosion large-scale asymmetries of
the SN is fundamental for the formation of the structures observed in
the inter-shock region of the remnant. The main aspects characterizing
the shocked ejecta morphology of \casa\ (Fe-rich regions, ring- and
crown-like features, inversion of ejecta layers, etc.) form during
the passage of the reverse shock from the environmental interaction,
as a consequence of the fast growth of RT fingers in the reverse-shock
heated material. The thick-disk appearance of the ejecta, the
``bubble-like'' structure of the unshocked ejecta, the distribution
of Fe-group elements and Ti interior and exterior to the reverse
shock, the asymmetries of the forward and reverse shocks, all these
features encode the fingerprints of a neutrino-driven SN explosion.

However, a crucial aspect of the \casa\ morphology was not accounted
for by our simulations: the development of two Si-rich wide-angle
jet-like features, visible in most of the wavelength bands. Some
authors have claimed that these features might originate in magnetic
jet-driven SNe (e.g.~\citealt{1999ApJ...524L.107K, 2003ApJ...598.1163M,
2009ApJ...696..953C, 2017RAA....17..113S, 2017MNRAS.468.1226G,
2018MNRAS.478..682B}). In the case of \casa, we suggest that they
might originate from a post-explosion phenomenon that is not included
in our simulations. For instance, an accretion disk of fallback
matter could well have formed around the new-born neutron star
(e.g.~\citealt{1989ApJ...346..847C, 2011ApJ...736..108P}). This
possibility is supported by the inferred surface composition of the
neutron star associated to \casa, which is rich in carbon
(\citealt{2009Natur.462...71H}); a surface composition rich of light
elements suggests a significant accretion after neutron-star
formation.  In this case, the fallback matter might have produced
the jets by a mechanism similar to that responsible for the formation
of jets in pulsars (e.g.~\citealt{2004ApJ...601L..71B}) or the early
afterglows of $\gamma$-ray bursts (e.g.~\citealt{2012ApJ...759...58D}).
In this context it is important to note that there are several
$10^{-2}\,M_{\odot}$ of fallback in the SN model with plenty of
angular momentum above the disk formation limit.

New simulations accounting for these additional phenomena are,
therefore, necessary to shed light on the still pending questions
of the origin of the many features observed in \casa\ and in other
SNRs. Finally, the analysis of multi-wavelength observations aimed
at reconstructing the 3D chemical distribution and structure of
stellar debris in SNRs (e.g. \citealt{2010ApJ...725.2038D,
2013ApJ...772..134M, 2014Natur.506..339G, 2015Sci...347..526M,
2017ApJ...834...19G, 2017ApJ...842L..24A, 2019ApJ...886...51C,
2020ApJ...894...73L}) and the comparison of model results with
observations are essential steps to constrain the models and,
therefore, to advance our understanding of the physical processes
associated with SNe.

\begin{acknowledgements}

We thank the anonymous referee for useful suggestions that have
allowed us to improve the paper. SO is grateful to A. Mignone for
his support with the PLUTO code and his advises to consider a code
configuration that minimizes the numerical diffusivity.  The PLUTO
code is developed at the Turin Astronomical Observatory (Italy) in
collaboration with the Department of General Physics of Turin
University (Italy) and the SCAI Department of CINECA (Italy).  SO,
MM, FB acknowledge financial contribution from the INAF mainstream
program and from the agreement ASI-INAF n.2017-14-H.O. We acknowledge
the ``Accordo Quadro INAF-CINECA (2017)'' and CINECA ISCRA Award
N.HP10BARP6Y for the availability of high performance computing
resources and support at the infrastructure Marconi based in Italy
at CINECA. Additional computations were carried out at the SCAN
(Sistema di Calcolo per l'Astrofisica Numerica) facility for high
performance computing at INAF-Osservatorio Astronomico di Palermo.
At Garching, funding by the European Research Council through grant
ERC-AdG no. 341157-COCO2CASA and by the Deutsche Forschungsgemeinschaft
(DFG, German Research Foundation) through Sonderforschungsbereich
(Collaborative Research Centre) SFB-1258 ``Neutrinos and Dark Matter
in Astro- and Particle Physics (NDM)'' and under Germany’s Excellence
Strategy through Cluster of Excellence ORIGINS (EXC-2094)-390783311
is acknowledged. Computer resources for this project have been
provided by the Max Planck Computing and Data Facility (MPCDF) on
the HPC systems Cobra and Draco. S.N. is partially supported by
``JSPS Grants-in-Aid for Scientific Research <KAKENHI> (A) 19H00693'',
``Pioneering Program of RIKEN for Evolution of Matter in the Universe
(r-EMU)'', and ``Interdisciplinary Theoretical and Mathematical
Sciences Program of RIKEN''. The navigable 3D graphics have
been developed in the framework of the project 3DMAP-VR (3-Dimensional
Modeling of Astrophysical Phenomena in Virtual Reality;
\citealt{2019RNAAS...3..176O}) at INAF-Osservatorio Astronomico di
Palermo.

\end{acknowledgements}

\appendix 

\section{Effects of radioactive decay and magnetic field on the mass
distributions in velocity space}

\label{app:vel_distr}

\begin{figure}[!t]
  \begin{center}
    \leavevmode
        \epsfig{file=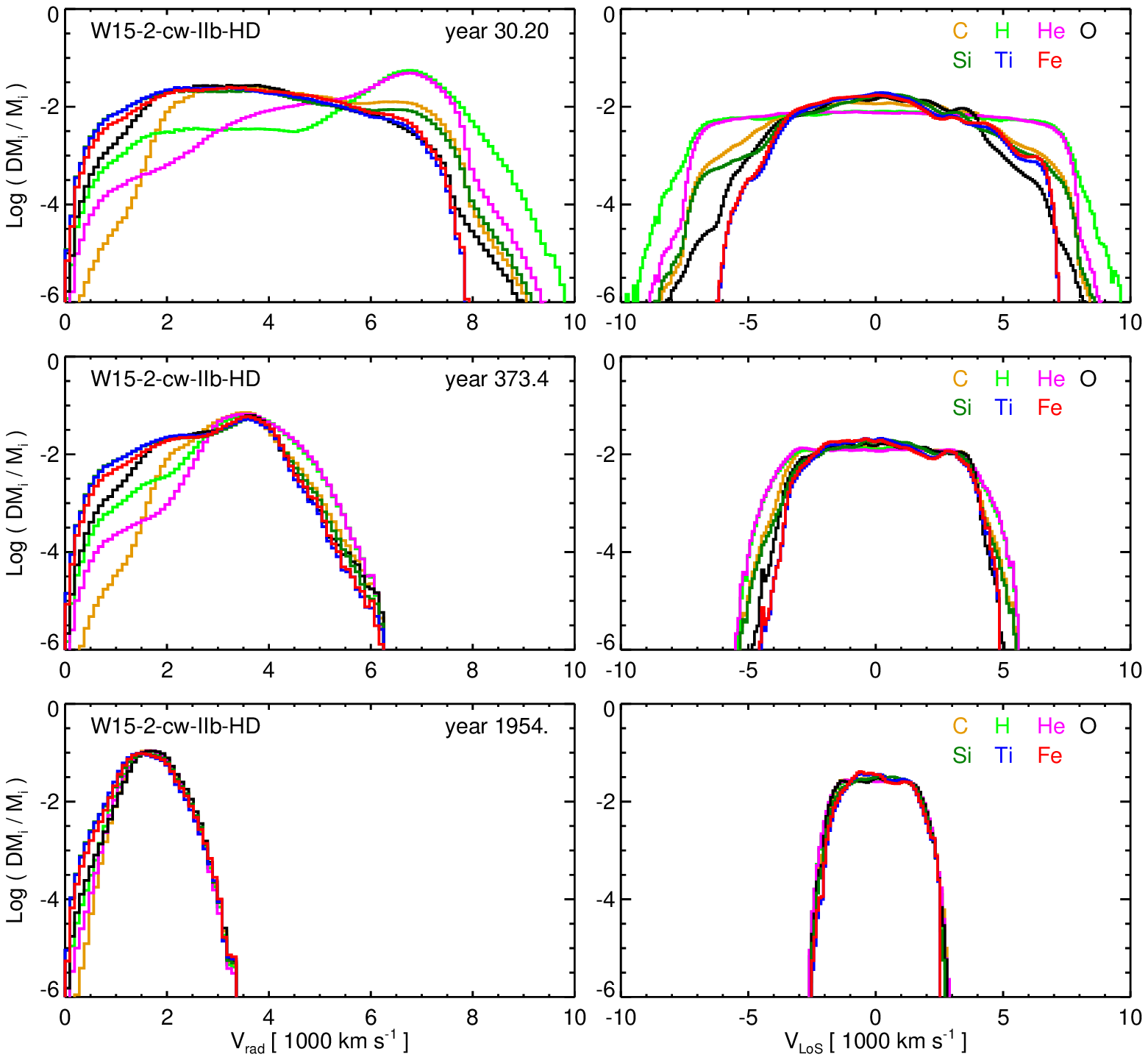, width=9cm}
	\caption{As in Fig.~\ref{prof_vel_hd_dec} but for model
	W15-2-cw-IIb-HD.}
  \label{prof_vel_hd}
\end{center} \end{figure}

The effects of radioactive decay on the overall remnant evolution
can be investigated by comparing the mass distributions of elements
versus the radial and LoS velocities derived from models either
with (model W15-2-cw-IIb-HD+dec) or without (model W15-2-cw-IIb-HD)
decay heating. Figure~\ref{prof_vel_hd} shows these distributions
at three epochs for model W15-2-cw-IIb-HD. The comparison
of this figure with the analogous one derived from model
W15-2-cw-IIb-HD+dec (Fig.~\ref{prof_vel_hd_dec}) reveals that the
fraction of Fe and Ti populating the high-velocity tail of
their distributions is the highest if the decay heating is taken into
account (compare left panels in Figs.~\ref{prof_vel_hd} and
\ref{prof_vel_hd_dec}). Also, in model W15-2-cw-IIb-HD+dec, the
tails of the fastest Fe and Ti stretch out the distributions to
slightly higher velocities than in model W15-2-cw-IIb-HD.

The effects of decay heating are more evident in ejecta propagating
away from us than in those propagating toward the observer. An
inspection of the mass distributions versus the LoS velocities
(right panels in Figs.~\ref{prof_vel_hd} and \ref{prof_vel_hd_dec})
shows that the main differences between models W15-2-cw-IIb-HD
and W15-2-cw-IIb-HD+dec are concentrated in the redshifted part of
the distributions. In fact the heating by radioactive decay provides
an additional pressure source which inflates regions dominated by the
decaying elements against their surroundings. As a result, the
ejecta in the inflated regions expand faster than if the decay was
not taken into account. The effect of decay are visible till the
end of the simulations at the age of $\approx 2000$~years.

\begin{figure}[!t]
  \begin{center}
    \leavevmode
        \epsfig{file=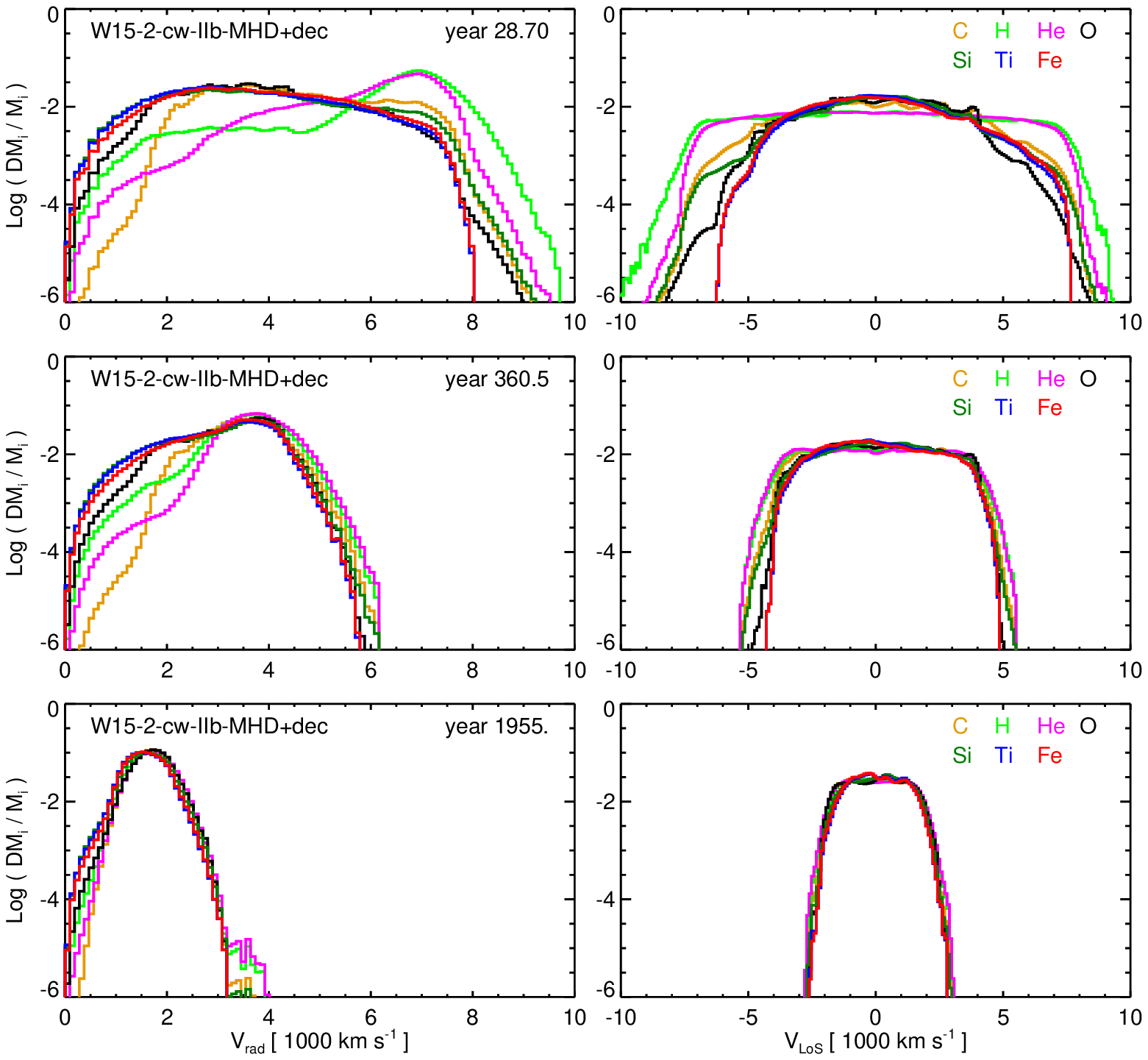, width=9cm}
	\caption{As in Fig.~\ref{prof_vel_hd_dec} but for model
	W15-2-cw-IIb-MHD+dec.}
  \label{prof_vel_mhd_dec}
\end{center} \end{figure}

The effects of an ambient magnetic field on the remnant evolution
can be investigated by comparing the mass distributions of elements
derived from models either with (model W15-2-cw-IIb-MHD+dec) or
without (model W15-2-cw-IIb-HD+dec) an ambient magnetic field. The
comparison between Figs.~\ref{prof_vel_hd_dec} and \ref{prof_vel_mhd_dec}
shows that the distributions in models W15-2-cw-IIb-HD+dec and
W15-2-cw-IIb-MHD+dec are very similar with no relevant differences,
indicating that the effects of magnetic field are negligible for
the overall remnant evolution.

\section{Effects of radioactive decay and magnetic field on the
spatial distribution of metal-rich ejecta}
\label{app:ej_distr}

\begin{figure*}[!t]
  \begin{center}
    \leavevmode
        \epsfig{file=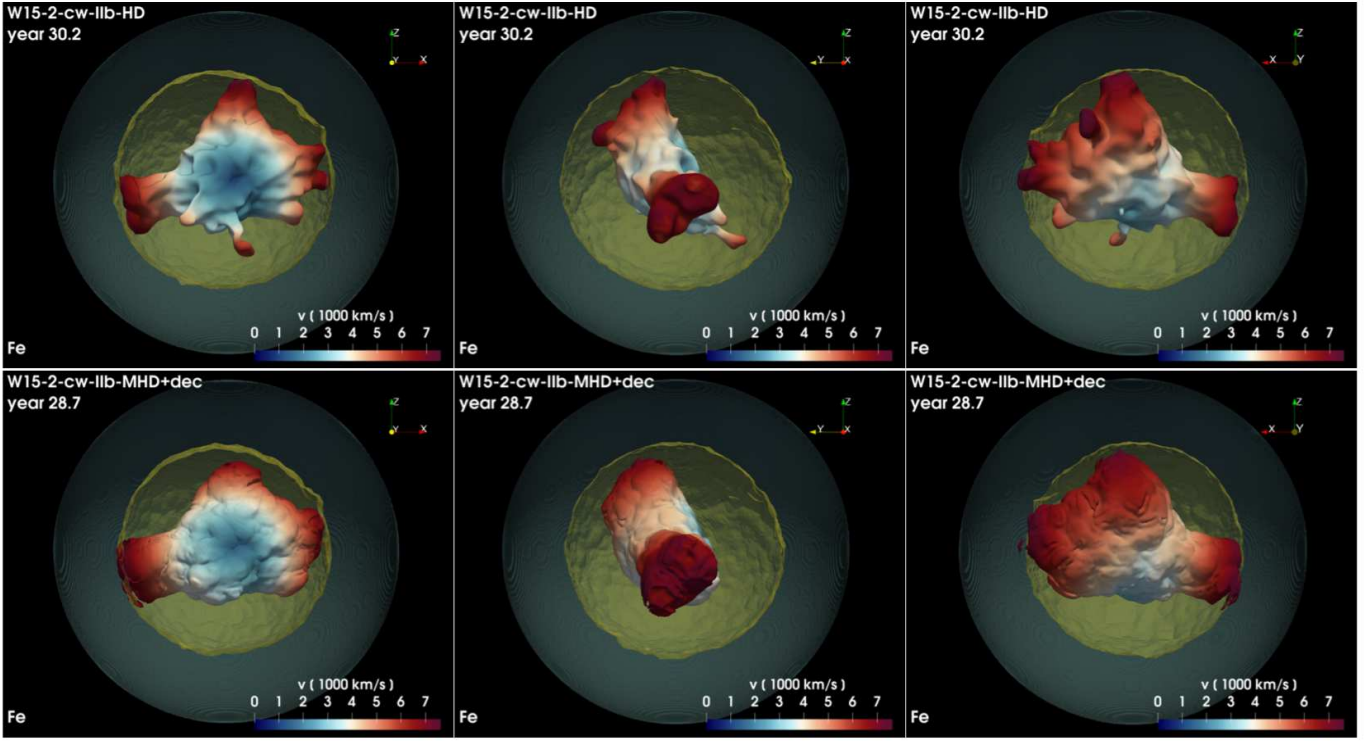, width=18cm}
	\caption{As the upper panels in Fig.~\ref{distr_fe_time}, but 
	for models W15-2-cw-IIb-HD (upper panels) and W15-2-cw-IIb-MHD+dec
	(lower panels).}
  \label{distr_fe_early}
\end{center}
\end{figure*}

We investigated the effects of radioactive decay on spatial
distribution of metal-rich ejecta by comparing model W15-2-cw-IIb-HD
(without decay heating) with model W15-2-cw-IIb-HD+dec (including
decay heating). Fig.~\ref{distr_fe_early} shows the spatial
distribution of Fe at the time when the reverse shock reached the
regions of Fe- and Ti-rich ejecta, about 30 years after the SN
event, for models W15-2-cw-IIb-HD (upper panels) and W15-2-cw-IIb-MHD+dec
(lower panels). In this phase of evolution the Fe-rich plumes were
not perturbed by the interaction with the reverse shock and the
ejecta expanded almost homologously. However, by comparing
the Fe distribution derived from model W15-2-cw-IIb-HD (upper
panel in Fig.~\ref{distr_fe_early}) with those derived from the
other two models (upper panel in Fig.~\ref{distr_fe_time} and lower
panels in Fig.~\ref{distr_fe_early}), we note that, in the latter
two cases, the Fe-rich ejecta appear to be more swollen than in
model W15-2-cw-IIb-HD (in which the Fe distribution is almost the
same as in the initial condition). This difference is due to heating
by radioactive decay of $^{56}$Ni and $^{56}$Co (species which
populate the post-explosion instability-driven structures) that
provide an additional gas pressure during the first year of evolution
which inflates the Fe-rich plumes of ejecta.

\begin{figure}[!t]
  \begin{center}
    \leavevmode
        \epsfig{file=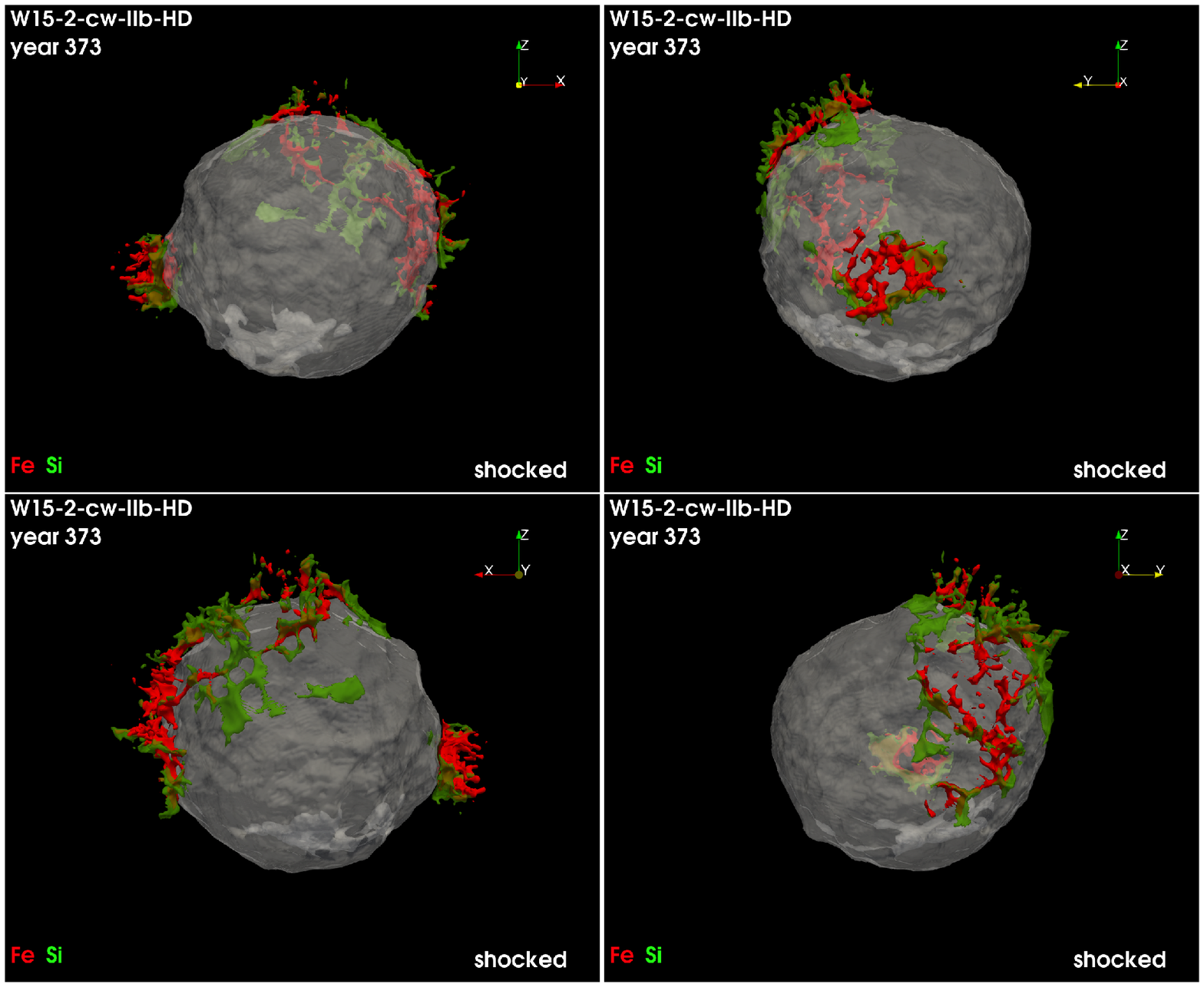, width=9cm}
	\caption{As in Fig.~\ref{sh_fe_si_casa} but for model
	W15-2-cw-IIb-HD.}
  \label{sh_fe_si_HD_casa}
\end{center} \end{figure}

\begin{figure}[!t]
  \begin{center}
    \leavevmode
        \epsfig{file=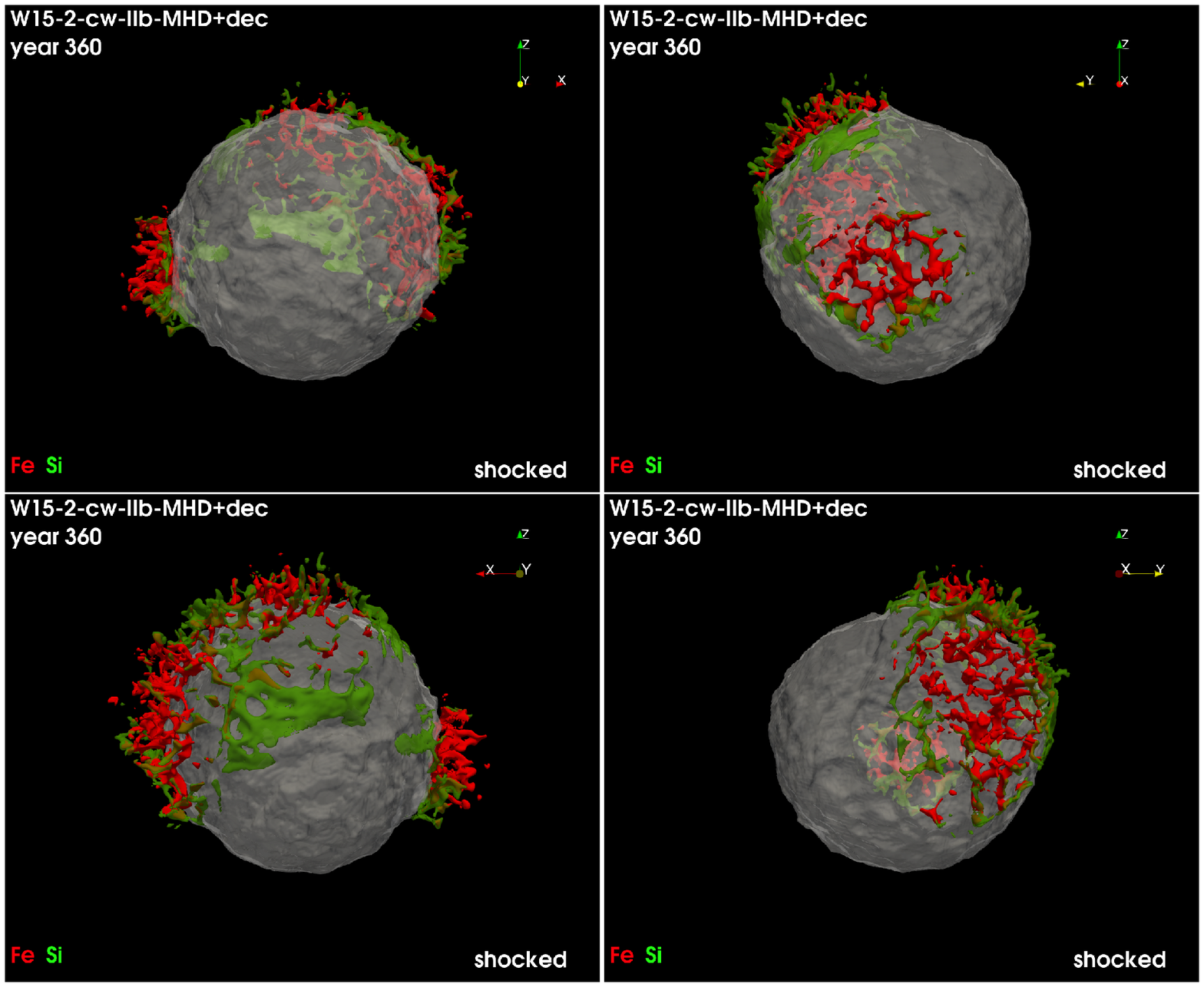, width=9cm}
	\caption{As in Fig.~\ref{sh_fe_si_casa} but for model
	W15-2-cw-IIb-MHD+dec.}
  \label{sh_fe_si_MHD+dec_casa}
\end{center}
\end{figure}

At later times, the ejecta distribution keeps memory of the decay
heating occurred during the first year of evolution. At the age of
\casa, the amount of shock-heated Fe is slightly higher if the
decay heating is taken into account (see Fig.~\ref{shocked_mass})
due to the additional boost to the ejecta expansion provided by the
heating. Furthermore the ring- and crown-like features of
shocked ejecta are more developed in models W15-2-cw-IIb-HD+dec and
W15-2-cw-IIb-MHD+dec than in model W15-2-cw-IIb-HD (compare
Figs.~\ref{sh_fe_si_casa} and \ref{sh_fe_si_MHD+dec_casa} with
Fig.~\ref{sh_fe_si_HD_casa}). In fact, the decay heating increases
the density contrast of Fe-rich fingers before they start to interact
with the reverse shock. As a result they can penetrate more efficiently
in the mixing region between the reverse and forward shocks (see
also \citealt{2012ApJ...749..156O, 2013MNRAS.430.2864M,
2016ApJ...822...22O, 2020A&A...642A..67T}).

\begin{figure*}[!t]
  \begin{center}
    \leavevmode
        \epsfig{file=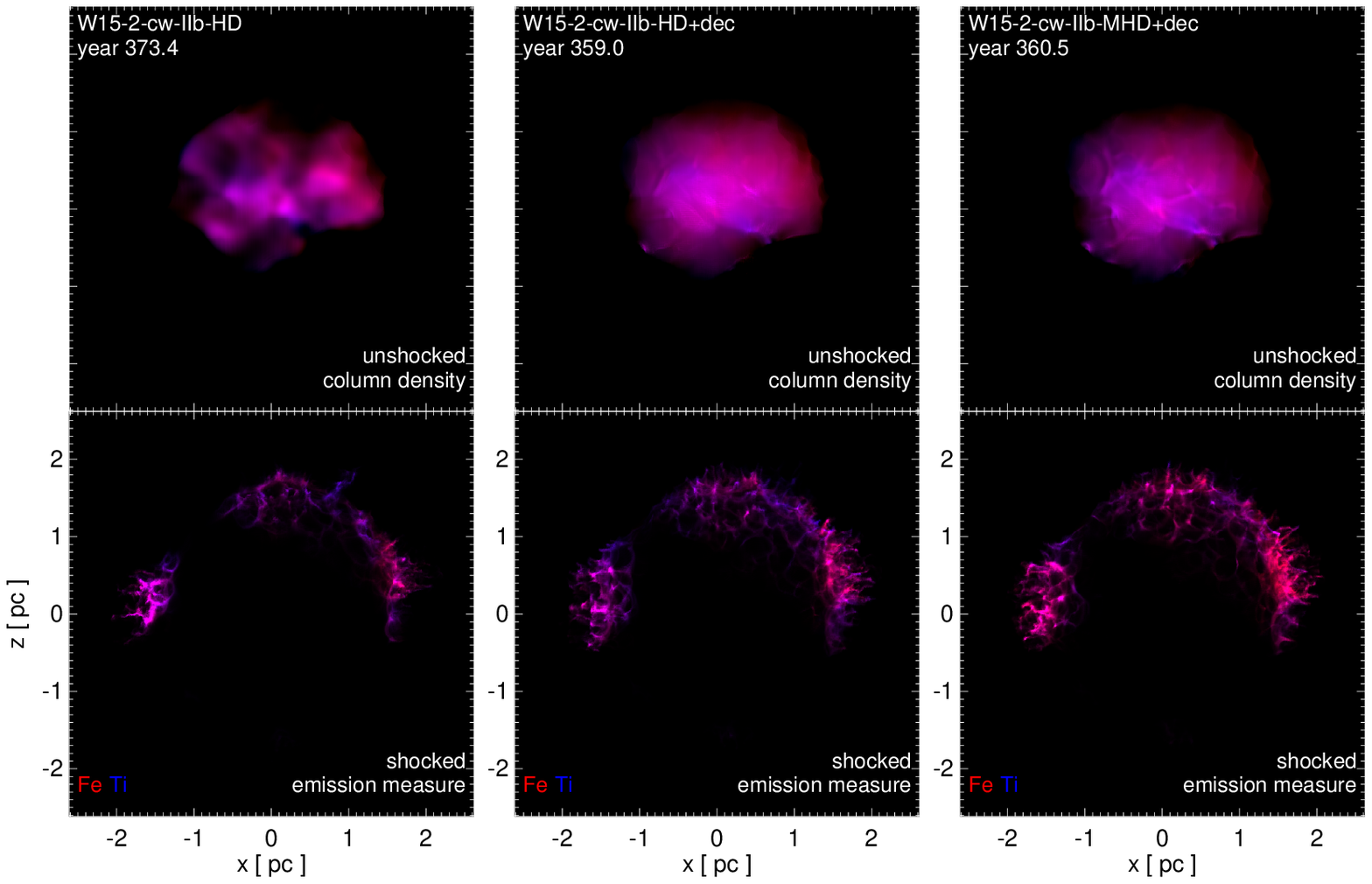, width=19cm}
	\caption{Upper panels: column density maps of unshocked Fe
	(red) and Ti (blue) at the age of \casa\ for our three
	models. Lower panels: corresponding EM distributions of
	shocked Fe (red) and Ti (blue) integrated along the $y$-axis
	(i.e. the LoS assuming the vantage point at Earth).}
  \label{maps_em_nh}
\end{center}
\end{figure*}

The shock-heated Fe-rich structures appear to be slightly more
defined and extended in model W15-2-cw-IIb-MHD+dec than in
W15-2-cw-IIb-HD+dec (compare Fig.~\ref{sh_fe_si_MHD+dec_casa} with
Fig.~\ref{sh_fe_si_casa}). In the former, the magnetic field limits
the growth of HD instabilities that would develop at the borders
of shocked clumps of ejecta through the tension of field lines which
maintain a more laminar flow around the clump borders
(e.g.~\citealt{1994ApJ...433..757M, 2005ApJ...619..327F,
2008ApJ...678..274O}). As a result, the magnetic field limits the
fragmentation by HD instabilities of the ring- and crown-like
features of ejecta and makes them to survive for a longer time than
those in model W15-2-cw-IIb-HD+dec, thus increasing their extension
toward the forward shock (e.g.~\citealt{2012ApJ...749..156O}).

\section{Column density and emission measure distributions of
$^{44}$Ti and $^{56}$Fe}
\label{app:nh_em}

Fig.~\ref{maps_em_nh} displays the column density maps of unshocked
$^{56}$Fe and $^{44}$Ti (upper panels) and the corresponding
distributions of EM of shocked species integrated along the LoS
(lower panels) at the age of \casa\ for our three models. The maps
of shocked and unshocked ejecta are clearly dominated by Fe which,
as expected, is more abundant than Ti. We note, however, that
some regions show a prominent blue color both in shocked and unshocked
ejecta, thus indicating that, along the LoS, either the EM of shocked
Ti integrated along the LoS or the column density of unshocked Ti
dominates over the corresponding distribution of Fe. This is
particularly evident in the EM distributions of all models. In
general, Ti-rich regions are located at the periphery of Fe-rich
regions. There it would be easier to detect shock-heated Ti
in high-energy observations. This is in agreement with NuSTAR
observations that detected $^{44}$Ti exterior to the reverse shock
in regions where shock-heated Fe is also detected
(e.g.~\citealt{2017ApJ...834...19G}). Conversely, the cores of
Fe-rich regions of shocked ejecta have a predominant red color,
indicating that there the abundance of Fe is largely dominant over
Ti and this may explain the evidence of regions where Fe
but not $^{44}$Ti is detected in NuSTAR observations
(e.g.~\citealt{2017ApJ...834...19G}).

We also note that the structure of shock-heated Fe and Ti is the
richest in models including the decay heating, because of the higher
density contrast of Fe-rich plumes when they start to interact with
the reverse shock (see Fig.~\ref{inversion}). The heating is
also responsible for the differences in the maps of column density
of unshocked Fe and Ti: they appear much smoother in models with
radioactive decay that in the model without. More bluish regions
are evident in model W15-2-cw-IIb-HD, probably, because of the more
clumpy structure of unshocked ejecta that allows to intercept regions
with a higher concentration of $^{44}$Ti more easily along the LoS.
We also note that  model W15-2-cw-IIb-MHD+dec shows the highest
abundance of Fe in the cores of Fe-rich regions of shocked ejecta
(the regions appear more reddish). This is due, most likely, to a
less mixing between layers of different chemical composition due
to the presence of the magnetic field which limits the growth of
HD instabilities which are responsible for the mixing (see Appendices
\ref{app:vel_distr} and \ref{app:ej_distr} for the effects of decay
heating and magnetic field on the evolution and structure of the
ejecta).

\bibliographystyle{aa}
\bibliography{references}

\end{document}